\documentclass[useAMS,usenatbib]{mn2e}
\usepackage{epsfig}
\usepackage{amsmath}
\usepackage{amssymb}
\usepackage{ifthen}
\usepackage{txfonts}
\usepackage{rotating}
\usepackage{url}
\usepackage{varioref}
\usepackage{verbatim}
\usepackage{latexsym}
\DeclareSymbolFont{cmletters}{OML}{cmm}{m}{it}
\DeclareMathSymbol{v}{\mathord}{cmletters}{"76}

\usepackage{subfigure}
\usepackage{ifthen}
\usepackage[usenames,dvipsnames]{color}
\usepackage{graphicx}

\usepackage{hyperref}

\newcommand\apjl{\rmfamily{ApJ}}%
\newcommand\apjs{\rmfamily{ApJS}}%
\newcommand\ao{\rmfamily{Appl.~Opt.}}%
\newcommand\aap{\rmfamily{A\&A}}%
\newcommand{\alf}{Alfv\'en }
\newcommand{\xc}{{x^{(2)}}}

\defcitealias{bz77}{BZ77}

\newcommand{\emath}{{\rm e}}
\newcommand{\eff}{{\tilde{e}}}

\newcommand{\detg}{{\sqrt{-g}}}

\newcommand{\cut}[1]{\hbox{}}

\title[Simulations of Magnetized Disks Around Black Holes]
{Simulations of Magnetized Disks Around Black Holes: Effects of Black Hole Spin, Disk Thickness, and Magnetic Field Geometry}

\author[R.~F. Penna, J.~C. McKinney, R. Narayan, A. Tchekhovskoy, R. Shafee \& J.~E. McClintock]
{%
Robert F. Penna,$^1$\thanks{\hbox{E-mail: penna@cfa.harvard.edu~(RFP);} \hbox{jmckinne@stanford.edu~(JCM);} \hbox{rnarayan@cfa.harvard.edu~(RN);} \hbox{atchekho@cfa.harvard.edu~(AT);} \hbox{shafee@fas.harvard.edu~(RS);} \hbox{jmcclintock@cfa.harvard.edu~(JEM)}}
Jonathan C. McKinney$^2$$^3$\footnotemark[1],
Ramesh Narayan$^1$\footnotemark[1],
  \newauthor 
Alexander Tchekhovskoy$^1$\footnotemark[1],
Rebecca Shafee$^4$\footnotemark[1],
Jeffrey E. McClintock$^1$\footnotemark[1] \\
  $^1$Harvard-Smithsonian Center for Astrophysics, 60 Garden Street, Cambridge, MA 02138, USA \\
  $^2$Department of Physics and Kavli Institute for Particle Astrophysics and Cosmology, Stanford University, Stanford, CA 94305-4060, USA \\
  $^3$Chandra Fellow \\
  $^4$Center for Brain Science, Harvard University, 52 Oxford Street, Cambridge, MA 02138, USA \\
}

\begin{document}
\date{Accepted 2010 June 9.  Received 2010 May 18; in original form 2010 March 4.}
\pagerange{\pageref{firstpage}--\pageref{lastpage}} \pubyear{2010}
\maketitle

\label{firstpage}

\begin{abstract}

The standard general relativistic model of a razor-thin accretion disk
around a black hole, developed by Novikov \& Thorne (NT) in 1973,
assumes the shear stress vanishes at the radius of the innermost
stable circular orbit (ISCO) and that, outside the ISCO, the shear
stress is produced by an effective turbulent viscosity.  However,
astrophysical accretion disks are not razor-thin, it is uncertain
whether the shear stress necessarily vanishes at the ISCO, and the
magnetic field, which is thought to drive turbulence in disks, may
contain large-scale structures that do not behave like a simple local
scalar viscosity.  We describe three-dimensional general relativistic
magnetohydrodynamic simulations of accretion disks around black holes
with a range of spin parameters, and we use the simulations to assess
the validity of the NT model.
Our fiducial initial magnetic field consists
of multiple (alternating polarity) poloidal field loops
whose shape is roughly isotropic in the disk in order
to match the isotropic turbulence expected in the poloidal plane.
For a thin disk with an aspect ratio $|h/r|\sim0.07$
around a non-spinning black hole, we find a decrease in the accreted
specific angular momentum of $2.9\%$ relative to the NT model
and an excess luminosity from
inside the ISCO of $3.5\%$.  The deviations in the case of spinning
black holes are also of the same order.  In addition, the deviations
decrease with decreasing $|h/r|$.  We therefore conclude that
magnetized thin accretion disks in x-ray binaries in the
thermal/high-soft spectral state ought to be well-described by the NT
model, especially at luminosities below $30\%$ of Eddington where we
expect a very small disk thickness $|h/r| \lesssim 0.05$.
We use our results to determine the spin equilibrium of black hole
accretion disks with a range of thicknesses
and to determine how electromagnetic stresses within the ISCO depend upon
black hole spin and disk thickness.
We find that the electromagnetic stress and the luminosity inside the ISCO
depend on the assumed initial magnetic field geometry.
We consider a second geometry
with field lines following density contours,
which for thin disks leads to highly radially-elongated magnetic field lines.
This gives roughly twice larger deviations from NT for both the
accreted specific angular momentum and the luminosity inside the ISCO.
Lastly, we find that the disk's corona (including any wind or jet)
introduces deviations from NT in the specific angular momentum
that are comparable to those contributed by the disk component,
while the excess luminosity of bound gas from within the ISCO
is dominated by only the disk component.
Based on these indications, we suggest that differences in results
between our work and other similar work
are due to differences in the assumed initial magnetic field geometry
as well as the inclusion of disk gas versus all the gas when comparing the
specific angular momentum from the simulations with the NT model.

\end{abstract}

\begin{keywords}
accretion, accretion discs, black hole physics, hydrodynamics,
(magnetohydrodynamics) MHD, methods: numerical, gravitation
\end{keywords}

\section{Introduction}
\label{sec_intro}

Accreting black holes (BHs) are among the most powerful astrophysical
objects in the Universe.  Although they have been the target of
intense research for a long time, many aspects of black hole accretion
theory remain uncertain to this day.  Pioneering work by
\citet{Bardeen:1970:SCO,sha73,nt73,pt74} and others indicated that
black hole accretion through a razor-thin disk can be highly
efficient, with up to $42\%$ of the accreted rest-mass-energy being
converted into radiation.  These authors postulated the existence of a
turbulent viscosity in the disk, parameterized via the famous
$\alpha$-prescription \citep{sha73}.  This viscosity causes outward
transport of angular momentum; in the process, it dissipates energy and
produces the radiation.  The authors also assumed that, within the inner-most
stable circular orbit (ISCO) of the black hole, the viscous torque
vanishes and material plunges into the black hole with constant energy
and angular momentum flux per unit rest-mass flux.
This is the so-called ``zero-torque'' boundary condition.

Modern viscous
hydrodynamical calculations of disks with arbitrary thicknesses
suggest that the zero-torque condition is a good
approximation when the height ($h$) to radius ($r$) ratio of the accreting gas is small:
$|h/r|\lesssim 0.1$ \citep{pac00,ap03,snm08,sadowski_slim_disks_2009,abr10}.
Radiatively efficient disks in
active galactic nuclei (AGN) and x-ray binaries are expected to have
disk thickness $|h/r| < 0.1$ whenever the luminosity is limited
to less than about $30\%$ of the Eddington luminosity \citep{mcc06}.
The above hydrodynamical studies thus suggest that systems in this
limit should be described well by the standard relativistic thin disk
theory as originally developed by \citet{nt73}, hereafter NT.

In parallel with the above work, it has for long been recognized
that the magnetic field could be a complicating factor
that may significantly modify accretion dynamics near and inside the ISCO \citep{thorne74}.
This issue has become increasingly important with the realization
that angular momentum transport in disks is
entirely due to turbulence generated
via the magnetorotational instability (MRI) \citep{bal91,bh98}.
However, the magnetic field does not necessarily behave
like a local viscous hydrodynamical stress.  Near the black hole,
the magnetic field may have large-scale structures
\citep{macdonald84}, which can induce stresses across the
ISCO \citep{krolik99,gammie99,ak00} leading to changes in, e.g.,
the density, velocity, and amount of dissipation and emission.
Unlike turbulence, the magnetic field can transport angular momentum without
dissipation (e.g. \citealt{li02}), or it can dissipate in current
sheets without transporting angular momentum.  In \citet{ak00}, the
additional electromagnetic stresses are treated simply as a freely tunable
model parameter on top of an otherwise hydrodynamical model.  A more
complete magnetohydrodynamical (MHD) model of a magnetized thin disk
has been developed by \citet{gammie99}.
In this model, the controlling free parameter is the
specific magnetic flux, i.e., magnetic flux per unit rest-mass flux.
Larger values of this parameter lead to larger deviations from NT
due to electromagnetic stresses, but the exact
value of the parameter for a given accretion disk is unknown.
For instance, it is entirely possible that electromagnetic stresses
become negligible in the limit when the disk thickness $|h/r|\to 0$.
The value of the specific magnetic flux
is determined by the nonlinear turbulent saturation of the magnetic field,
so accretion disk simulations are the best way to establish its magnitude.

The coupling via the magnetic field between a spinning black hole and
an external disk, or between the hole and the corona, wind and jet
(hereafter, corona-wind-jet),
might also play an important role
in modifying the accretion flow near the black hole.
The wind or jet (hereafter, wind-jet)
can transport angular momentum and energy away from the
accretion disk and black hole
\citep{blandford_accretion_disk_electrodynamics_1976,bz77,mg04,mck06jf,mck07b,km07}.
The wind-jet power depends upon factors such as
the black hole spin \citep{mck05,hk06,kom07},
disk thickness \citep{meier01,tmn08,tmn09,tnm09,tnm09b},
and the strength and large-scale behavior of the magnetic field
\citep{mg04,bhk08,mb09}, and these can affect the angular
momentum transport through an accretion disk.  In this context, we
note that understanding how such factors affect disk structure may be
key in interpreting the distinct states and complex behaviors observed
for black hole X-ray binaries \citep{fend04a,remm06}.
These factors also affect the black hole spin history
\citep{gammie_bh_spin_evolution_2004},
and so must be taken into account when considering
the effect of accretion on the cosmological evolution of black hole spin
\citep{Hughes:2003:BHM,gammie_bh_spin_evolution_2004,bv08}.

Global simulations of accretion disks using general relativistic
magnetohydrodynamical (GRMHD) codes (e.g. \citealt{gam03,dev03})
currently provide
the most complete understanding of how turbulent magnetized accretion
flows around black holes work.  Most simulations have studied
thick ($|h/r|\gtrsim 0.15$) disks without radiative cooling.
Such global simulations of the inner accretion flow
have shown that fluid crosses the ISCO without any clear evidence that
the torque vanishes at the ISCO, i.e., there is no
apparent ``stress edge''
\citep{mg04,kro05,beckwith08b}.  Similar results were previously found with
a pseudo-Newtonian potential for the black hole \citep{kh02}.  In
these studies, a plot of the radial profile of the normalized
stress within the ISCO appears to indicate a significant deviation
from the NT thin disk theory \citep{kro05,beckwith08b}, and it was thus
expected that much thinner disks might also deviate significantly from NT.
A complicating factor in drawing firm conclusions from
such studies is that the assumed initial
global magnetic field geometry and strength can significantly
change the magnitude of electromagnetic stresses
and associated angular momentum transport inside the ISCO \citep{mg04,bhk08}.

The implications of the above studies for truly thin
disks ($|h/r|\lesssim 0.1$) remain uncertain.  Thin disks are difficult
to resolve numerically, and simulations have been attempted only
recently.  Simulations of thin disks using a
pseudo-Newtonian potential for the black hole reveal good
agreement with standard thin disk theory \citep{rf08}.  The first
simulation of a thin ($|h/r|\approx 0.05$) disk using a full GRMHD model
was by \citet{shafee08}, hereafter S08.  They
considered a non-spinning ($a/M=0$) black hole
and an initial field geometry consisting of multiple opposite-polarity
bundles of poloidal loops
within the disk.
They found that, although the stress profile appears to indicate
significant torques inside the ISCO, the actual angular momentum flux
per unit rest-mass flux through the disk component deviates from the
NT prediction by only $2\%$, corresponding to an estimated deviation
in the luminosity of only about $4\%$.
The study by S08 was complemented by
\citet{noble09}, hereafter N09, who considered a thin ($|h/r|\approx 0.1$)
disk around an $a/M=0.9$ black hole and an initial field geometry
consisting of a single highly-elongated poloidal loop bundle
whose field lines follow the density contours of the thin disk.
They found $6\%$ more luminosity than predicted by NT.  More
recently, \citet{nkh10}, hereafter N10, considered
a thin ($|h/r|\approx 0.07$) disk around a non-spinning ($a/M=0$)
black hole and reported up to $10\%$
deviations from NT in the specific angular momentum accreted
through the accretion flow.

In this paper, we extend the work of S08 by considering a range of
black hole spins, disk thicknesses, field geometries,
box sizes, numerical resolutions, etc.
Our primary result is that we confirm S08, viz., geometrically thin disks
are well-described by the NT model.  We show that there are important
differences between the dynamics of the gas in the disk and in
the corona-wind-jet.
In addition, we find that the torque and luminosity within the ISCO
can be significantly affected by the geometry and strength of the
initial magnetic field, a result that should be considered when
comparing simulation results to thin disk theory.  In this context, we
discuss likely reasons for the apparently different conclusions
reached by N09 and N10.

The equations we solve are given in \S\ref{sec:goveqns}, diagnostics
are described in \S\ref{sec:diagnostics}, and our numerical setup is
described in \S\ref{sec:nummodels}.  Results for our fiducial thin
disk model for a non-rotating black hole are given in
\S\ref{sec:fiducialnonrot}, and we summarize convergence studies in
\S\ref{sec:convergence}.  Results for a variety of
black hole spins and disk thicknesses are presented in
\S\ref{sec:thicknessandspin} and for thin disks with different
magnetic field geometries and strengths in \S\ref{sec:magneticfield}.
We compare our results with previous studies in
\S\ref{sec:comparison}, discuss the implications of our results in
\S\ref{sec:discussion}, and conclude with a summary of the salient
points in \S\ref{sec:conclusions}.

\section{Governing Equations}
\label{sec:goveqns}

The system of interest to us is a magnetized accretion disk
around a rotating black hole.
We write the black hole Kerr metric in Kerr-Schild (KS,
horizon-penetrating) coordinates \citep{fip98,pf98}, which can be mapped to
Boyer-Lindquist (BL) coordinates or an orthonormal basis in any frame
\citep{mg04}.  We work with Heaviside-Lorentz units, set the speed of
light and gravitational constant to unity ($c=G=1$), and let $M$ be
the black hole mass.  We solve the general relativistic
magnetohydrodynamical (GRMHD) equations of motion for rotating black
holes \citep{gam03} with an additional cooling term designed to keep the
simulated accretion disk at a desired thickness.

Mass conservation gives
\begin{equation}
\nabla_\mu (\rho_0 u^\mu) = 0 ,
\end{equation}
where $\rho_0$ is the rest-mass density,
corresponding to the mass density in the fluid frame,
and $u^\mu$ is the contravariant 4-velocity.
Note that we write the orthonormal 3-velocity as $v_i$
(the covariant 3-velocity is never used below).
Energy-momentum conservation gives
\begin{equation}\label{emomeq}
\nabla_\mu  T^\mu_\nu = S_\nu ,
\end{equation}
where the stress energy tensor $T^\mu_\nu$ includes both matter and
electromagnetic terms,
\begin{equation}
T^\mu_\nu = (\rho_0 + u_g + p_g + b^2) u^\mu u_\nu + (p_g + b^2/2)\delta^\mu_\nu - b^\mu b_\nu ,
\end{equation}
where $u_g$ is the internal energy density and $p_g=(\Gamma-1)u_g$ is
the ideal gas pressure with $\Gamma=4/3$ \footnote{Models with
$\Gamma=5/3$ show some minor differences compared to models with
$\Gamma=4/3$ \citep{mg04,mm07}.}.  The contravariant fluid-frame magnetic 4-field
is given by $b^\mu$, and is related to the lab-frame 3-field via $b^\mu
= B^\nu h^\mu_\nu/u^t$ where $h^\mu_\nu = u^\mu u_\nu +
\delta^\mu_\nu$ is a projection tensor,
and $\delta^\mu_\nu$ is the Kronecker delta function.
We write the orthonormal 3-field as $B_i$ (the covariant 3-field is never used below).
The magnetic energy density ($u_b$)
and magnetic pressure ($p_b$) are then given by $u_b=p_b=b^\mu b_\mu/2 = b^2/2$.
Note that the angular velocity of the gas is $\Omega=u^\phi/u^t$.
Equation (\ref{emomeq}) has a source term
\begin{equation}\label{eq:cooling_term}
S_\nu = \left(\frac{dU}{d\tau}\right) u_\nu ,
\end{equation}
which is a radiation 4-force corresponding to a simple isotropic
comoving cooling term given by $dU/d\tau$.  We ignore radiative
transport effects such as heat currents, viscous stresses, or other
effects that would enter as additional momentum sources in the
comoving frame.  In order to keep the accretion disk thin, we employ
the same ad hoc cooling function as in S08:
\begin{equation}\label{cooling}
\frac{dU}{d\tau} = - u_g  \frac{\log\left(K/K_c\right)}{\tau_{\rm cool}} S[\theta] ,
\end{equation}
where $\tau$ is the fluid proper time,
$K=p_g/\rho_0^\Gamma$ is the entropy constant,
$K_c=0.00069$ is set to be the same entropy constant as the torus atmosphere
and is the entropy constant we cool the disk towards,
and $K_0\gtrsim K_c$ is the entropy constant of the initial torus\footnote{We intended
to have a constant $K$ everywhere at $t=0$, but
a normalization issue led to $K_c\lesssim K_0$.
Because of this condition,
the disk cools toward a slightly thinner equilibrium at the start of the simulation,
after which the cooling proceeds as originally desired by cooling towards
the fiducial value $K=K_c$.
Our models with $|h/r|\approx 0.07$ are least affected by this issue.
Also, since we do not make use of the cooling-based luminosity near $t=0$,
this issue does not affect any of our results.
We confirmed that this change leads to no significant issues
for either the magnitude or scaling of quantities with thickness
by repeating some simulations with the intended $K_c=K_0$.
The otherwise similar simulations have thicker disks as expected
(very minor change for our thin disk model as expected),
and we find consistent results
for a given measured thickness in the saturated state.}.
The gas cooling time is set to $\tau_{\rm cool}=2\pi/\Omega_{\rm K}$,
where $\Omega_{\rm K} = (1/M)/[ (a/M) +(R/M)^{3/2}]$ is the Keplerian angular frequency
and $R=r\sin\theta$ is the cylindrical radius
(We consider variations in the cooling timescale in section~\ref{sec_fluxdiskcorona}.).
We use a shaping function given by the
quantity $S[\theta] = \exp[-(\theta-\pi/2)^2 / (2(\theta_{\rm nocool})^2)]$,
where we set $\theta_{\rm nocool}=\{0.1,0.3,0.45,0.45\}$
for our sequence of models with target thickness of $|h/r|=\{0.07, ~0.1, ~0.2, ~0.3\}$,
although we note that the thickest model with target $|h/r|=0.3$ has no cooling turned on.
The shaping function $S[\theta]$ is used to avoid cooling in the low density
funnel-jet region where the thermodynamics is not accurately evolved
and where the gas is mostly unbound
(see Figure~\ref{taperoff} in section~\ref{sec_fluxdiskcorona}).
In addition, we set the cooling function $dU/d\tau=0$
if 1) the timestep, $dt$, obeys $dt>\tau_{\rm cool}$,
which ensures that the cooling does not create negative entropy gas ;
or 2) $\log(K/K_c)<0$, which ensures the gas is only cooled, never heated.
Photon capture by the black hole is not included,
so the luminosity based upon this cooling function is an upper limit
for radiation from the disk.
The above cooling function drives the specific entropy of the gas
toward the reference specific entropy $K_c$.
Since specific entropy always increases due to dissipation,
this cooling function explicitly tracks dissipation.
Hence, the luminosity generated from the cooling function
should not be considered as the true luminosity,
but instead should be considered as representing the emission
rate in the limit that all dissipated energy is lost as radiation.
Any other arbitrary cooling function that does not track dissipation
would require full radiative transfer to obtain the true luminosity.

Magnetic flux conservation is given by the induction equation
\begin{equation}
\partial_t(\detg B^i) = -\partial_j[\detg(B^i v^j - B^j v^i)] ,
\end{equation}
where $v^i=u^i/u^t$, and $g={\rm Det}(g_{\mu\nu})$ is the determinant of the
metric.  In steady-state, the cooling is balanced by heating from
shocks, grid-scale reconnection, and grid-scale viscosity.  No
explicit resistivity or viscosity is included.

\section{Diagnostics}
\label{sec:diagnostics}

In this section, we describe several important diagnostics
that we have found useful in this study.
First, since we regulate the disk height via an ad hoc cooling function,
we check the scale height of the simulated disk
as a function of time and radius
to ensure that our cooling function operates properly.
Second, the equations we solve consist of
seven time-dependent ideal MHD equations,
corresponding to four relevant conserved
quantities\footnote{The energy-momentum of the fluid is not strictly conserved
because of radiative cooling; however, the fluid component of the
energy-momentum equations still proves to be useful.
Only energy conservation of the fluid is strongly affected for our types of models.}.
Using these quantities we construct three dimensionless flux ratios
corresponding to
the accreted specific energy,
specific angular momentum, and specific magnetic flux.
Third, we check what the duration of the simulations should be
in order to reach a quasi-steady state (``inflow equilibrium'') at any given radius.
Finally, we describe how we compute the luminosity.

When the specific fluxes are computed as a spatial or temporal average/integral,
we always take the ratio of averages/integrals of fluxes (i.e. $\int dx F_1/\int dx F_2$)
rather than the average/integral of the ratio of fluxes (i.e. $\int dx (F_1/F_2)$).
The former is more appropriate for capturing the mean behavior,
while the latter can be more appropriate when investigating
fluxes with significant phase shifted correlations between each other.
As relevant for this study, the accretion disk has significant
vertical stratification and the local value of the ratio of fluxes
can vary considerably without any affect on the bulk accretion flow.
Similarly, potentially one flux can (e.g.) nearly vanish over short periods,
while the other flux does not, which leads to unbounded values for the ratio of fluxes.
However, the time-averaged behavior of the flow is not greatly affected by such short
periods of time.
These cases demonstrate why the ratio of averages/integrals is always
performed for both spatial and temporal averages/integrals.

When comparing the flux ratios or luminosities from
the simulations against the NT model,
we evaluate the percent relative difference $D[f]$ between
a quantity $f$ and its NT value as follows:
\begin{equation}
D[f] \equiv 100\frac{f-f[{\rm NT}]}{f[{\rm NT}]} .
\end{equation}
For a density-weighted time-averaged value of $f$, we compute
\begin{equation}\label{meantheta}
\langle f \rangle_{\rho_0} \equiv
\frac{\int\int\int f \,\rho_0(r,\theta,\phi) dA_{\theta\phi}dt}
{\int\int\int \rho_0(r,\theta,\phi) dA_{\theta\phi}dt} ,
\end{equation}
where $dA_{\theta\phi}\equiv \sqrt{-g} d\theta d\phi$ is an area
element in the $\theta-\phi$ plane, and the integral over $dt$
is a time average over the duration of interest,
which corresponds to the period when the disk is in steady-state.
For a surface-averaged value of $f$, we compute
\begin{equation}
\langle f \rangle \equiv \frac{\int\int f\; dA_{\theta\phi}}{\int\int dA_{\theta\phi}} .
\end{equation}

\subsection{Disk Thickness Measurement}
\label{sec:diskthick1}

We define the dimensionless
disk thickness per unit radius, $|h/r|$,
as the density-weighted mean angular deviation
of the gas from the midplane,
\begin{equation}\label{thicknesseq}
\left|\frac{h}{r}\right| \equiv \left\langle \left|\theta-\frac{\pi}{2}\right| \right\rangle_{\rho_0} .
\end{equation}
(This quantity was called ${\Delta\theta}_{\rm abs}$ in S08.)
Notice that we assume the accretion disk plane is on the equator
(i.e. we assume $\langle\theta\rangle_{\rho_0}=\pi/2$).
As defined above, $|h/r|$ is a function of $r$.  When we wish
to characterize the disk by means of a single estimate of its
thickness, we use the value of $|h/r|$ at $r=2r_{\rm ISCO}$, where
$r_{\rm ISCO}$ is the ISCO radius ($r_{\rm ISCO}=6M$ for a
non-spinning BH
and $r_{\rm ISCO}=M$ for a maximally-spinning BH; \citealt{shapirobook83}).
As we show in \S\ref{sec:diskthick2}, this choice is quite reasonable.
An alternative thickness measure,
given by the root-mean-square thickness $(h/r)_{\rm rms}$,
allows us to estimate how accurate we can be about our definition of thickness.
This quantity is defined by
\begin{equation}\label{thicknessrms}
\left(\frac{h}{r}\right)_{\rm rms} \equiv \left\langle \left(\theta-\frac{\pi}{2}\right)^2\right\rangle_{\rho_0}^{1/2} .
\end{equation}
The range of $\theta$ for the disk thickness integrals in the above
equations is from $0$ to $\pi$.

\subsection{Fluxes of Mass, Energy, and Angular Momentum}

The mass, energy and angular momentum conservation equations give
the following fluxes,
\begin{eqnarray}\label{Dotsmej}
\dot{M}(r,t) &=& -\int_\theta \int_\phi \rho_0 u^r dA_{\theta\phi}, \\
\emath \equiv \frac{\dot{E}(r,t)}{\dot{M}(r,t)} &=& \frac{\int_\theta\int_\phi T^r_t dA_{\theta\phi}}{\dot{M}(r,t)} , \\
\jmath \equiv \frac{\dot{J}(r,t)}{\dot{M}(r,t)} &=& -\frac{\int_\theta\int_\phi T^r_\phi dA_{\theta\phi}}{\dot{M}(r,t)} .
\end{eqnarray}
The above relations define
the rest-mass accretion rate (sometimes just referred to as the mass accretion rate), $\dot{M}$;
the accreted energy flux per unit rest-mass flux, or {\it specific energy}, $\emath$;
and the accreted angular momentum flux per unit rest-mass flux,
or {\it specific angular momentum}, $\jmath$.
Positive values of these quantities
correspond to an inward flux through the black hole horizon.

The black hole spin evolves due to the accretion of mass, energy, and angular momentum,
which can be described by the dimensionless spin-up parameter s,
\begin{equation}\label{spinevolve}
s \equiv \frac{d(a/M)}{dt}\frac{M}{\dot{M}}  =  \jmath - 2\frac{a}{M}\emath ,
\end{equation}
where the angular integrals used to compute $\jmath$ and $\emath$
include all $\theta$ and $\phi$ angles \citep{gammie_bh_spin_evolution_2004}.
For $s=0$ the black hole is in so-called ``spin equilibrium,''
corresponding to when the dimensionless black hole spin, $a/M$,
does not change in time.

The ``nominal'' efficiency, corresponding
to the total loss of specific energy from the fluid,
is obtained by removing the rest-mass term from the accreted specific energy:
\begin{equation}
\eff \equiv  1- \emath .
\end{equation}
The time-averaged value of $\eff$ at the horizon ($r=r_{\rm H}$)
gives the total nominal efficiency: $\langle\eff(r_{\rm H})\rangle$,
which is an upper bound on the total photon radiative efficiency.

The range of $\theta$ over which the flux density integrals in the above equations
are computed depends on the situation.  In S08, we limited the
$\theta$ range to $\delta\theta=\pm 0.2$ corresponding
to 2--3 density scale heights in order to focus on the disk
and to avoid including the disk wind or black hole jet.
In this paper, we are interested in studying how the contributions to the
fluxes vary as a function of height above the equatorial plane.  Our
expectation is that the disk and corona-wind-jet contribute differently to
these fluxes.  Thus, we consider different ranges of $\theta$ in the
integrals, e.g., from $(\pi/2)-2|h/r|$ to
$(\pi/2)+2|h/r|$, $(\pi/2)-4|h/r|$ to
$(\pi/2)+4|h/r|$, or $0$ to $\pi$.  The first and third
are most often used in later sections.

\subsection{Splitting Angular Momentum Flux into Ingoing and Outgoing Components}

For a more detailed comparison of the simulation results with the
NT model, we decompose the flux of angular momentum into
an ingoing (``in'') term which is related to the advection of
mass-energy into the black hole
and an outgoing (``out'') term which is related to the forces and
stresses that transport angular momentum radially outward.
These ingoing and outgoing components of the specific angular momentum
are defined by
\begin{eqnarray}\label{Dotssplit}
{\jmath}_{\rm in}(r,t) &\equiv& \frac{\langle(\rho_0 + u_g + b^2/2) u^r \rangle \langle u_\phi \rangle}{\langle -\rho_0 u^r\rangle}, \\
{\jmath}_{\rm out}(r,t) &\equiv& \jmath - {\jmath}_{\rm in}(r,t) .
\end{eqnarray}
By this definition, the ``in'' quantities correspond to inward transport of the
comoving mass-energy density of the disk, $u^\mu u^\nu
T_{\mu\nu}=\rho_0 + u_g + b^2/2$.  Note that ``in'' quantities are products
of the mean velocity fields $\langle u^r \rangle$ and $\langle u_\mu \rangle$
and not the combination $\langle u^r u_\mu \rangle$; the latter
includes a contribution from
correlated fluctuations in $u^r$ and $u_\mu$, which corresponds to the
Reynolds stress.
The residual of the total flux minus the ``in''
flux gives the outward, mechanical transport by Reynolds stresses
and electromagnetic stresses.
One could also consider a similar splitting for the specific energy.
The above decomposition most closely
matches our expectation that the inward flux should agree with the NT result
as $|h/r|\to 0$.  Note, however, that our conclusions in
this paper do not require any
particular decomposition.
This decomposition is different
from S08 and N10 where the entire magnetic term ($b^2 u^r u_\phi - b^r b_\phi$)
is designated as the ``out'' term.
Their choice overestimates the effect of electromagnetic stresses,
since some magnetic energy is simply advected into the black hole.
Also, the splitting used in S08 gives non-monotonic ${\jmath}_{\rm in}$
vs. radius for some black hole spins,
while the splitting we use gives monotonic values for all black hole
spins.

\subsection{The Magnetic Flux}
\label{magneticfluxdiag}

The no-monopoles constraint implies that the total magnetic flux
($\Phi = \int_S \vec{B}\cdot \vec{dA}$)
vanishes through any closed surface
or any open surface penetrating a bounded flux bundle.
The magnetic flux conservation equations require that
magnetic flux can only be transported to the black hole
or through the equatorial plane by advection.
The absolute magnetic flux ($\int_S |\vec{B}\cdot \vec{dA}|$)
has no such restrictions and can grow arbitrarily due to the MRI.
However, the absolute magnetic flux can saturate when the
electromagnetic field comes into force balance with the matter.
We are interested in such a saturated state of the magnetic field
within the accretion flow and threading the black hole.

We consider the absolute magnetic flux
penetrating a spherical surface and an equatorial surface given, respectively, by
\begin{eqnarray}
\Phi_r(r,\theta,t) &=& \int_\theta \int_\phi |B^r| dA_{\theta'\phi} , \\
\Phi_\theta(r,\theta,t) &=& \int_{r'=r_{\rm H}}^{r'=r} \int_\phi |B^\theta| dA_{r'\phi} .
\end{eqnarray}
Nominally, $\Phi_r$ has an integration range of $\theta'=0$ to $\theta'=\theta$
when measured on the black hole horizon,
while when computing quantities around the equatorial plane $\theta'$
has the range $\langle\theta\rangle\pm\theta$.
One useful normalization of the magnetic fluxes is
by the total flux through one hemisphere of the black hole plus through the equator
\begin{equation}
\Phi_{\rm tot}(r,t) \equiv \Phi_r(r'=r_{\rm H},\theta'=0\ldots \pi/2,t) + \Phi_\theta(r,\theta'=\pi/2,t) ,
\end{equation}
which gives the normalized absolute radial magnetic flux
\begin{equation}
\tilde{\Phi}_r(r,\theta,t) \equiv \frac{\Phi_r(r,\theta,t)}{\Phi_{\rm tot}(r=R_{\rm out},t=0)} ,
\end{equation}
where $R_{\rm out}$ is the outer radius of the computational box.
The normalized absolute magnetic flux measures
the absolute magnetic flux on the black hole horizon
or radially through the equatorial disk per unit absolute flux
that is initially available.

The \citet{gammie99} model of a magnetized thin accretion flow
suggests another useful normalization of the magnetic flux is
by the absolute mass accretion rate
\begin{equation}\label{massg}
\dot{M}_G(r,t) \equiv \int_\theta \int_\phi \rho_0 |u^r| dA_{\theta\phi} ,
\end{equation}
which gives the normalized specific absolute magnetic fluxes
\begin{eqnarray}\label{Dotsgammie}
\Xi(r,t) &=& \frac{\Phi_r(r,t)}{\dot{M}_G(r,t)}  , \\
\Upsilon(r,t) &\equiv&  \sqrt{2} \left|\frac{\Xi(r,t)}{M}\right| \sqrt{\left|\frac{\dot{M}_G(r=r_{\rm H},t)}{{\rm SA}_{\rm H}}\right|} \label{equpsilon} ,
\end{eqnarray}
where ${\rm SA} = (1/r^2)\int_\theta \int_\phi dA_{\theta\phi}$ is the local solid angle,
${\rm SA}_{\rm H}={\rm SA}(r=r_{\rm H})$ is the local solid angle on the horizon,
$\Xi(r,t)$ is the radial magnetic flux per unit rest-mass flux
(usually specific magnetic flux),
and $\Upsilon(r,t) c^{3/2}/G$ is a particular dimensionless
normalization of the specific magnetic flux
that appears in the MHD accretion model developed by \citet{gammie99}.
Since the units used for the magnetic field are arbitrary,
any constant factor can be applied to $\Xi$
and one would still identify the quantity as the specific magnetic flux.
So to simplify the discussion we henceforth call
$\Upsilon$ the specific magnetic flux.
To obtain Equation~(\ref{equpsilon}),
all involved integrals should have a common $\theta$ range around the equator.
These quantities all have absolute magnitudes
because a sign change does not change the physical effect.
The quantities $\jmath$, $\emath$, $\eff$, $\Xi$, and $\Upsilon$
are each conserved along poloidal field-flow lines
for stationary ideal MHD solutions \citep{bekenstein_new_conservation_1978,tntt90}.

Gammie's (1999) model of a magnetized accretion flow within the ISCO assumes:
1) a thin equatorial flow ;
2) a radial poloidal field geometry (i.e., $|B_\theta|\ll |B_r|$) ;
3) a boundary condition at the ISCO corresponding to zero radial velocity ;
and 4) no thermal contribution.
The model reduces to the NT solution within the ISCO for $\Upsilon\to 0$,
and deviations from NT's solution are typically small
(less than $12\%$ for $\jmath$ across all black hole spins;
see Appendix~\ref{sec_gammie}) for $\Upsilon\lesssim 1$.
We have defined the quantity $\Upsilon$ in equation~(\ref{Dotsgammie})
with the $\sqrt{2}$ factor,
the square root of the total mass accretion rate through the horizon per unit solid angle,
and Heaviside-Lorentz units for $B^r$
so that the numerical value of $\Upsilon$ at the horizon is identically
equal to the numerical value of the free parameter in \citet{gammie99},
i.e., their $F_{\theta\phi}$ normalized by $F_{\rm M}=-1$.
As shown in that paper, $\Upsilon$ directly controls deviations
of the specific angular momentum and specific energy
away from the non-magnetized thin disk theory values of the NT model.
Even for disks of finite thickness, the parameter shows how electromagnetic stresses
control deviations between the horizon and the ISCO.
Note that the flow can deviate from NT at the ISCO
simply due to finite thermal pressure \citep{mg04}.
In Appendix~\ref{tbl_gammie} Table~\ref{sec_gammie},
we list numerical values of $\jmath$ and $\eff$ for Gammie's (1999) model,
and show how these quantities deviate from NT
for a given black hole spin and $\Upsilon$.

We find $\Upsilon$ to be more useful as a measure of the importance of the magnetic field
within the ISCO than our previous measurement in S08 of
the $\alpha$-viscosity parameter,
\begin{equation}\label{alphaeq}
\alpha=\frac{T^{\hat{\phi}\hat{r}}}{p_g+p_b} ,
\end{equation}
where $T^{\hat{\phi}\hat{r}} = {e^{\hat{\phi}}}_{\mu} {e^{\hat{r}}}_{\nu} T^{\mu\nu}$
is the orthonormal stress-energy tensor components in the comoving frame,
and ${e^{\hat{\nu}}}_{\mu}$ is the contravariant tetrad system in the local fluid-frame.
This is related to the normalized stress by
\begin{equation}\label{stress}
\frac{W}{\dot{M}} = \frac{\int\int T^{\hat{\phi}\hat{r}} dA'_{\theta\phi}}{\dot{M}\int_\phi dL'_{\phi}} ,
\end{equation}
where
$dA'_{\theta\phi} = {e^{\hat{\theta}}}_{\mu} {e^{\hat{\phi}}}_{\nu} d\theta^\mu d\phi^\nu$
is the comoving area element,
$dL'_{\phi} = {e^{\hat{\phi}}}_{\nu} d\phi^\nu$ evaluated at $\theta=\pi/2$
is the comoving $\phi$ length element,
$\theta^\mu=\{0,0,1,0\}$, and $\phi^\nu=\{0,0,0,1\}$.
This form for $W$ is a simple generalization of Eq. 5.6.1b in NT73,
and note that the NT solution for $W/\dot{M}$ is given by Eq. 5.6.14a in NT73.
In S08, $W$ was integrated over fluid satisfying
$-u_t (\rho_0 + u_g + p_g + b^2)/\rho_0 < 1$
(i.e., only approximately gravitationally bound fluid and no wind-jet).
We use the same definition of bound in this paper.
As shown in S08, a plot of the radial profile
of $W/\dot{M}$ or $\alpha$ within the ISCO does not necessarily quantify
how much the magnetic field affects the accretion flow properties,
since even apparently large values of this quantity within
the ISCO do not cause a significant deviation from NT
in the specific angular momentum accreted.
On the other hand, the \citet{gammie99} parameter $\Upsilon$
does directly relate to the electromagnetic stresses within the ISCO
and is an ideal MHD invariant (so constant vs. radius) for a stationary flow.
One expects that appropriately time-averaged simulation data
could be framed in the context of this stationary model
in order to measure the effects of electromagnetic stresses.

\subsection{Inflow Equilibrium}
\label{sec_infloweq}

When the accretion flow has achieved steady-state inside a given radius, the
quantity $\dot M(r,t)$ will (apart from normal fluctuations due to turbulence) be
independent of time,
and if it is integrated over all $\theta$ angles will be constant
within the given radius\footnote{If we
integrate over a restricted range of $\theta$, then
there could be additional mass flow through the boundaries in the
$\theta$ direction and $\dot{M}(r,t)$ will no longer be independent of
$r$, though it would still be independent of $t$.}.  The energy and
angular momentum fluxes have a non-conservative contribution due to
the cooling function and therefore are not strictly constant.
However, the cooling is generally a minor contribution (especially in
the case of the angular momentum flux), and so we may still measure the
non-radiative terms to verify inflow equilibrium.

The radius out to which inflow equilibrium can be achieved in a given
time can be estimated by calculating the mean radial velocity $v_r$
and then deriving from it a viscous timescale $-r/v_r$.  From standard
accretion disk theory and using the definition of $\alpha$ given in
Eq.~(\ref{stress}), the mean radial velocity is given by
\begin{equation}\label{eqvr}
v_r \sim -\alpha \left|\frac{h}{r}\right|^2 v_{\rm K} ,
\end{equation}
where $v_{\rm K}\approx(r/M)^{-1/2}$ is the Keplerian speed at radius $r$
and $\alpha$ is the standard viscosity parameter
given by equation~(\ref{alphaeq}) \citep{fkr92}.
Although the viscous timescale is the nominal
time needed to achieve steady-state, in practice it takes several viscous times
before the flow really settles down, e.g., see the calculations reported in
\citet{shapiro2010}.  In the present paper, we assume that inflow equilibrium
is reached after two viscous times, and hence we
estimate the inflow equilibrium time, $t_{\rm ie}$, to be
\begin{equation}\label{tie}
t_{\rm ie} \sim -2\frac{r}{v_r} \sim 2 \left(\frac{r}{M}\right)^{3/2} \left(\frac{1}{\alpha |h/r|^2}\right) \sim 5000 \left(\frac{r}{M}\right)^{3/2} ,
\end{equation}
where, in the right-most relation,
we have taken a typical value of $\alpha\sim 0.1$ for the gas in  the disk proper (i.e., outside the ISCO)
and we have set $|h/r|\approx 0.064$,
as appropriate for our thinnest disk models.

A simulation must run until $t\sim t_{\rm ie}$ before we can expect
inflow equilibrium at radius $r$.  According to the above Newtonian estimate, a thin
disk simulation with $|h/r| \sim 0.064$ that has been run for a time of $30000M$ will achieve
steady-state out to a radius of only $\sim3M$.  However, this estimate is inaccurate since
it does not allow for the boundary condition on the flow at the ISCO.  Both
the boundary condition as well
as the effects of GR are included in the formula for the radial velocity
given in Eq.~5.9.8 of NT, which we present for completeness in Appendix~\ref{sec_inflow}.
That more accurate result, which is what we use for all our plots and numerical estimates,
shows that our thin disk simulations should
reach inflow equilibrium
within $r/M=9,~7,~5.5,~5$, respectively, for $a/M=0,~0.7,~0.9,~0.98$.
These estimates are roughly consistent with the radii out to which
we have a constant $\jmath$ vs. radius in the simulations
discussed in \S\ref{sec:thicknessandspin}.

\subsection{Luminosity Measurement}

We measure the radiative luminosity of the accreting gas directly from
the cooling function $dU/d\tau$.  At a given radius, $r$, in
the steady region of the flow,
the luminosity per unit rest-mass accretion rate
interior to that radius is given by
\begin{equation}\label{lum}
\frac{{L}(<r)}{\dot{M}(r,t)} = \frac{1}{\dot{M}(r,t)(t_f-t_i)} \int_{t=t_i}^{t_f}
\int_{r'=r_{\rm H}}^{r}\int_{\theta=0}^\pi\int_\phi \left(\frac{dU}{d\tau}\right)u_t\,
dV_{t r'\theta\phi}  ,
\end{equation}
where $dV_{t r'\theta\phi} = \detg dt dr' d\theta d\phi$
and the 4D integral goes from the initial time $t_i$ to the final
time $t_f$ over which the simulation results are time-averaged, from the
radius $r_{\rm H}$ of the horizon to the radius $r$ of interest,
and usually over all $\theta$ and $\phi$.
We find it useful to compare the simulations
with thin disk theory by computing the ratio of the luminosity emitted inside the ISCO
(per unit rest-mass accretion rate) to the total radiative efficiency of the NT model:
\begin{equation}\label{Lin}
\tilde{L}_{\rm in} \equiv \frac{L(<r_{\rm ISCO})}{\dot{M}\eff[{\rm NT}]} .
\end{equation}
This ratio measures the excess radiative luminosity from inside the ISCO in the
simulation relative to the total luminosity in the NT model (which predicts zero luminosity here).
We also consider the excess luminosity over the entire inflow equilibrium region
\begin{equation}\label{Leq}
\tilde{L}_{\rm eq} \equiv \frac{L(r<r_{\rm eq})-L(r<r_{\rm eq})[{\rm NT}]}{\dot{M}\eff[{\rm NT}]} ,
\end{equation}
which corresponds to the luminosity (per unit mass accretion rate)
inside the inflow equilibrium region (i.e. $r<r_{\rm eq}$, where
$r_{\rm eq}$ is the radius out to which inflow equilibrium has been established)
subtracted by the NT luminosity all divided by the total NT efficiency.
Large percent values of $\tilde{L}_{\rm in}$ or $\tilde{L}_{\rm eq}$
would indicate large percent deviations from NT.

\section{Physical Models and Numerical Methods}
\label{sec:nummodels}

This section describes our physical models and numerical methods.
Table~\ref{tbl_models} provides a list of all our simulations
and shows the physical and numerical parameters that we vary.
Our primary models are labeled by names of the form AxHRy, where x is
the value of the black hole spin parameter and y is approximately
equal to the disk thickness $|h/r|$.  For instance, our fiducial model
A0HR07 has a non-spinning black hole ($a/M=0$) and a geometrically
thin disk with $|h/r| \sim 0.07$.  We discuss this particular model in
detail in \S\ref{sec:fiducialnonrot}.  Table~\ref{tbl_models} also
shows the time span (from $T_i/M$ to $T_f/M$) used to perform the
time-averaging, and the last column shows the actual value of $|h/r|$
in the simulated model as measured during inflow equilibrium, e.g.,
$|h/r|= 0.064$ for model A0HR07.

\subsection{Physical Models}
\label{sec:modelsetup}

This study considers black hole accretion disk systems
with a range of black hole spins: $a/M=0, ~0.7, ~0.9, ~0.98$,
and a range of disk thicknesses: $|h/r|=0.07, ~0.13, ~0.2, ~0.3$.
The initial mass distribution is an isentropic equilibrium torus \citep{chak85a,chak85b,dev03}.
All models have an initial inner torus edge at $r_{\rm in}=20M$,
while the torus pressure maximum for each model
is located between $R_{\rm max}=35M$ and $R_{\rm max}=65M$.
We chose this relatively large radius for the initial torus
because S08 found that placing the torus at smaller radii caused the
results to be sensitive to the initial mass distribution.
We initialize the solution so that $\rho_0=1$ is the maximum rest-mass density.
In S08, we set $q=1.65$ ($\Omega\propto r^{-q}$ in non-relativistic limit)
and $K=0.00034$ with $\Gamma=4/3$,
while in this paper we adjust the initial angular momentum profile
such that the initial torus has the target value of $|h/r|$ at the
pressure maximum.
For models with $|h/r|=0.07, ~0.13, ~0.2, ~0.3$,
we fix the specific entropy of the torus by setting, respectively,
$K=K_0\equiv\{0.00034, ~0.0035, ~0.009, ~0.009\}$
in the initial polytropic equation of state given by $p=K_0\rho_0^\Gamma$.
The initial atmosphere surrounding the torus has
the same polytropic equation of state
with nearly the same entropy constant of $K=0.00069$,
but with an initial rest-mass density of $\rho_0=10^{-6} (r/M)^{-3/2}$,
corresponding to a Bondi-like atmosphere.

Recent GRMHD simulations of thick disks
indicate that the results for the disk (but not the wind-jet)
are roughly independent of the initial field geometry
\citep{mck07a,mck07b,bhk08}.
However, a detailed study for thin disks has yet to be performed.

We consider a range of magnetic field geometries
described by the vector potential $A_\mu$ which is related
to the Faraday tensor by $F_{\mu\nu} = A_{\nu,\mu} - A_{\mu,\nu}$.
As in S08, we consider a general multiple-loop field geometry corresponding
to $N$ separate loop bundles stacked radially within the initial disk.
The vector potential we use is given by
\begin{equation}\label{vectorpot}
A_{\phi,\rm N} \propto Q^2\sin\left(\frac{\log(r/S)}{\lambda_{\rm field}/(2\pi r)}\right) \left[1 + w({\rm ranc}-0.5)\right] ,
\end{equation}
where ${\rm ranc}$ is a random number generator for the domain $0$ to $1$
(see below for a discussion of perturbations.).
All other $A_\mu$ are initially zero.
All our multi-loop and 1-loop simulations have $S=22M$, and the values of
$\lambda_{\rm field}/(2\pi r)$ are listed in Table~\ref{tbl_models}.
For multi-loop models, each additional field loop bundle has opposite polarity.
We use $Q = (u_g/u_{g,\rm max} - 0.2) (r/M)^{3/4}$, and set $Q=0$ if either $r<S$ or $Q<0$,
and $u_{g,\rm max}$ is the maximum value of the internal energy density in the torus.
By comparison, in S08, we set $S = 1.1 r_{\rm in}$, $r_{\rm in}=20M$,
$\lambda_{\rm field}/(2\pi r)=0.16$,
such that there were two loops centered at $r=28M$ and $38M$.
The intention of introducing multiple loop bundles
is to keep the aspect ratio of the bundles roughly 1:1 in the poloidal plane,
rather than loop(s) that are highly elongated in the radial direction.
For each disk thickness, we tune $\lambda_{\rm field}/(2\pi r)$
in order to obtain initial poloidal loops that are roughly isotropic.

As in S08, the magnetic field strength is set such that
the plasma $\beta$ parameter satisfies
$\beta_{\rm maxes}\equiv p_{g,\rm max}/p_{b,\rm max}=100$,
where $p_{g,\rm max}$ is the maximum thermal pressure
and $p_{b,\rm max}$
is the maximum magnetic pressure in the entire torus.
Since the two maxima never occur at the same location,
$\beta=p_g/p_b$ varies over a wide range of values within the disk.
This approach is similar to how the magnetic field was
normalized in other studies \citep{gam03,mg04,mck06jf,mck07b,km07}.
It ensures that the magnetic field is weak throughout the disk.
Care must be taken with how one normalizes any given initial magnetic field geometry.
For example, for the 1-loop field geometry used by \citet{mg04},
if one initializes the field with a {\it mean}
(volume-averaged) $\bar{\beta}=100$,
then the inner edge of the initial torus has $\beta\sim 1$
and the initial disk is not weakly magnetized.

For most models, the vector potential at all grid points
was randomly perturbed by $2\%$ ($w$ in equation~\ref{vectorpot})
and the internal energy density at all grid points
was randomly perturbed by $10\%$ \footnote{In S08,
we had a typo saying we perturbed the field by $50\%$,
while it was actually perturbed the same as these models, i.e.:
$2\%$ vector potential perturbations and $10\%$ internal energy perturbations.}.
If the simulation starts with perturbations of the vector potential,
then we compute $\Phi_{\rm tot}$ (used to obtain $\tilde{\Phi}_r$)
using the pre-perturbed magnetic flux
in order to gauge how much flux is dissipated due to the perturbations.
Perturbations should be large enough to excite the non-axisymmetric MRI
in order to avoid the axisymmetric channel solution,
while they should not be so large as to induce significant dissipation
of the magnetic energy due to grid-scale magnetic dissipation just after the evolution begins.
For some models, we studied different amplitudes for the initial perturbation in order to ensure
that the amplitude does not significantly affect our results.
For a model with $|h/r|\sim 0.07$, $a/M=0$, and a single polarity field loop,
one simulation was initialized with $2\%$ vector potential perturbations
and $10\%$ internal energy perturbations,
while another otherwise similar simulation was given no seed perturbations.
Both become turbulent at about the same time $t\sim 1500M$.
The magnetic field energy at that time is negligibly different,
and there is no evidence for significant differences in any quantities during inflow equilibrium.

\begin{table*}
\caption{Simulation Parameters}
\begin{center}
\begin{tabular}[h]{|r|r|r|r|r|r|r|r|r|r|r|r|r|r|r|}
\hline
Model Name & $T_i/M$-$T_f/M$ & $\frac{a}{M}$ & $N_r$ & $N_\theta$ & $N_\phi$ & $\frac{R_{\rm in}}{r_{\rm H}}$ & $Y$ & $\Delta\phi$ & $\frac{R_{\rm max}}{M}$ & $\frac{R_{\rm out}}{M}$ & $q$ & cooling & $\frac{\lambda_{\rm field}}{2\pi r}$ & $\frac{h}{r}$ \\
\hline
A0HR07 & 12500-27350 & 0 & 256 & 64 & 32 & 0.9 $$ &  0.13 & $\pi/2$ & 35 & 50 & 1.65 & yes & 0.065      & 0.064\\
A7HR07 & 12500-20950 & 0.7 & 256 & 64 & 32 & 0.92 $$ &  0.13 & $\pi/2$ & 35 & 50 & 1.65 & yes & 0.065   & 0.065\\
A9HR07 & 14000-23050 & 0.9 & 256 & 64 & 32 & 0.92 $$ &  0.13 & $\pi/2$ & 35 & 50 & 1.65 & yes & 0.065   & 0.054\\
A98HR07 & 14000-19450 & 0.98 & 256 & 64 & 32 & 0.92 $$ &  0.13 & $\pi/2$ & 35 & 50 & 1.65 & yes & 0.065 & 0.059\\
A0HR1 & 5050-14150 & 0 & 256 & 64 & 32 & 0.9 $$ &  0.37 & $3\pi/4$ & 45 & 120 & 1.94 & yes & 0.25       & 0.12 \\
A7HR1 & 5050-12550 & 0.7 & 256 & 64 & 32 & 0.9 $$ &  0.37 & $3\pi/4$ & 45 & 120 & 1.92 & yes & 0.25     & 0.09 \\
A9HR1 & 5050-10000 & 0.9 & 256 & 64 & 32 & 0.9 $$ &  0.37 & $3\pi/4$ & 45 & 120 & 1.91 & yes & 0.25     & 0.13 \\
A98HR1 & 12000-13600 & 0.98 & 256 & 64 & 32 & 0.9 $$ &  0.37 & $3\pi/4$ & 45 & 120 & 1.91 & yes & 0.25  & 0.099  \\
A0HR2 & 6000-13750 & 0 & 256 & 64 & 32 & 0.9 $$ &  0.65 & $\pi$ & 65 & 200 & 1.97 & yes & 0.28          & 0.18 \\
A7HR2 & 12000-17900 & 0.7 & 256 & 64 & 32 & 0.9 $$ &  0.65 & $\pi$ & 65 & 200 & 1.97 & yes & 0.28       & 0.16 \\
A9HR2 & 12000-15200 & 0.9 & 256 & 64 & 32 & 0.88 $$ &  0.65 & $\pi$ & 65 & 200 & 1.97 & yes & 0.28      & 0.21 \\
A98HR2 & 6100-7100 & 0.98 & 256 & 64 & 32 & 0.9 $$ &  0.65 & $\pi$ & 65 & 200 & 1.97 & yes & 0.28       & 0.18 \\
A0HR3 & 4700-7900 & 0 & 256 & 64 & 32 & 0.9 $$ &  0.65 & $\pi$ & 65 & 200 & 1.97 & no & 0.28            & 0.35 \\
A7HR3 & 10000-11900 & 0.7 & 256 & 64 & 32 & 0.9 $$ &  0.65 & $\pi$ & 65 & 200 & 1.97 & no & 0.28          & 0.34 \\
A9HR3 & 4700-7900 & 0.9 & 256 & 64 & 32 & 0.88 $$ &  0.65 & $\pi$ & 65 & 200 & 1.97 & no & 0.28         & 0.341 \\
A98HR3 & 4700-7900 & 0.98 & 256 & 64 & 32 & 0.9 $$ &  0.65 & $\pi$ & 65 & 200 & 1.97 & no & 0.28        & 0.307 \\
C0 &  6000-10000 & 0 & 512 & 128 & 32 & 0.9 $$ &  0.15 & $\pi/4$ & 35 & 50 & 1.65 & yes & 0.16          & 0.052 \\
C1 & 12500-18900 & 0 & 256 & 64 & 16 & 0.9 $$ &  0.13 & $\pi/4$ & 35 & 50 & 1.65 & yes & 0.065          & 0.063\\
C2 & 12500-22500 & 0 & 256 & 64 & 64 & 0.9 $$ &  0.13 & $\pi$ & 35 & 50 & 1.65 & yes & 0.065            & 0.062\\
C3 & 12500-19500 & 0 & 256 & 64 & 16 & 0.9 $$ &  0.13 & $\pi/2$ & 35 & 50 & 1.65 & yes & 0.065          & 0.061\\
C4 & 12500-21700 & 0 & 256 & 64 & 64 & 0.9 $$ &  0.13 & $\pi/2$ & 35 & 50 & 1.65 & yes & 0.065          & 0.061\\
C5 & 12500-20000 & 0 & 256 & 32 & 32 & 0.9 $$ &  0.13 & $\pi/2$ & 35 & 50 & 1.65 & yes & 0.065          & 0.052\\
C6 & 12500-20800 & 0 & 256 & 128 & 32 & 0.9 $$ &  0.13 & $\pi/2$ & 35 & 50 & 1.65 & yes & 0.065         & 0.065\\
A0HR07LOOP1 & 17000-22100 & 0 & 256 & 64 & 32 & 0.9 $$ &  0.13 & $\pi/2$ & 35 & 50 & 1.65 & yes & 0.25        & 0.048\\
A0HR3LOOP1 & 3000-8000 & 0 & 256 & 64 & 32 & 0.9 $$ &  0.65 & $\pi$ & 65 & 200 & 1.97 & no & 0.5            & 0.377 \\
\hline
\end{tabular}
\end{center}
\label{tbl_models}
\end{table*}

\subsection{Numerical Methods}
\label{sec:nummethods}

We perform simulations using the GRMHD code HARM
that is based upon a conservative shock-capturing Godunov scheme.
One key feature of our code is that
we use horizon-penetrating Kerr-Schild coordinates
for the Kerr metric \citep{gam03,mg04,mck06ffcode,nob06,mm07,tch_wham07},
which avoids any issues with the coordinate singularity in Boyer-Lindquist coordinates.
Even with Kerr-Schild coordinates, one must ensure
that the inner-radial boundary of the computational domain
is outside the so-called inner horizon (at $r/M\equiv 1-\sqrt{1-(a/M)^2}$)
so that the equations remain hyperbolic,
and one must ensure that there are plenty of grid cells
spanning the region near the horizon in order to avoid
numerical diffusion out of the horizon.

Another key feature of our code is the use of a 3rd order accurate (4th order error)
PPM scheme for the interpolation of primitive quantities
(i.e. rest-mass density, 4-velocity relative to a ZAMO observer, and
lab-frame 3-magnetic field) \citep{mck06jf}.
The interpolation is similar to that described in \citet{cw84},
but we modified it to be consistent with interpolating through point values of primitives
rather than average values.
We do not use the PPM steepener, but we do use the PPM flattener
that only activates in strong shocks
(e.g. in the initial bow shock off the torus surface, but rarely elsewhere).
The PPM scheme attempts to fit a monotonic 3rd order polynomial directly through the grid
face where the dissipative flux enters in the Godunov scheme.
Only if the polynomial is non-monotonic does the interpolation reduce order
and create discontinuities at the cell face, and so only then
does it introduce dissipative fluxes.
It therefore leads to extremely small dissipation
compared to the original schemes used in HARM,
such as the 1st order accurate (2nd order error)
minmod or monotonized central (MC) limiter type schemes that always
create discontinuities (and so dissipative fluxes) at the cell face regardless
of the monotonicity for any primitive quantity that is not linear in space.

Simulations of fully three-dimensional models of accreting black holes producing jets
using our 3D GRMHD code show that this PPM scheme
leads to an improvement in effective resolution by at least factors of roughly two
per dimension as compared to the original
HARM MC limiter scheme for models with resolution $256\times 128\times 32$ \citep{mb09}.
The PPM method is particularly well-suited for resolving turbulent flows since they
rarely have strong discontinuities
and have most of the turbulent power in long wavelength modes.
Even moving discontinuities are much more accurately resolved by PPM than minmod or MC.
For example, even without a steepener, a simple moving contact
or moving magnetic rotational discontinuity
is sharply resolved within about 4 cells using the PPM scheme as
compared to being diffusively resolved within about 8-15
cells by the MC limiter scheme.

A 2nd order Runge-Kutta method-of-lines scheme is used to step forward in time,
and the timestep is set by using the fast magnetosonic wavespeed
with a Courant factor of $0.8$.
We found that a 4th order Runge-Kutta scheme does not significantly improve accuracy,
since most of the flow is far beyond the grid cells
inside the horizon that determine the timestep.
The standard HARM HLL scheme is used for the dissipative fluxes,
and the standard HARM T\'oth scheme is used for the magnetic field evolution.

\subsection{Numerical Model Setup}
\label{sec:numsetup}

The code uses uniform internal coordinates $(t,x^{(1)},x^{(2)},x^{(3)})$
mapped to the physical coordinates $(t,r,\theta,\phi)$.
The radial grid mapping is
\begin{equation}
r(x^{(1)}) = R_0 + \exp{(x^{(1)})} ,
\end{equation}
which spans from $R_{\rm in}$ to $R_{\rm out}$.
The parameter $R_0=0.3M$ controls the resolution near the horizon.
Absorbing (outflow, no inflow allowed) boundary conditions are used.
The $\theta$-grid mapping is
\begin{equation}
\theta(x^{(2)}) = [Y(2\xc-1) + (1-Y)(2\xc-1)^7 +1](\pi/2) ,
\end{equation}
where $x^{(2)}$ ranges from $0$ to $1$ (i.e. no cut-out at the poles)
and $Y$ is an adjustable parameter that can be used to concentrate grid
zones toward the equator as $Y$ is decreased from $1$ to $0$.
Roughly half of the $\theta$ resolution is concentrated
in the disk region within $\pm 2|h/r|$ of the midplane.
The HR07 and HR2 models listed in Table~\ref{tbl_models} have $11$ cells per $|h/r|$,
while the HR1 and HR3 models have $7$ cells per $|h/r|$.
The high resolution run, C6, has $22$ cells per $|h/r|$,
while the low resolution model, C5, has $5$ cells per $|h/r|$.
For $Y=0.15$ this grid gives roughly $6$ times more angular resolution compared
to the grid used in \citet{mg04} given by equation~(8) with $h=0.3$.
Reflecting boundary conditions are used at the polar axes.

The $\phi$-grid mapping is given by $\phi(x^{(3)}) = 2\pi x^{(3)}$,
such that $x^{(3)}$ varies from $0$ to $1/8,1/4,3/8,1/2$
for boxes with $\Delta\phi = \pi/4,\pi/2,3\pi/4,\pi$, respectively.
Periodic boundary conditions are used in the $\phi$-direction.
In all cases, the spatial integrals are renormalized to refer to the full $2\pi$ range in $\phi$,
even if our computational box size is limited in the $\phi$-direction.
We consider various $\Delta\phi$ in order to check whether this changes our results.
Previous GRMHD simulations with the full $\Delta\phi=2\pi$ extent
suggest that $\Delta\phi=\pi/2$ is sufficient since coherent
structures only extend for about one radian (see Fig.~12 in \citealt{skh06}).
However, in other GRMHD studies with $\Delta\phi=2\pi$,
the $m=1$ mode was found to be dominant,
so this requires further consideration \citep{mb09}.
Note that S08 used $\Delta\phi=\pi/4$, while both N09 and N10 used $\Delta\phi=\pi/2$.

The duration of our simulations with the thinnest disks
varies from approximately $20000M$ to $30000M$
in order to reach inflow equilibrium
and to minimize fluctuations in time-averaged quantities.
We ensure that each simulation runs for a couple of viscous times
in order to reach inflow equilibrium over a reasonable range of radius.
Note that the simulations cannot be run for a duration
longer than $t_{\rm acc}\sim M_{\rm disk}(t=0)/\dot{M}\sim 10^5M$,
corresponding to the time-scale for accreting
a significant fraction of the initial torus.  We are always well
below this limit.

Given finite computational resources,
there is a competition between duration and resolution of a simulation.
Our simulations run for relatively long durations,
and we use a numerical resolution of $N_r\times N_\theta \times N_\phi
=256\times 64\times 32$ for all models (except those used for convergence testing).
In S08 we found this resolution to be sufficient to obtain convergence compared
to a similar $512\times 128\times 32$ model with $\Delta\phi=\pi/4$.
In this paper, we explicitly confirm that our resolution is sufficient
by convergence testing our results (see section~\ref{sec:convergence}).
Near the equatorial plane at the ISCO,
the grid aspect ratio in $dr:r d\theta:r\sin\theta d\phi$
is 2:1:7, 1:1:4, 1:1:3, and 1:1:3, respectively, for
our HR07, HR1, HR2, and HR3 models.
The 2:1:7 grid aspect ratio for the HR07
model was found to be sufficient in S08.
A grid aspect ratio of 1:1:1 would be preferable
in order to ensure the dissipation is isotropic in Cartesian coordinates,
since in Nature one would not expect highly anisotropic dissipation
on the scale resolved by our grid cells.
However, finite computational resources require a balance
between a minimum required resolution, grid aspect ratio,
and duration of the simulation.

As described below, we ensure that the MRI is resolved in each simulation
both as a function of space and as a function of time by measuring
the number of grid cells per fastest growing MRI mode:
\begin{equation}
Q_{\rm MRI} \equiv \frac{\lambda_{\rm MRI}}{\Delta_{\hat{\theta}}} \approx 2\pi \frac{|v^{\hat{\theta}}_{\rm A}|/|\Omega(r,\theta)|}{\Delta_{\hat{\theta}}},
\end{equation}
where $\Delta_{\hat{\theta}} \equiv |{e^{\hat{\theta}}}_{\mu} dx^\mu|$
is the comoving orthonormal $\theta$-directed grid cell length,
${e^{\hat{\nu}}}_{\mu}$ is the contravariant tetrad system in the local fluid-frame,
$|v^{\hat{\theta}}_{\rm A}|=\sqrt{b_{\hat{\theta}} b^{\hat{\theta}}/(b^2 + \rho_0 + u_g + p_g)}$
is the~\alf speed,
$b^{\hat{\theta}} \equiv {e^{\hat{\theta}}}_{\mu} b^\mu$
is the comoving orthonormal $\theta$-directed 4-field,
and $|\Omega(r,\theta)|$ is the temporally and azimuthally averaged absolute value
of the orbital frequency.

During the simulation,
the rest-mass density and internal energy densities
can become quite low beyond the corona,
but the code only remains accurate and stable
for a finite value of $b^2/\rho_0$, $b^2/u_g$, and $u_g/\rho_0$ for any given resolution.
We enforce $b^2/\rho_0\lesssim 10^4$, $b^2/u_g\lesssim 10^4$, and $u_g/\rho_0\lesssim 10^4$
by injecting a sufficient amount of mass or internal energy
into a fixed zero angular momentum observer (ZAMO)
frame with 4-velocity $u_\mu=\{-\alpha,0,0,0\}$,
where $\alpha=1/\sqrt{-g^{tt}}$ is the lapse.
In some simulations, we have to use stronger limits
given by $b^2/\rho_0\lesssim 10$, $b^2/u_g\lesssim 10^2$, and $u_g/\rho_0\lesssim 10$,
in order to maintain stability and accuracy.
Compared to our older method of injecting mass-energy into the comoving frame,
the new method avoids run-away injection of energy-momentum in the low-density regions.
We have confirmed that this procedure of injecting mass-energy
does not contaminate our results for the accretion rates and other diagnostics.

\section{Fiducial Model of a Thin Disk Around a Non-Rotating Black Hole}
\label{sec:fiducialnonrot}

Our fiducial model, A0HR07, consists of a magnetized thin accretion disk around a
non-rotating ($a/M=0$) black hole.  This is similar to the model
described in S08;
however, here we consider a larger suite of diagnostics,
a resolution of $256\times64\times32$,
and a computational box with $\Delta\phi=\pi/2$.
As mentioned in section~\ref{sec:modelsetup},
the initial torus parameters are set so that the
inner edge is at $r=20M$,
the pressure maximum is at $r=35M$,
and $|h/r|\lesssim 0.1$ at the pressure maximum (see Figure~\ref{initialtorus}).

The initial torus is threaded with magnetic field in the multi-loop
geometry as described in section~\ref{sec:modelsetup}.
For this model, we use four loops in order to ensure that
the loops are roughly circular in the poloidal plane.  Once
the simulation begins, the MRI leads to MHD
turbulence which causes angular momentum transport and
drives the accretion flow to a quasi-steady state.

The fiducial model is evolved for a total time of $27350M$.  We
consider the period of steady-state to be from $T_i=12500M$ to
$T_f=27350M$ and of duration $\Delta T=14850M$.
All the steady-state results described below are obtained
by time-averaging quantities over this steady-state period,
which corresponds to about $160$ orbital periods at the ISCO,
$26$ orbits at the inner edge of the initial torus ($r=20M$),
and $11$ orbits at the pressure maximum of the initial torus ($r=35M$).

\subsection{Initial and Evolved Disk Structure}

\begin{figure}
\centering
\includegraphics[width=3.3in,clip]{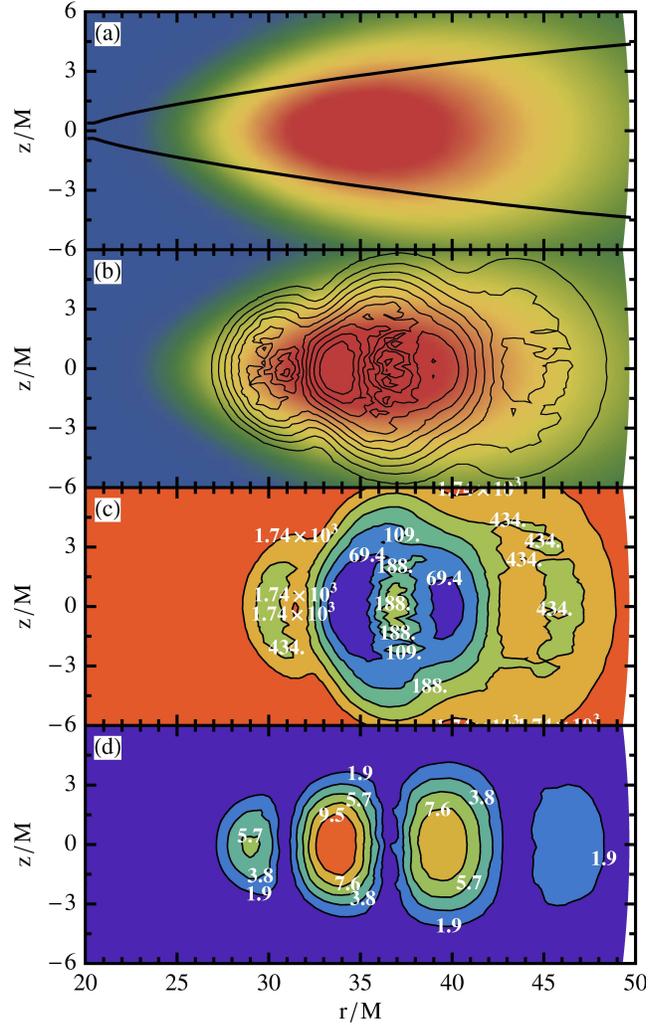}
\caption{The initial state of the fiducial model (A0HR07) consists of weakly magnetized
gas in a geometrically thin torus around a non-spinning ($a/M=0$) black hole.
Color maps have red as highest values and blue as lowest values.
Panel (a): Linear color map of rest-mass density,
with solid lines showing the thickness $|h/r|$ of the initial torus.  Note
that the black hole horizon is at $r=2M$, far to the left of the plot, so
the torus is clearly geometrically thin.
Near the pressure maximum $|h/r| \lesssim 0.1$, and elsewhere $|h/r|$ is even smaller.
Panel (b): Contour plot of $b^2$ overlaid on linear color map of rest-mass density
shows that the initial field consists of four poloidal loops centered at $r/M=
29,$ $34$, $39$, $45$.
The wiggles in $b^2$ are due to the initial perturbations.
Panel (c): Linear color map of the plasma $\beta$ shows that the disk is
weakly magnetized throughout the initial torus.
Panel (d): Linear color map of the number of grid cells per fastest growing MRI wavelength,
$Q_{\rm MRI}$,
shows that the MRI is properly resolved for the primary two loops at the center of the disk.
}
\label{initialtorus}
\end{figure}

Figure~\ref{initialtorus} shows contour plots of various quantities in
the initial solution projected on the
($R$,\;$z$) $=(r\sin\theta$,\;$r\cos\theta$)-plane.  Notice the relatively small
vertical extent of the torus.
The disk has a thickness of $|h/r|\sim0.06-0.09$ over the radius range
containing the bulk of the mass.  The four magnetic loops are clearly
delineated.  The plot of $Q_{\rm MRI}$ indicates
that the MRI is well-resolved within the two primary loops.
The left-most and right-most loops are marginally under-resolved, so a slightly slower-growing MRI mode
is expected to control the dynamics in this region.
However, the two primary loops tend to dominate the overall evolution of the gas.

\begin{figure}
\centering
\includegraphics[width=3.3in,clip]{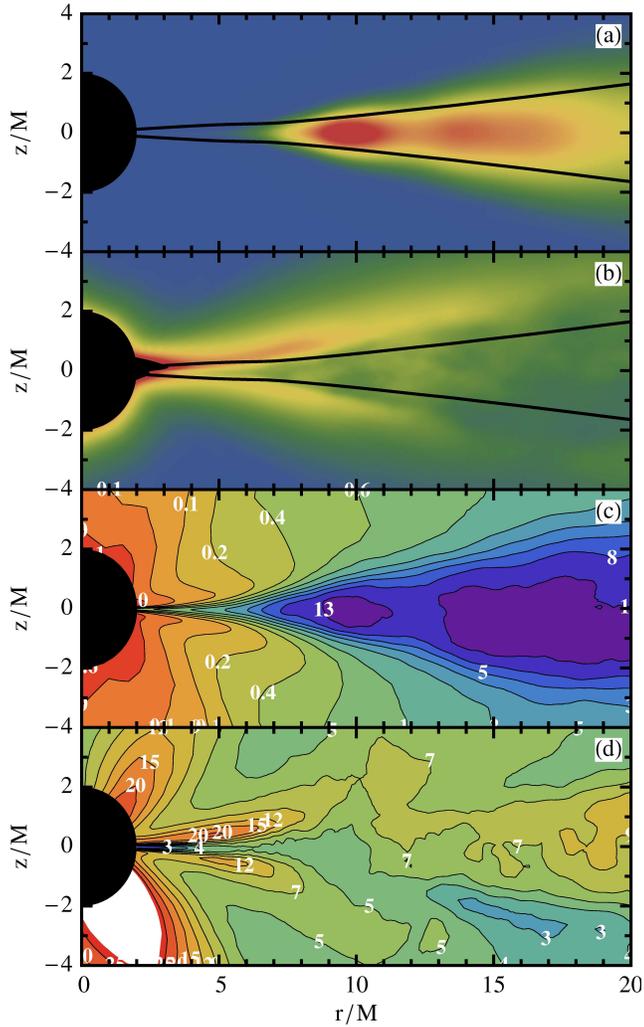}
\caption{The evolved state of the fiducial model (A0HR07)
consists of a weakly magnetized thin disk surrounded by a strongly magnetized corona.
All plots show quantities that have been time-averaged over the period $12500M$ to $27350M$.
Color maps have red as highest values and blue as lowest values.
Panel (a): Linear color map of rest-mass density, with solid lines showing
the disk thickness $|h/r|$.  Note that the
rest-mass density drops off rapidly inside the ISCO.
Panel (b): Linear color map of $b^2$ shows that a strong magnetic field is present
in the corona above the equatorial disk.
Panel (c): Linear color map of plasma $\beta$
shows that the $\beta$ values are much lower than in
the initial torus.  This indicates that considerable
field amplification has occurred via the MRI.
The gas near the equatorial plane has $\beta\sim 10$ far outside the ISCO and approaches
$\beta\sim 1$ near the black hole.
Panel (d): Linear color map of the number of grid cells per fastest
growing MRI wavelength, $Q_{\rm MRI}$,
shows that the MRI is properly resolved within most of the accretion flow.
Note that $Q_{\rm MRI}$ (determined by the vertical magnetic field strength)
is not expected to be large inside the plunging region
where the field is forced to become mostly radial
or above the disk within the corona where the field is mostly toroidal.
}
\label{finaltorus}
\end{figure}

Figure~\ref{finaltorus} shows the time-averaged solution during the
quasi-steady state period from $T_i=12500M$ to $T_f=27350M$.  We refer to the disk
during this period as being ``evolved'' or ``saturated.''
The evolved disk is in steady-state up to $r\sim 9M$,
as expected for the duration of our simulation.
The rest-mass density is concentrated in the disk midplane within $\pm 2|h/r|$,
while the magnetic energy density is concentrated above the disk in a corona.
The MRI is properly resolved with $Q_{\rm MRI}\approx 6$ in the
disk midplane\footnote{\citet{sano04} find that having about
$6$ grid cells per wavelength of the fastest growing MRI mode
during saturation leads to convergent behavior for the electromagnetic stresses,
although their determination of $6$ cells was based upon a 2nd order van Leer scheme
that is significantly more diffusive than our PPM scheme.
Also, the (time-averaged or single time value of) vertical field is already
(at any random spatial position)
partially sheared by the axisymmetric MRI, and so may be less relevant
than the (e.g.) maximum vertical field per unit orbital time
at any given point that is not yet sheared and so represents
the vertical component one must resolve.  These issues imply
we may only need about $4$ cells per wavelength of the fastest growing mode
(as defined by using the time-averaged absolute vertical field strength).}.
The gas in the midplane has plasma $\beta\sim 10$
outside the ISCO and $\beta\sim 1$ near the black hole,
indicating that the magnetic field has been amplified
beyond the initial minimum of $\beta\sim 100$.

\begin{figure}
\centering
\includegraphics[width=3.3in,clip]{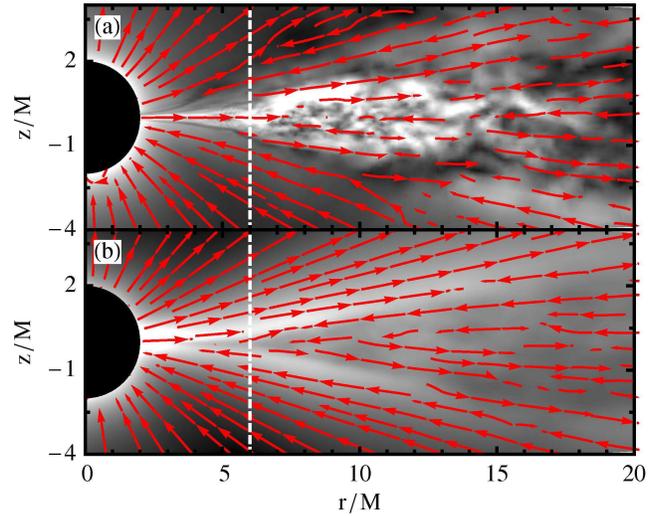}
\caption{Magnetic field lines (red vectors)
and magnetic energy density (greyscale map) are shown for the fiducial
model (A0HR07).
Panel (a): Snapshot of the magnetic field structure at time $27200M$
shows that the disk is highly turbulent for $r > r_{\rm ISCO}=6M$
and laminar for $r<r_{\rm ISCO}$.
Panel (b): Time-averaged magnetic field in the saturated state
shows that for $r\lesssim 9M$, viz., the region of the flow that we expect to have
achieved inflow equilibrium, the geometry of the time-averaged magnetic field
closely resembles that of a split-monopole.
The dashed, vertical line marks the position of the ISCO.
}
\label{magnetictorus}
\end{figure}

Figure~\ref{magnetictorus} shows the time-averaged structure of the
magnetic field during the quasi-steady state period.  The field has a
smooth split-monopole structure near and inside the ISCO.
Beyond $r\sim 9M$, however, the field becomes irregular, reversing direction more
than once.  At these radii, the simulation has not reached inflow equilibrium.

\subsection{Velocities and the Viscous Time-Scale}\label{sec:velvisc}

\begin{figure}
\centering
\includegraphics[width=3.3in,clip]{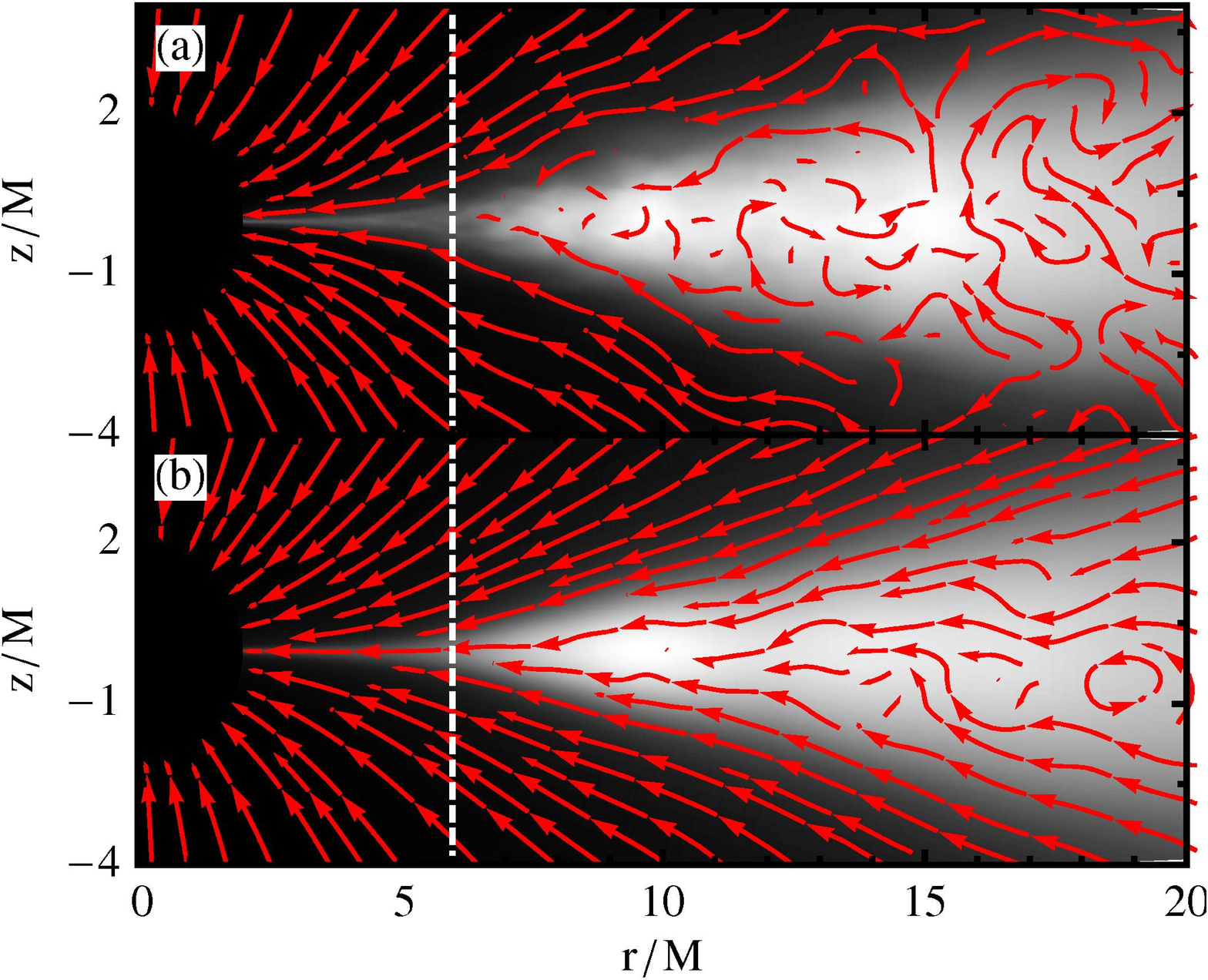}
\caption{Flow stream lines (red vectors) and rest-mass density (greyscale map)
are shown for the fiducial model (A0HR07).
Panel (a): Snapshot of the velocity structure and rest-mass density at time $27200M$
clearly shows MRI-driven turbulence in the interior of the disk.
The rest-mass density appears more diffusively distributed
than the magnetic energy density shown in Figure~\ref{magnetictorus}a.
Panel (b): Time-averaged streamlines and rest-mass density
show that for $r\lesssim 9M$ the velocity field is mostly radial
with no indication of a steady outflow.
Time-averaging smooths out the turbulent fluctuations
in the velocity.
The dashed, vertical line marks the position of the ISCO.
}
\label{velocitytorus}
\end{figure}

\begin{figure}
\centering
\includegraphics[width=3.3in,clip]{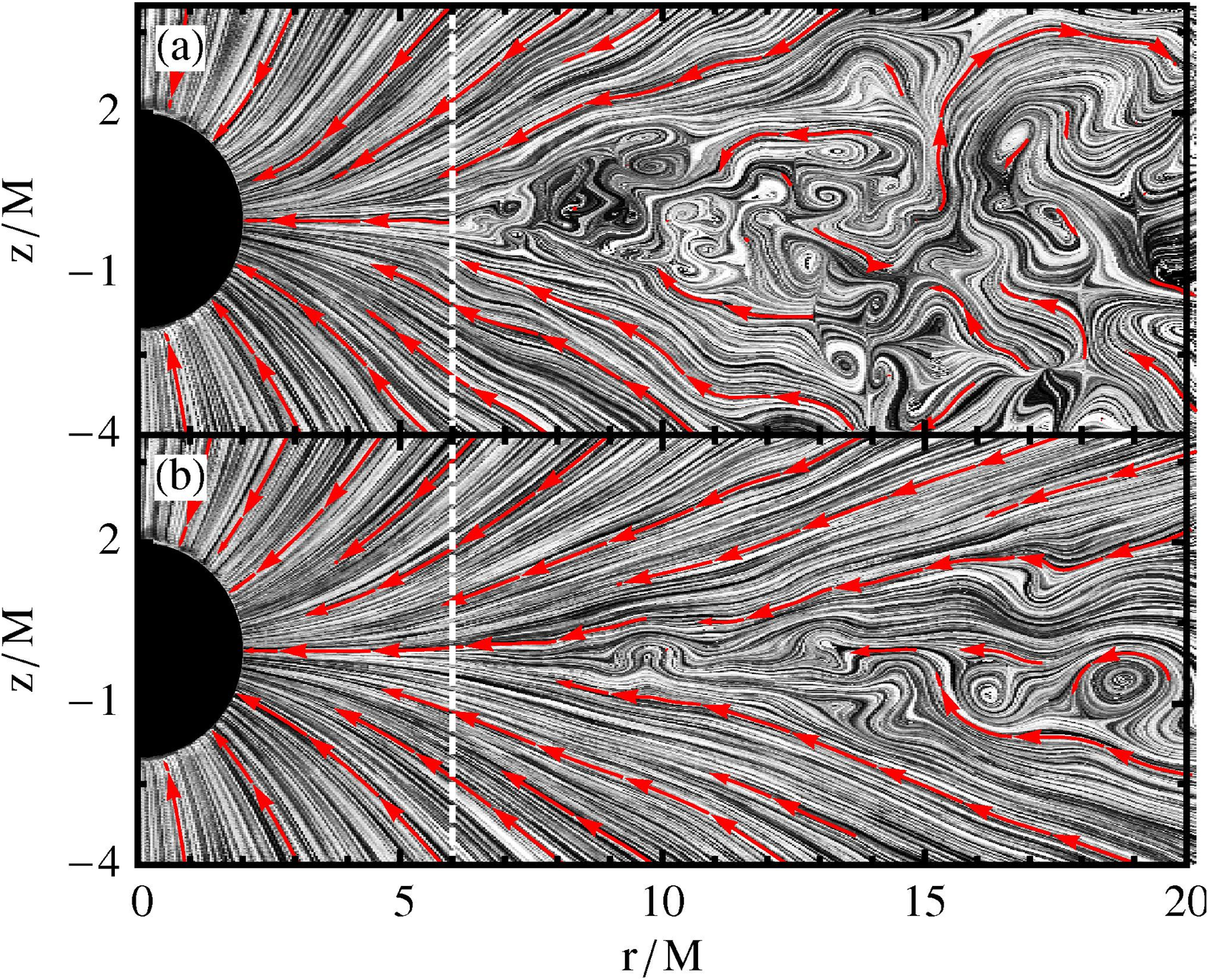}
\caption{Flow stream lines are shown for the fiducial model (A0HR07).
  Panel (a): Snapshot of the velocity structure
  at time $27200M$ clearly shows MRI-driven turbulence in the interior
  of the disk.
Panel (b): Time-averaged streamlines
  show that for $r\lesssim 9M$ the velocity field is mostly radial.
  The dashed, vertical line marks the position of the ISCO.
}
\label{streamtorus}
\end{figure}

Figure~\ref{velocitytorus} shows the velocity structure in the evolved
model.  The snapshot indicates well-developed
turbulence in the interior of the disk at radii beyond the ISCO
($r>6M$), but laminar flow inside the ISCO and over most of the
corona.  The sudden transition from turbulent to laminar behavior at
the ISCO, which is seen also in the magnetic field
(Figure~\ref{magnetictorus}a), is a clear sign that the flow dynamics
are quite different in the two regions.  Thus the ISCO
clearly has an affect on the accreting gas.  The time-averaged flow shows
that turbulent fluctuations are smoothed out within $r\sim 9M$.
Figure~\ref{streamtorus} shows the velocity stream lines
using the line integral convolution method to illustrate vector fields.  This figure again
confirms that the accretion flow is turbulent at radii larger than $r_{\rm ISCO}$
but it becomes laminar inside the ISCO,
and it again shows that time-averaging
smooths out turbulent fluctuations out to $r\sim 9M$.

\begin{figure}
\centering
\includegraphics[width=3.3in,clip]{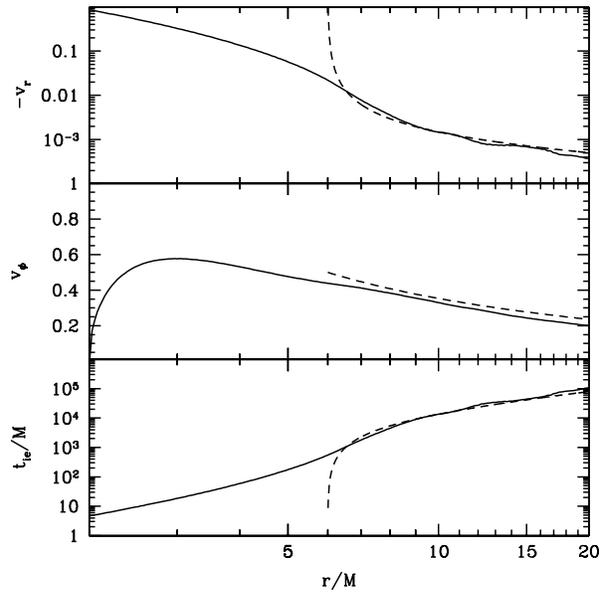}
\caption{The time-averaged, angle-averaged, rest-mass density-weighted
3-velocities and viscous timescale in the fiducial model (A0HR07)
are compared with the NT model.
Angle-averaging is performed over the disk gas lying within $\pm 2|h/r|$ of the midplane.
Top Panel: The orthonormal radial 3-velocity (solid line),
and the analytical GR estimate given in Eq. \ref{eq:inflow} of Appendix~\ref{sec_inflow} (dashed line).
Agreement for $r>r_{\rm ISCO}$ between the simulation
and NT model is found when we set $\alpha|h/r|^2\approx 0.00033$.
At smaller radii, the gas dynamics is no longer determined by viscosity
and hence the two curves deviate.
Middle Panel: Shows the orthonormal azimuthal 3-velocity $v_\phi$ (solid line)
and the corresponding Keplerian 3-velocity (dashed line).
Bottom Panel: The inflow equilibrium time scale $t_{\rm ie}\sim -2r/v_r$ (solid line)
of the disk gas is compared to the analytical GR thin disk estimate (dashed line).
At $r\sim 9M$, we see that $t_{\rm ie}\sim 2\times10^4M$.  Therefore,
the simulation needs to be run for this time period (which we do)
before we can reach inflow equilibrium at this radius.
}
\label{velvsr}
\end{figure}

Figure~\ref{velvsr} shows components of the time-averaged velocity
that are angle-averaged over $\pm 2|h/r|$ around the midplane (thick dashed lines
in Figure~\ref{horvsradius}).  By limiting the range of the
$\theta$ integral, we
focus on the gas in the disk, leaving out the corona-wind-jet.
Outside the ISCO, the radial velocity from the simulation agrees well with the analytical GR
estimate (Eq. \ref{eq:inflow} in Appendix~\ref{sec_inflow}).
By making this comparison, we found $\alpha |h/r|^2 \approx 0.00033$.
For our disk thickness
$|h/r| = 0.064$, this corresponds to
$\alpha\approx 0.08$, which is slightly smaller than the nominal
estimate $\alpha\sim0.1$ we assumed in \S\ref{sec_infloweq}.
As the gas approaches the ISCO, it accelerates rapidly in
the radial direction and finally free-falls into the black hole.  This
region of the flow is not driven by viscosity and hence the dynamics
here are not captured
by the analytical formula.

Figure~\ref{velvsr} also shows the inflow equilibrium time $t_{\rm ie}$,
which we take to be twice the GR version of the viscous time: $t_{\rm ie} = -2r/v_r$.
This is our estimate of the time it will take for the gas at a given radius to
reach steady-state.  We see that, in a time of $\sim 27350M$,
the total duration of our simulation, the solution can be in steady-state
only inside a radius of $\sim9M$.  Therefore, in the time-averaged
results described below, we consider the results to be reliable only over
this range of radius.

\subsection{Fluxes vs. Time}

\begin{figure}
\centering
\includegraphics[width=3.3in,clip]{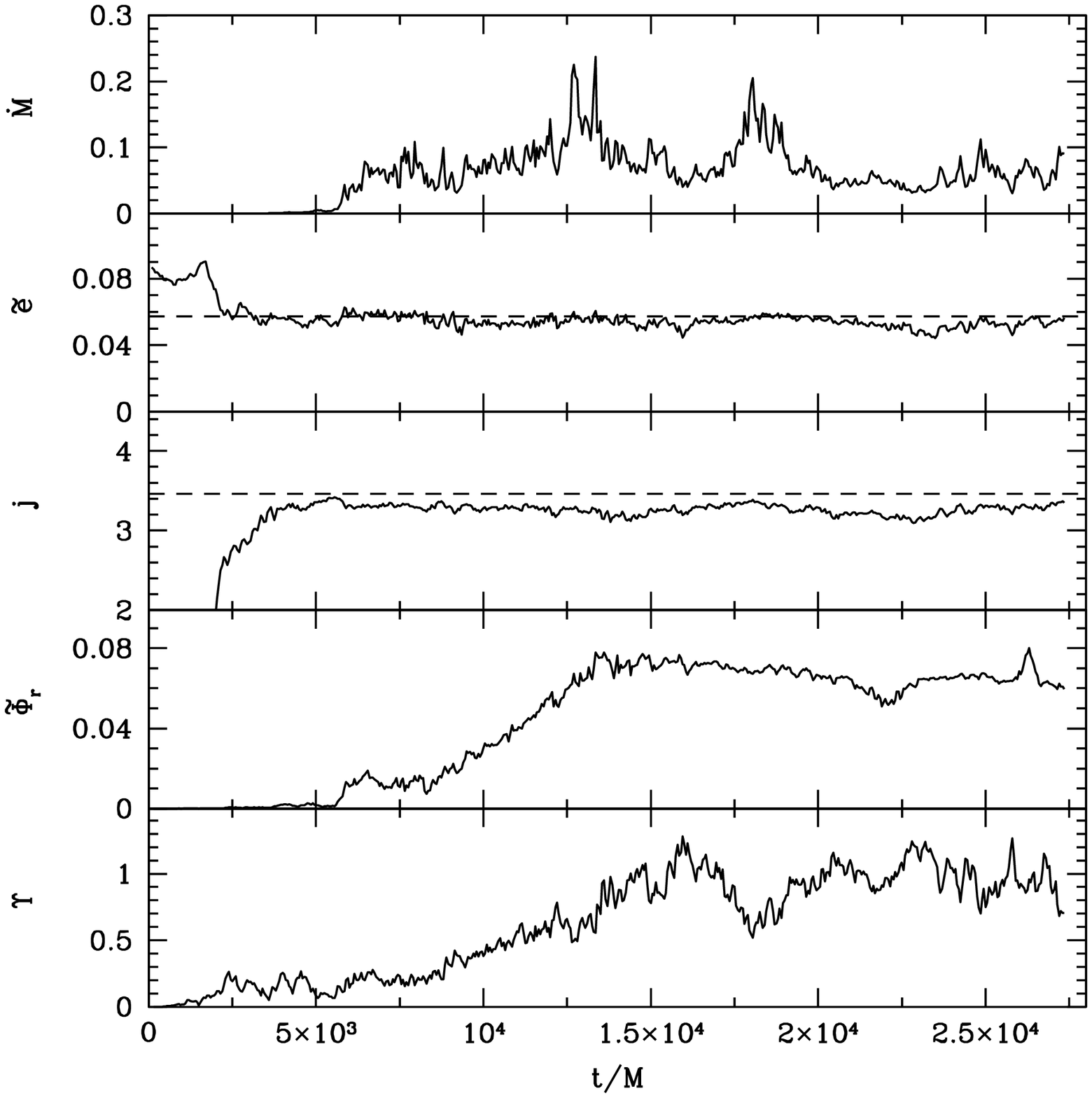}
\caption{Shows for the fiducial model (A0HR07)
the time-dependence at the horizon of the mass accretion rate, $\dot{M}$ (top panel);
nominal efficiency, $\eff$,
with dashed line showing the NT value (next panel);
accreted specific angular momentum, $\jmath$,
with dashed line showing the NT value (next panel);
absolute magnetic flux relative to
the initial absolute magnetic flux, $\tilde{\Phi}_r$ (next panel);
and dimensionless specific magnetic flux, $\Upsilon$ (bottom panel).
All quantities have been integrated over all angles.
The mass accretion rate varies by factors of up to four
during the quasi-steady state phase.
The nominal efficiency is close to, but on average slightly lower than,
the NT value.
This means that the net energy loss through photons, winds, and jets
is below the radiative efficiency of the NT model.
The specific angular momentum is clearly lower than the NT value,
which implies that some stresses are present inside the ISCO.
The absolute magnetic flux at the black hole horizon grows
until it saturates due to local force-balance.
The specific magnetic flux $\Upsilon\lesssim 1$,
indicating that electromagnetic stresses inside the ISCO are weak
and cause less than $7\%$ deviations from NT in $\jmath$.
}
\label{5dotpanel}
\end{figure}

Figure~\ref{5dotpanel} shows various fluxes vs. time that should
be roughly constant once inflow equilibrium has been reached.
The figure shows
the mass flux, $\dot{M}(r_{\rm H},t)$,
nominal efficiency, $\eff(r_{\rm H},t)$,
specific angular momentum, $\jmath(r_{\rm H},t)$,
normalized absolute magnetic flux, $\tilde{\Phi}_r(r_{\rm H},t)$,
(normalized using the unperturbed initial total flux),
and specific magnetic flux, $\Upsilon(r_{\rm H},t)$,
all measured at the event horizon ($r=r_H$).
These fluxes have been integrated over the
entire range of $\theta$ from 0 to $\pi$.  The quantities $\dot{M}$,
$\eff$ and $\jmath$ appear to saturate already at $t\sim7000M$.
However, the magnetic field parameters saturate only at $\sim12500M$.
We consider the steady-state period of the disk to begin
only after all these quantities reach their saturated values.

The mass accretion rate is quite variable, with root-mean-square (rms)
fluctuations of order two.
The nominal efficiency $\eff$ is fairly close to the NT efficiency,
while the specific angular momentum $\jmath$ is clearly below the NT value.
The results indicate that torques are present within the ISCO,
but do not dissipate much energy or cause significant energy to be transported
out of the ISCO.
The absolute magnetic flux per unit initial absolute flux, $\tilde{\Phi}_r$,
threading the black hole grows to about $1\%$,
which indicates that the magnetic field strength near the black hole is not
just set by the amount of magnetic flux in the initial torus.
This suggests our results are insensitive to the total
absolute magnetic flux in the initial torus.
The specific magnetic flux, $\Upsilon\approx 0.86$ on average.
Magnetic stresses are relatively weak since $\Upsilon\lesssim 1$,
which implies the magnetic field contributes
no more than $7\%$ to deviations from NT in $\jmath$ \citep{gammie99} ;
see Appendix~\ref{sec_gammie}.
During the quasi-steady state period,
the small deviations from NT in $\jmath$
are correlated in time with the magnitude of $\Upsilon$.
This is consistent with the fact that the specific magnetic flux controls these deviations.
Also, notice that $\tilde{\Phi}_r$ is roughly constant in time
while $\Upsilon$ varies in time.  This is clearly because $\dot{M}$
is varying in time and also consistent with the fact that
$\Upsilon$ and $\dot{M}$ are anti-correlated in time.

\subsection{Disk Thickness and Fluxes vs. Radius}
\label{sec:diskthick2}

\begin{figure}
\centering
\includegraphics[width=3.3in,clip]{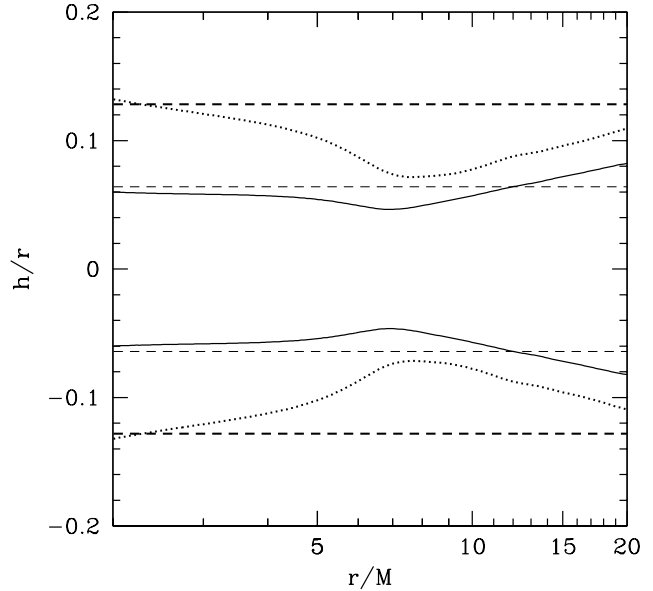}
\caption{The time-averaged scale-height, $|h/r|$, vs. radius in the fiducial model
(A0HR07) is shown by the solid lines.
The above-equator and below-equator values
of the disk thickness are $|h/r|\sim 0.04$--$0.06$
in the inflow equilibrium region $r<9M$.
We use the specific value of $|h/r|=0.064$ as measured at $r=2r_{\rm ISCO}$
(light dashed lines) as a representative thickness for the entire flow.
Twice this representative thickness (thick dashed lines)
is used to fix the $\theta$ range of integration for averaging
when we wish to focus only on the gas in the disk
instead of the gas in the corona-wind-jet.
The root mean square thickness $(h/r)_{\rm rms}\sim 0.07$--$0.13$
is shown by the dotted lines.
}
\label{horvsradius}
\end{figure}

Figure~\ref{horvsradius} shows the time-averaged disk thickness of the
fiducial model as a function of radius.  Both
measures of thickness defined in \S\ref{sec:diskthick1} are shown; they
track each other.  As expected, our primary thickness
measure, $|h/r|$, is the
smaller of the two.  This thickness measure varies by a small amount
across the disk, but it is generally consistent with the following fiducial
value, viz., the value $|h/r|=0.064$ at $r=2r_{\rm ISCO}=12M$.

\begin{figure}
\centering
\includegraphics[width=3.3in,clip]{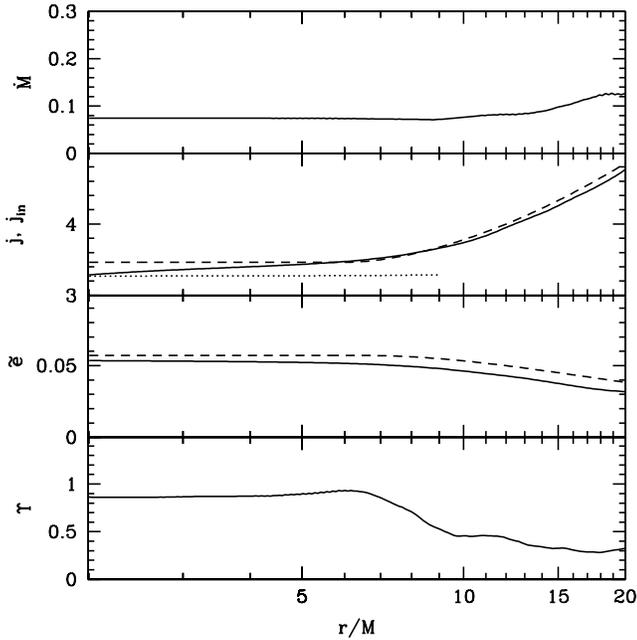}
\caption{Mass accretion rate and specific fluxes are shown
as a function of radius for the fiducial model (A0HR07).
From top to bottom the panels show:
Top Panel: mass accretion rate;
Second Panel: the accreted specific angular momentum, $\jmath$ (dotted line),
$\jmath_{\rm in}$ (solid line), and the NT profile (dashed line);
Third Panel: the nominal efficiency $\eff$ (solid line)
and the NT profile (dashed line);
Bottom Panel: the specific magnetic flux $\Upsilon$.
For all quantities the integration range includes all $\theta$.
The mass accretion rate and $\jmath$ are roughly constant
out to $r\sim 9M$, as we would expect for inflow equilibrium.
The profile of $\jmath_{\rm in}$ lies below the
NT value at large radii because we include gas in the slowly rotating corona.
At the horizon, $\jmath$ and $\eff$ are modestly below the corresponding NT values.
The quantity $\Upsilon\sim 0.86$ and is roughly constant out to $r\sim 6M$,
indicating that electromagnetic stresses are weak inside the ISCO.
}
\label{fluxesvsradius1}
\end{figure}

Figure~\ref{fluxesvsradius1} shows the behavior of various fluxes
versus radius for the full $\theta$ integration range ($0$ to $\pi$).
We see that the mass accretion rate, $\dot{M}$,
and the specific angular momentum flux, $\jmath$,
are constant up to a radius $r \sim 9M$.  This is exactly the
distance out to which we expect inflow equilibrium to have been
established, given the inflow velocity and viscous time scale results
discussed in \S\ref{sec:velvisc}.  The consistency of these two
measurements gives us confidence that the simulation has truly
achieved steady-state conditions inside $r=9M$.  Equally clearly,
and as also expected, the simulation is not in steady-state at larger radii.

The second panel in Figure~\ref{fluxesvsradius1} shows that
the inward angular momentum flux, $\jmath_{\rm in}$,
agrees reasonably well with the NT prediction.  It falls
below the NT curve at large radii, i.e., the gas there is
sub-Keplerian.  This is not surprising since we have included
the contribution of the corona-wind-jet gas which, being at high latitude,
does not rotate at the Keplerian rate.  Other quantities, described below,
show a similar effect due to the corona.  At the horizon, $\jmath_{\rm
in}=3.286$, which is $5\%$ lower than the NT value.  This deviation is
larger than that found by S08.  Once again, it is because
we have included the gas in the corona-wind-jet,
whereas S08 did not.

The third panel in Figure~\ref{fluxesvsradius1} shows that
the nominal efficiency $\eff$ at the horizon lies below the NT prediction.
This implies that the full accretion flow (disk+corona+wind+jet) is radiatively
less efficient than the NT model.
However, the overall shape of the
curve as a function of $r$ is similar to the NT curve.
The final panel in Figure~\ref{fluxesvsradius1} shows the value of $\Upsilon$
vs. radius.  We see that $\Upsilon\approx 0.86$ is constant out to $r\sim 6M$.
A value of $\Upsilon\sim 1$ would have led to $7\%$ deviations from NT in $\jmath$,
and only for $\Upsilon\sim 6.0$ would deviations become $50\%$ (see Appendix~\ref{sec_gammie}).
The fact that $\Upsilon\sim 0.86\lesssim 1$ indicates that electromagnetic stresses are weak
and cause less than $7\%$ deviations from NT in $\jmath$.
Note that one does not expect $\Upsilon$ to be constant\footnote{
We also find that the ideal MHD invariant related to the
``isorotation law'' of field lines,
$\Omega_F(r)\equiv \left(\int\int d\theta d\phi \detg |v^r B^\phi - v^\phi B^r|\right)/\left(\int\int d\theta d\phi\detg |B^r|\right)$,
is Keplerian outside the ISCO and is (as predicted by the \citealt{gammie99} model)
roughly constant from the ISCO to the horizon (see also \citealt{mg04,mck07a}).}
outside the ISCO where the magnetic field is dissipating due to MHD turbulence
and the gas is forced to be nearly Keplerian despite a sheared magnetic field.

\begin{figure}
\centering
\includegraphics[width=3.3in,clip]{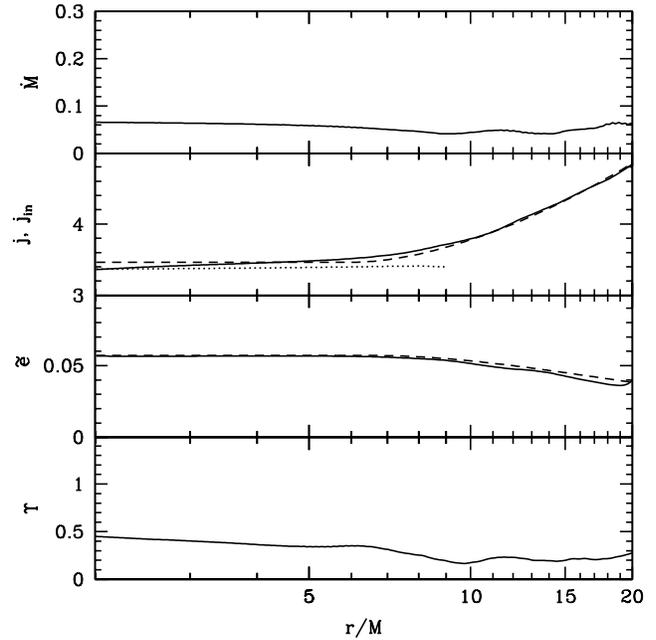}
\caption{Similar to Figure~\ref{fluxesvsradius1},
but here the integration range only includes angles within
$\pm 2|h/r|=\pm0.128$ radians of the midplane.  This allows us to focus on the disk gas.
The mass accretion rate is no longer constant
because streamlines are not precisely radial.
The quantities shown in the second and third panels are not
affected by the non-constancy of $\dot{M}$
because they are ratios of time-averaged fluxes
within the equatorial region and are related to ideal MHD invariants.
As compared to Figure~\ref{fluxesvsradius1},
here we find that $\jmath$, $\jmath_{\rm in}$, and $\eff$ closely follow the NT model.
For example, $\jmath(r_{\rm H})=3.363$ is only $2.9\%$ less than NT.
This indicates that the disk and coronal regions behave quite differently.
As one might expect, the disk region behaves like the NT model,
while the corona-wind-jet does not.
The specific magnetic flux is even smaller than in
Figure~\ref{fluxesvsradius1} and is  $\Upsilon\sim 0.45$,
which indicates that electromagnetic stresses
are quite weak inside the disk near the midplane.
}
\label{fluxesvsradius2}
\end{figure}

As we have hinted above, we expect
large differences between the properties of the gas that
accretes in the disk proper, close to the midplane, and that which
flows in the corona-wind-jet region.  To focus just on the disk gas, we show
in Figure~\ref{fluxesvsradius2} the same fluxes as in Figure~\ref{fluxesvsradius1},
except that we have restricted the $\theta$ range to
$\pi/2 \pm 2|h/r|$.  The mass accretion rate is
no longer perfectly constant for $r<9M$.  This is simply a consequence
of the fact that the flow streamlines do not perfectly follow the
particular constant $2|h/r|$ disk boundary we have chosen.  The
non-constancy of $\dot{M}$ does not significantly affect the other quantities
plotted in this figure since they are all normalized by the local $\dot{M}$.

The specific angular momentum, specific energy, and specific magnetic flux
are clearly shifted closer to the NT values when we restrict the angular integration range.
Compared to the NT value, viz., $\jmath_{\rm NT}(r_{\rm H})=3.464$, the fiducial model
gives $\jmath(r_{\rm H})=3.363$ ($2.9\%$ less than NT) when integrating
over $\pm 2|h/r|$ around the midplane (i.e., only over the disk gas)
and gives $\jmath(r_{\rm H})=3.266$ ($5.7\%$ less than NT) when integrating over all $\theta$
(i.e., including the corona-wind-jet).
Even though the mass accretion rate through the corona-wind-jet
is much lower than in the disk, still this gas
contributes essentially as much to the deviation of the specific angular momentum
as the disk gas does.
In the case of the specific magnetic flux, integrating over
$\pm 2|h/r|$ around the midplane we find $\Upsilon\approx 0.45$,
while when we integrate over all angles $\Upsilon\approx 0.86$.
The \citet{gammie99} model of an equatorial (thin) magnetized flow
within the ISCO shows that deviations in the specific angular
momentum are determined by the value of $\Upsilon$.
We find that the measured values of $\Upsilon$ are able to roughly predict
the measured deviations from NT in $\jmath$.

In summary, a comparison of Figure~\ref{fluxesvsradius1} and
Figure~\ref{fluxesvsradius2} shows
that all aspects of the accretion flow in the fiducial simulation
agree much better with the NT prediction
when we restrict our attention to regions close to the midplane.
In other words, the gas in the disk proper,
defined here as the region lying within $\pm 2|h/r|$ of the midplane, is well
described by the NT model.  The deviation of the angular momentum flux
$\jmath_{\rm in}$ or $\jmath$ at the horizon relative to NT is $\lesssim 3\%$,
similar to the deviation found by S08\footnote{The quantities $\jmath_{\rm in}$ and
$\jmath$ are nearly equal at the horizon in the calculations reported
here whereas they were different in S08.  This is because S08 used an
alternate definition of $\jmath_{\rm in}$.  If we had used that definition
here, we would have found a deviation of $\sim2\%$ in $\jmath_{\rm in}$,
just as in S08}, while the nominal efficiency $\eff$ agrees to within $\sim1\%$.

\subsection{Comparison with Gammie (1999) Model}
\label{sec_gammiecompare}

\begin{figure}
\centering
\includegraphics[width=3.3in,clip]{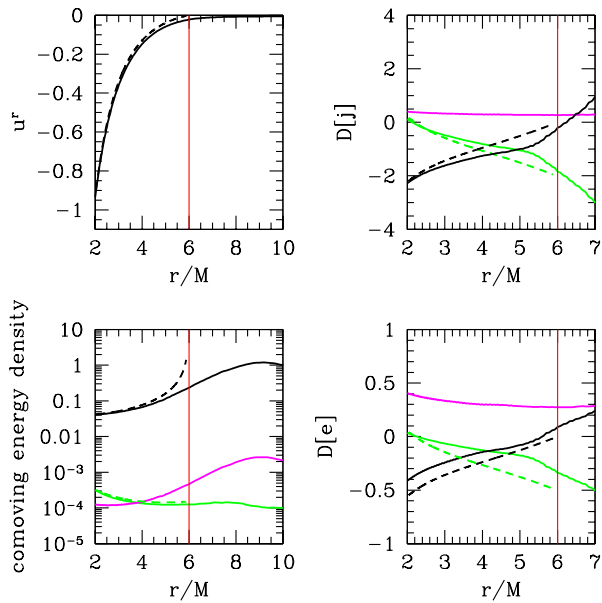}
\caption{Comparison between the accretion flow
(within $\pm 2|h/r|$ around the midplane) in the fiducial model (A0HR07), shown by solid lines,
and the model of a magnetized thin accretion disk
(inflow solution) within the ISCO by \citet{gammie99}, shown by dashed lines.
In all cases the red vertical line shows the location of the ISCO.
Top-left panel: Shows
the radial 4-velocity, where the Gammie solution assumes $u^r=0$ at the ISCO.
Finite thermal effects lead to non-zero $u^r$ at the ISCO for the simulated disk.
Bottom-left panel: Shows the rest-mass density ($\rho_0$, black line),
the internal energy density ($u_g$, magenta line),
and magnetic energy density ($b^2/2$, green line).
Top-right and bottom-right panels: Show the percent deviations
from NT for the simulations and Gammie solution
for the specific particle kinetic flux ($u_\mu$, black line),
specific enthalpy flux ($(u_g + p_g) u_\mu/\rho_0$, magenta line),
and specific electromagnetic flux ($(b^2 u^r u_\mu - b^r b_\mu)/(\rho_0 u^r)$, green line),
where for $\jmath$ we use $\mu=\phi$ and for $\emath$ we use $\mu=t$.
As usual, the simulation result for the specific fluxes
is obtained by a ratio of flux integrals instead of the direct ratio of flux densities.
The total specific flux is constant vs. radius and is a sum of
the particle, enthalpy, and electromagnetic terms.
This figure is comparable to Fig.~10 for a thick ($|h/r|\sim 0.2$--$0.25$)
disk in \citet{mg04}.
Finite thermal pressure effects cause the fiducial model to deviate
from the inflow solution near the ISCO, but the solutions rapidly
converge inside the ISCO and the differences between the simulation
result and the Gammie model (relative to the total specific angular momentum or energy)
are less than $0.5\%$.
}
\label{gammie4panel}
\end{figure}

Figure~\ref{gammie4panel} shows a comparison between the fiducial model
and the \citet{gammie99} model of a magnetized thin accretion flow within the ISCO
(see also Appendix~\ref{sec_gammie}).
Quantities have been integrated within $\pm 2|h/r|$ of the midplane
and time-averaged over a short period from $t=17400M$ to $t=18400M$.
Note that time-averaging $b^2$, $\rho_0$, etc.
over long periods can lead to no consistent comparable solution if the value of $\Upsilon$
varies considerably during the period used for averaging.
Also, note that the presence of vertical stratification,
seen in Figures~\ref{fluxesvsradius1} and~\ref{fluxesvsradius2}
showing that $\Upsilon$ depends upon height,
means the vertical-averaging used to obtain $\Upsilon$
can sometimes make it difficult to compare the simulations
with the \citet{gammie99} model which has no vertical stratification.
In particular, using equation~(\ref{Dotsgammie}) over this time period,
we find that $\Upsilon\approx 0.2~,0.3,~0.44,~0.7,~0.8$
for integrations around the midplane of, respectively,
$\pm0.01,~\pm0.05,~\pm2|h/r|,~\pm\pi/4,~\pm\pi/2$,
with best matches to the Gammie model (i.e. $b^2/2$ and other quantities match)
using an actual value of $\Upsilon=0.2,~0.33,~0.47,~0.8,~0.92$.
This indicates that stratification likely causes
our diagnostic to underestimate the best match with the Gammie model
once the integration is performed over highly-stratified regions.
However, the consistency is fairly good considering how much $\Upsilon$
varies with height.

Overall, Figure~\ref{gammie4panel} shows
how electromagnetic stresses control the deviations from NT within the ISCO.
The panels with $D[\jmath]$ and $D[\emath]$ show how the electromagnetic flux starts
out large at the ISCO and drops to nearly zero on the horizon.
This indicates the electromagnetic flux has been converted
into particle flux within the ISCO by ideal (non-dissipative) electromagnetic stresses\footnote{This
behavior is just like that seen in ideal MHD jet solutions, but inverted with radius.}.
The simulated magnetized thin disk agrees quite well with the Gammie solution,
in contrast to the relatively poor agreement found for thick disks \citep{mg04}.
Only the single parameter $\Upsilon$ determines the Gammie solution,
so the agreement with the value and radial dependence among multiple independent terms
is a strong validation that the Gammie model is working well.
Nevertheless, there are some residual deviations
near the ISCO where the thermal pressure dominates the magnetic pressure.
Even if deviations from NT are present right at the ISCO,
the total deviation of the particle flux between the ISCO and horizon
equals the deviation predicted by the \citet{gammie99} model,
as also found in \citet{mg04} for thick disks.
This indicates that the \citet{gammie99} model accurately predicts
the effects of electromagnetic stresses inward of the ISCO.

Finally, note that the electromagnetic stresses within the ISCO
are ideal and non-dissipative in the Gammie model.
Since the flow within the ISCO in the simulation
is mostly laminar leading to weak non-ideal (resistive or viscous) effects,
the dissipative contribution (which could lead to radiation) can be quite small.
An exception to this is the presence of extended current sheets,
present near the equator within the ISCO in the simulations,
whose dissipation requires a model of the (as of yet, poorly understood)
rate of relativistic reconnection.

\subsection{Luminosity vs. Radius}

\begin{figure}
\centering
\includegraphics[width=3.3in,clip]{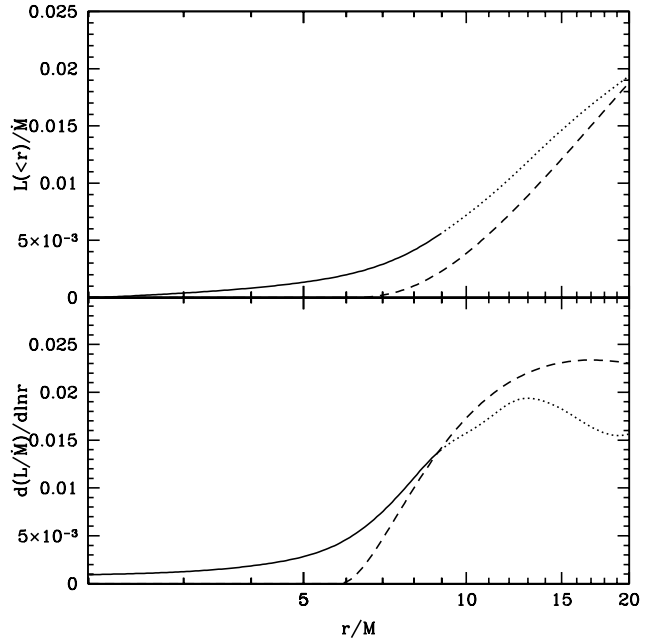}
\caption{Luminosity per unit rest-mass accretion rate vs. radius (top panel)
and the logarithmic derivative of this quantity (bottom panel)
are shown for the fiducial model (A0HR07).
The integration includes all $\theta$ angles.
The simulation result (solid lines, truncated into dotted lines
outside the radius of inflow equilibrium)
shows that the accretion flow emits more radiation
than the NT prediction (dashed lines) at small radii.
However, the excess luminosity within the ISCO
is only $\tilde{L}_{\rm in}\approx 3.5\%$,
where $\eff[{\rm NT}]$ is the NT efficiency at the horizon
(or equivalently at the ISCO).}
\label{luminosityvsradius}
\end{figure}

Figure~\ref{luminosityvsradius} shows radial profiles of two measures
of the disk luminosity: $L(<r)/\dot{M}$, which is the cumulative
luminosity inside radius $r$, and $d(L/\dot{M})/d\ln{r}$, which gives
the local luminosity at $r$.
We see that the profiles from the simulation are quite close to
the NT prediction, especially in the steady-state
region.  As a way of measuring the deviation of the simulation results from
the NT model, we estimate what fraction of the disk luminosity is
emitted inside the ISCO; recall that the NT model predicts zero luminosity here.
The fiducial simulation gives $L(<r_{\rm ISCO})/\dot{M} = 0.0021$,
which is $3.5\%$ of the nominal efficiency $\eff[{\rm NT}]=0.058$ of a thin NT disk
around a non-spinning black hole.
This shows that the excess luminosity radiated within the ISCO is quite small.
The relative luminosity within the ISCO is $\tilde{L}_{\rm in}=3.5\%$
and the relative luminosity within the inflow equilibrium region is
$\tilde{L}_{\rm out}=8.0\%$.
Hence, we conclude that, for accretion disks which are as thin as our
fiducial model, viz., $|h/r|\sim0.07$,
the NT model provides a good description of the luminosity profile.

\subsection{Luminosity from Disk vs. Corona-Wind-Jet}
\label{sec_fluxdiskcorona}

The fiducial model described so far includes a tapering of the cooling
rate as a function of height above the midplane, given by the
function $S[\theta]$ (see equation~\ref{cooling}).
We introduced this taper in order to only cool bound ($-u_t(\rho_0+u_g+ p_g+b^2)/\rho_0 < 1$)
gas and to avoid including the emission from the part of the corona-wind-jet
that is prone to excessive numerical dissipation due to the low resolution
used high above the accretion disk.
This is a common approach that others have also taken
when performing GRMHD simulations of thin disks (N09, N10).
However, since our tapering function does not explicitly
refer to how bound the gas is, we need to check that it
is consistent with cooling only bound gas.
We have explored this question by re-running the fiducial model with all
parameters the same except that we turned off the tapering function altogether,
i.e., we set $S[\theta]=1$.
This is the only model for which the tapering function is turned off.

\begin{figure}
\centering
\includegraphics[width=3.3in,clip]{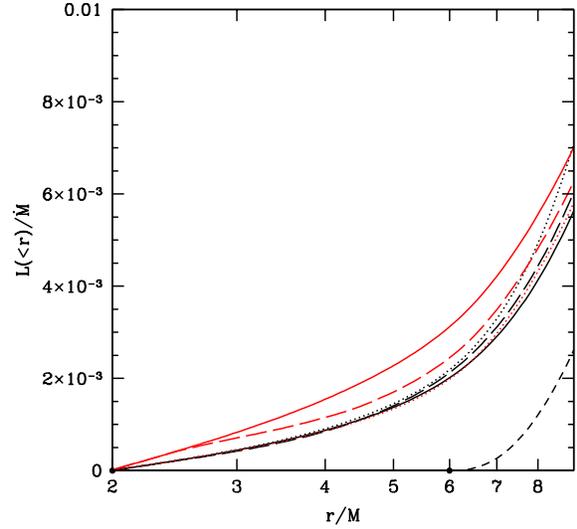}
\caption{Shows enclosed luminosity vs. radius for models
with different cooling prescriptions and $\theta$
integration ranges.  The black dashed line corresponds to the
NT model.  The luminosity for the fiducial model A0HR07,
which includes a tapering of the cooling with disk height as
described in \S\ref{sec:goveqns},
is shown integrated over $\pm2|h/r|$ from the midplane (black dotted line),
integrated over all bound gas (black long dashed line), and integrated over all fluid (black solid line).  Essentially all the gas is bound and so the black solid and long dashed lines are indistinguishable.
The red lines are for a model that is identical to the fiducial run,
except that no tapering is applied to the cooling.
For this model the lines are:
red solid line: all angles, all fluid;
red dotted line: $\pm 2|h/r|$ around the midplane;
red long dashed line: all bound gas.
The main result is that the luminosity from bound gas is
nearly the same (especially at the ISCO)
whether or not we include tapering
(compare the red long dashed line and the black long dashed line).
}
\label{taperoff}
\end{figure}

Figure~\ref{taperoff} shows a number of luminosity profiles
for the fiducial model and the no-tapering model.
This comparison shows that, whether or not we include a taper,
the results for the luminosity from all the bound gas is nearly the same.
Without a tapering, there is some luminosity at high latitudes above $\pm 8|h/r|$
corresponding to emission from the low-density jet region (black solid line).
This region is unbound and numerically inaccurate,
and it is properly excluded when we use the tapering function.
Another conclusion from the above test is that, as far as the
luminosity is concerned, it does not matter much whether we focus
on the midplane gas ($(\pi/2)\pm2|h/r|$) or include all the bound gas.
The deviations of the luminosity from NT in the two cases are similar --
{\it changes} in the deviation are less than $1\%$.

An important question to ask is whether the excess luminosity from
within the ISCO is correlated with, e.g., deviations from NT in
$\jmath$, since $D[\jmath]$ could then be used as a proxy for the
excess luminosity.  We investigate this in the context of the
simulation with no tapering.  For an integration over $\pm 2|h/r|$
around the midplane (which we identify with the disk component), or
over all bound gas, or over all the gas (bound and unbound), the
excess luminosity inside the ISCO is $\tilde{L}_{\rm
in}=3.3\%,~4.4\%,~5.4\%$, and the deviation from NT in $\jmath$ is
$D[\jmath]=-3.6\%, ~-6.7\%, ~-6.7\%$, respectively.  We ignore the
luminosity from unbound gas since this is mostly due to material in a
very low density region of the simulation where thermodynamics is not
evolved accurately.  Considering the rest of the results, we see that
$D[\jmath]$ is $100\%$ larger when we include bound gas outside the
disk compared to when we consider only the disk gas, whereas the
excess luminosity increases by only $32\%$.  Therefore, when we
compute $\jmath$ by integrating over all bound gas and then assess
the deviation of the simulated accretion flow from the NT model, we
strongly overestimate the excess luminosity of the bound gas relative
to NT.  A better proxy for the latter is the deviations from NT in
$\jmath$ integrated only over the disk component (i.e. over $\pm
2|h/r|$ around the midplane).

Furthermore, we note that the gas that lies beyond $\pm 2|h/r|$ from
the disk midplane consists of coronal gas, which is expected to be
optically thin and to emit a power-law spectrum of photons.  For many
applications, we are not interested in this component but rather care
only about the thermal blackbody-like emission from the
optically-thick region of the disk.  For such studies, the most
appropriate diagnostic from the simulations is the radiation emitted
within $\pm2|h/r|$ of the midplane.  According to this diagnostic, the
excess emission inside the ISCO is only $\tilde{L}_{\rm in}=3.4\%$ in
the model without tapering, and $3.5\%$ in the fiducial model that
includes tapering.

Lastly, we consider variations in the cooling timescale, $\tau_{\rm cool}$ ,
which is another free parameter of our cooling model that we generally set to $2\pi/\Omega_{\rm K}$.
However, we consider one model that is otherwise identical to the fiducial model
except we set $\tau_{\rm cool}$ to be five times shorter so that the
cooling rate is five times faster.
We find that $\tilde{L}_{\rm in}=4.2\%$,
which is slightly larger than the fiducial model with
$\tilde{L}_{\rm in}=3.5\%$.
Even though the cooling rate is five times faster than an orbital rate,
there is only $20\%$ more luminosity from within the ISCO.
This is likely due to the flow within the ISCO being mostly laminar
with little remaining turbulence to drive dissipation and radiation.

\section{Convergence with Resolution and Box Size}
\label{sec:convergence}

\begin{figure}
  \begin{center}
      \includegraphics[width=3.3in]{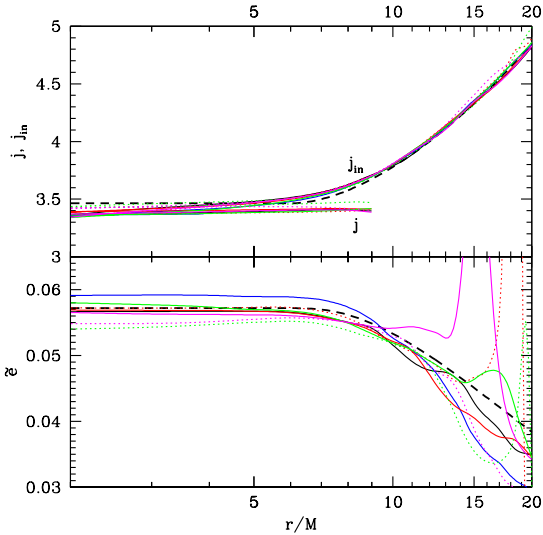}
  \end{center}
  \caption{This plot shows $\jmath$, $\jmath_{\rm in}$, and $\eff$
  for a sequence of simulations that are similar to the fiducial run
  (A0HR07), viz., $|h/r|\approx 0.07$, $a/M=0$, but use different
  radial resolutions, or  $\theta$ resolutions, or box sizes.
  The integration range in $\theta$ is over $\pm 2|h/r|$ around the midplane.
  Only the region of the flow in inflow equilibrium, $2M<r<9M$, is
  shown in the case of $\jmath$.
  The different lines are as follows:
  black dashed line: NT model; black solid line: fiducial model A0HR07;
  blue solid line: model C0 (S08);
  magenta dotted line: model C1;
  magenta solid line: model C2;
  red dotted line: model C3;
  red solid line: model C4;
  green dotted line: model C5;
  green solid line: model C6.
  Note that changes in the numerical resolution
  or other computational parameters lead to
  negligible changes in the values of $\jmath$, $\jmath_{\rm in}$,
  and $\eff$ in the region of the flow that is in inflow equilibrium, $r<9M$.
  For $r\gtrsim 9M$, the flow has not achieved steady state,
  which explains the large deviations in $\eff$.
  Only the lowest resolution models are outliers.
  }
  \label{fluxconvramesh}
\end{figure}

\begin{figure}
  \begin{center}
      \includegraphics[width=3.3in]{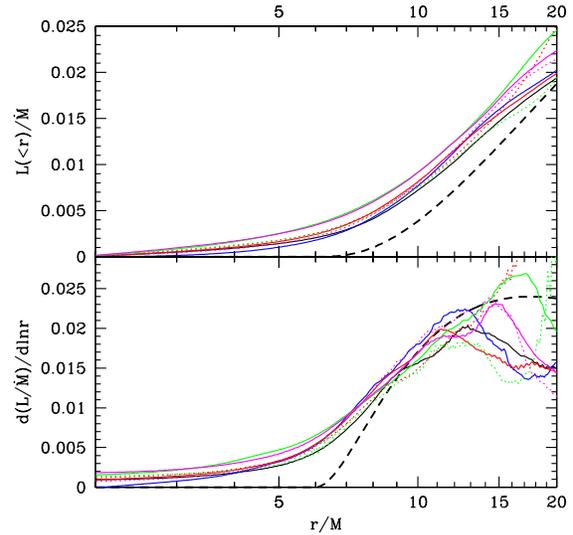}
  \end{center}
  \caption{Similar to Figure~\ref{fluxconvramesh},
  but for the normalized luminosity, $L(<r)/\dot{M}$,
  and its logarithmic derivative, $d(L/\dot{M})/d\ln{r}$, both shown vs. radius.
  We see that all the models used to test convergence show consistent
  luminosity profiles over the region that is in inflow equilibrium, $r<9M$.
  The well-converged models have
  $\tilde{L}_{\rm in}\lesssim 4\%$,
  which indicates only a low level of luminosity
  inside the ISCO.
  }
  \label{lumconvramesh}
\end{figure}

\begin{figure}
  \begin{center}
      \includegraphics[width=3.3in,clip]{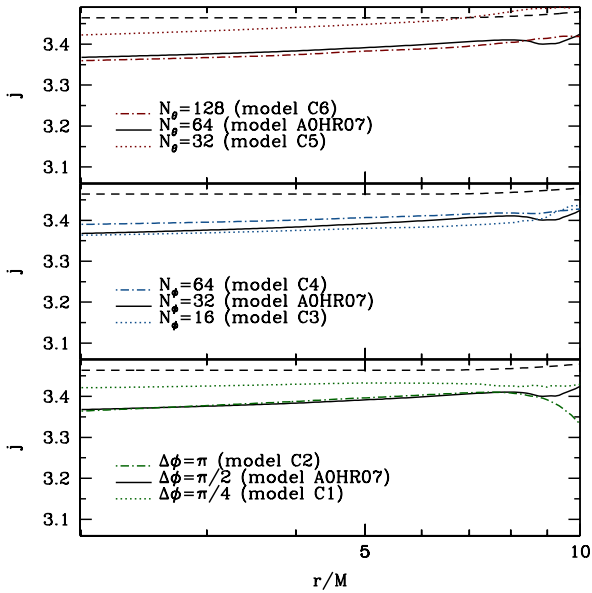}
  \end{center}
  \caption{This is a more detailed version of Figure~\ref{fluxconvramesh},
  showing $\jmath$ vs $r$ for individually labeled models.
  The models correspond to the fiducial resolution (solid lines),
  a higher resolution run (dot-dashed lines),
  and a lower resolution run (dotted lines).
  Generally, there are only minor differences between the fiducial and higher resolution models.
  }
  \label{jin4panel}
\end{figure}

\begin{table*}
\caption{Convergence Study}
\begin{center}
\begin{tabular}[h]{|r|r|r|r|r|r|r|r|r|r|r|r|r|r|}
\hline
Model Name & $|h/r|$ & $\dot{M}$ & $\eff$ & $D[\eff]$ & $\jmath$ & $D[\jmath]$ & $\jmath_{\rm in}$ & $D[\jmath_{\rm in}]$ & $s$ & $\tilde{L}_{\rm in}$  & $\tilde{L}_{\rm eq}$ & $100\tilde{\Phi}_r$ & $\Upsilon$ \\
\hline
\underline{$\pm 2|h/r|$} & \\
A0HR07 & 0.064  & 0.066   &  0.058  &  -0.829  &  3.363 &   -2.913  &  3.355 &   -3.153  &  3.363  &  0.035  & 0.080  &   1.355  &   0.450   \\
C0     & 0.052  & 0.064   &  0.059  &  -2.708  &  3.363 &   -2.905 &   3.351 &   -3.259  &   3.363  & 0.019  & 0.042  &   0.399  &   0.359\\
C1     & 0.063  &  0.041  &  0.056  &  2.204  &  3.415  &  -1.406  &  3.408  &  -1.621  &  3.415  &   0.034  & 0.072  &   0.584  &   0.223 \\
C2     & 0.062  &  0.063  &  0.058  &  -0.699  &  3.360 &   -3.016  &  3.333 &   -3.789  &  3.360  &  0.047  & 0.109  &   0.925  &   0.323  \\
C3     & 0.061  &  0.026  &  0.058  &  -1.969  &  3.358 &   -3.060  &  3.339 &   -3.610  &  3.358  &  0.039  & 0.087  &   0.727  &   0.449  \\
C4     & 0.061  &  0.054  &  0.058  &  -1.406  &  3.385 &   -2.286  &  3.378 &   -2.489  &  3.385  &  0.019  & 0.018  &   0.714  &   0.296  \\
C5     & 0.052  &  0.008  &  0.055  &  3.955  &  3.417  &  -1.364  &  3.427  &  -1.067  &  3.417  &   0.034  & 0.070  &   0.315  &   0.322  \\
C6     & 0.065  &  0.088  &  0.059  &  -3.355  &  3.355 &   -3.155  &  3.333 &   -3.778  &  3.355  &  0.054  & 0.103  &   0.933  &   0.256  \\
\underline{All $\theta$} & \\
A0HR07 &  0.064 &  0.074  &  0.054  &  4.723  &  3.266  &  -5.717  &  3.281  &  -5.275  &  3.266  &  0.035  & 0.053  &  6.677  &   0.863   \\
C0     & 0.052  &  0.071  &  0.057  &  0.738  &  3.312  &  -4.392  &  3.291  &  -5.002  &  3.312  &  0.032  & 0.049  &  2.18   &  0.480 \\
C1     & 0.063  &  0.042  &  0.055  &  3.680  &  3.398  &  -1.894  &  3.389 &   -2.178  &  3.398  &  0.037  & 0.062  &  0.940  &   0.142   \\
C2     & 0.062  &  0.069  &  0.056  &  1.664  &  3.307  &  -4.541  &  3.262 &   -5.846  &  3.307  &  0.053  & 0.080  &  5.757  &   0.710   \\
C3     & 0.061  &  0.029  &  0.057  &  0.893  &  3.325  &  -4.008  &  3.305 &   -4.580  &  3.325  &  0.042  & 0.064  &  1.401  &   0.309   \\
C4     & 0.061  &  0.057  &  0.057  &  0.811  &  3.358  &  -3.075  &  3.351 &   -3.255  &  3.358  &  0.020  & 0.009  &  2.690  &   0.359   \\
C5     & 0.052  &  0.009  &  0.053  &  7.687  &  3.338  &  -3.636  &  3.353 &   -3.218  &  3.338  &  0.039  & 0.059  &  0.726  &   0.243   \\
C6     & 0.065  &  0.092  &  0.058  & -1.813  &  3.334  &  -3.761  &  3.306 &   -4.560  &  3.334  &  0.057  & 0.086  &  5.091  &   0.534   \\
\hline
\end{tabular}
\end{center}
\label{tbl_resolution}
\end{table*}

The fiducial model described earlier was computed with a numerical
resolution of $256\times64\times32$, using an azimuthal wedge of
$\pi/2$.  This is to be compared with the simulation described in S08,
which made use of a $512\times128\times32$ grid and used a $\pi/4$ wedge.
These two runs give very similar results, suggesting that the details
of the resolution and wedge size are not very important.
To confirm this, we have run a number of simulations with
different resolutions and wedge angles.  The complete list
of runs is:
$256\times64\times32$ with $\Delta\phi=\pi/2$ (fiducial run, model A0HR07)),
$512\times128\times32$ with $\Delta\phi=\pi/4$ (S08, model C0),
$256\times128\times32$ with $\Delta\phi=\pi/2$ (model C6),
$256\times32\times32$ with $\Delta\phi=\pi/2$ (model C5),
$256\times64\times64$ with $\Delta\phi=\pi/2$ (model C4),
$256\times64\times64$ with $\Delta\phi=\pi$ (model C2),
$256\times64\times16$ with $\Delta\phi=\pi/2$ (model C3), and
$256\times64\times16$ with $\Delta\phi=\pi/4$ (model C1).

Figure~\ref{fluxconvramesh} shows the
accreted specific angular momentum, $\jmath$,
ingoing component of the specific angular momentum, $\jmath_{\rm in}$,
and the nominal efficiency $\eff$ as functions of radius
for all the models used for convergence testing.
Figure~\ref{lumconvramesh} similarly shows the cumulative
luminosity $L(<r)/\dot{M}$ and differential luminosity $d(L/\dot{M})/d\ln{r}$
as functions of radius.
The overwhelming impression from these plots is that the
sequence of convergence simulations agree with one another quite
well.  Also, the average of all the runs matches the NT model very
well; this is especially true for the steady-state region of the flow,
$r<9M$.  Thus, qualitatively, we conclude that our results
are well-converged.

For more quantitative comparison, Figure~\ref{jin4panel} shows the
profile of $\jmath$ vs $r$
for the various models, this time with each model separately identified.
It is clear that $\jmath$ has converged,
since there are very minor deviations from our
highest resolution/largest box size to our next
highest resolution/next largest box size.
All other quantities, including $\eff$, $\jmath_{\rm in}$, and $\Upsilon$
are similarly converged.
The model with $N_\phi=64$ shows slightly {\it less} deviations from NT
in $\jmath$ than our other models.
However, it also shows slightly higher luminosity than our other models.
This behavior is likely due to the stochastic temporal
behavior of all quantities vs. time,
but this could also be due to the higher $\phi$-resolution causing
a weaker ordered magnetic field to be present
leading to weaker ideal electromagnetic stresses,
smaller deviations from NT in $\jmath$ within the ISCO,
but with the remaining turbulent field being dissipated giving a higher luminosity.
The $N_\phi$ resolution appears to be the limit on our accuracy.

Further quantitative details are given in Table~\ref{tbl_resolution},
where we list numerical results for all the convergence test
models, with the $\theta$ integration
performed over both $\pm 2|h/r|$ around the midplane and over all angles.
We see that there are some trends as a
function of resolution and/or $\Delta\phi$.  Having only $32$ cells in
$\theta$ or $16$ cells in $\phi$ gives somewhat poor results, so these
runs are under-resolved.  However, even for these
runs, the differences are not large.
Note that $\Upsilon$ reaches a steady-state much later than all other
quantities, and our C? (where ? is 0 through 6)
models did not run as long as the fiducial model.
This explains why $\Upsilon$ is a bit lower for the C? models.
Overall, we conclude that our choice of resolution $256\times 64\times 32$
for the fiducial run (A0HR07) is adequate to reach convergence.

\section{Dependence on Black Hole Spin and Disk Thickness}
\label{sec:thicknessandspin}

In addition to the fiducial model and the convergence runs described
in previous sections, we have run a number of other simulations to
explore the effect of the black hole spin parameter $a/M$ and the disk
thickness $|h/r|$ on our various diagnostics: $\jmath$, $\jmath_{\rm
in}$, $\eff$, the luminosity, and $\Upsilon$.
We consider four values of the black hole
spin parameter, viz., $a/M=0, ~0.7, ~0.9, ~0.98$, and four disk
thicknesses, viz., $|h/r|=0.07, ~0.1, ~0.2, ~0.3$.  We summarize
here our results for this $4\times4$ grid of models.

Geometrically thick disks are expected on quite general grounds to
deviate from the standard thin disk model.  The inner edge of the
disk, as measured for instance by the location of the sonic point, is
expected to deviate from the ISCO, the shift scaling roughly as
$|r_{\rm in}-r_{\rm ISCO}| \propto (c_{\rm s}/v_{\rm K})^2$ ($c_{\rm s}$ is
sound speed, where $c_{\rm s}^2=\Gamma p_g/(\rho_0 + u_g + p_g)$).  This
effect is seen in hydrodynamic models of thick disks, e.g.
\citet{nkh97} and \citet{abr10}, where it is shown that $r_{\rm in}$
can move either inside or outside the ISCO; it moves inside when $\alpha$
is small and outside when $\alpha$ is large.  In either case, these
hydrodynamic models clearly show that, as $|h/r|\to 0$, i.e., as
$c_{\rm s}/v_{\rm K}\to 0$, the solution always tends toward the NT model \citep{snm08}.

While the hydrodynamic studies mentioned above have driven much of our
intuition on the behavior of thick and thin disks, it is an open
question whether or not the magnetic field plays a significant role.
In principle, magnetic effects may cause
the solution to deviate significantly from the NT model even in
the limit $|h/r|\to 0$ \citep{krolik99,gammie99}.  One of the major
goals of the present paper is to investigate this question.  We show
in this section that, as $|h/r|\to 0$, magnetized disks do tend toward
the NT model.  This statement appears to be true for a range of black
hole spins.  We also show that the specific magnetic flux $\Upsilon$
inside the ISCO decreases with decreasing $|h/r|$ and remains quite
small.  This explains why the magnetic field does not cause significant
deviations from NT in thin disks.

\begin{figure}
\centering
\includegraphics[width=3.3in,clip]{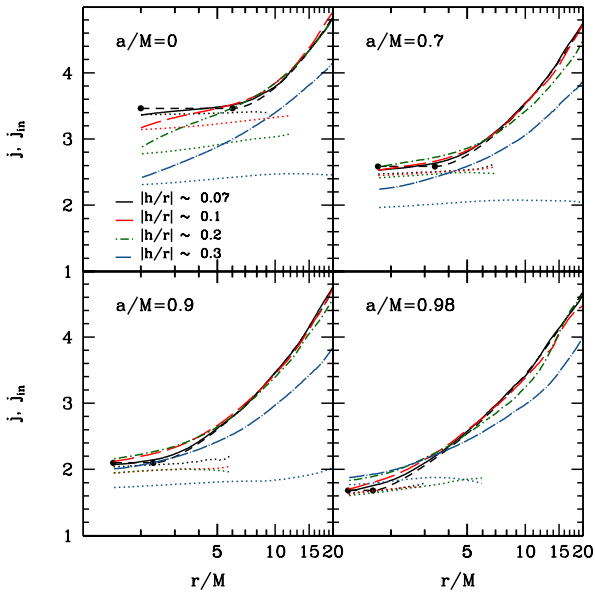}
\caption{The net accreted specific angular momentum, $\jmath$ (the nearly horizontal
dotted lines),
and the ingoing component of this quantity, $\jmath_{\rm in}$ (the sloping curved lines),
as a function of radius for the $4\times4$ grid of models.
Each panel corresponds to a single black hole spin, $a/M=0, ~0.7, ~0.9$, or
$0.98$, and shows models
with four disk thicknesses, $|h/r|\approx 0.07, ~0.1, ~0.2, ~0.3$ (see legend).
The $\theta$ integral has been taken over $\pm 2|h/r|$ around the midplane.
In each panel, the thin dashed black line, marked by two circles
which indicate the location of the horizon and the ISCO,
shows the NT solution for $\jmath_{\rm in}$.
As expected, we see that thicker disks exhibit larger deviations from NT.
However, as a function of spin, there is no indication that deviations from NT
become any larger for larger spins.  In the case of the thinnest models with $|h/r| \approx 0.07$,
the NT model works well for gas close to
the midplane for all spins.
}
\label{jspecvsspin}
\end{figure}

Figure~\ref{jspecvsspin} shows the specific angular momentum, $\jmath$,
and the ingoing component of this quantity,
$\jmath_{\rm in}$, vs. radius for the $4\times 4$ grid of models.
The $\theta$ integral has been taken over $\pm 2|h/r|$ around the midplane
in order to focus on the equatorial disk properties.
The value of $\jmath$ is roughly constant out to a radius
well outside the ISCO,
indicating that we have converged solutions in inflow equilibrium
extending over a useful range of radius.
As discussed in section~\ref{sec_infloweq},
inflow equilibrium is expected within $r/M=9,~7,~5.5,~5$,
respectively, for $a/M=0,~0.7,~0.9,~0.98$.
This is roughly consistent with the radius out to which
the quantity $\jmath$ (integrated over all angles) is constant,
and this motivates why in all such plots we only show $\jmath$ over the
region where the flow is in inflow equilibrium.
The four panels in Figure~\ref{jspecvsspin}
show a clear trend, viz., deviations from NT are larger
for thicker disks, as expected.
Interestingly, for higher black hole spins,
the relative deviations from NT actually decrease.

\begin{figure}
\centering
\includegraphics[width=3.3in,clip]{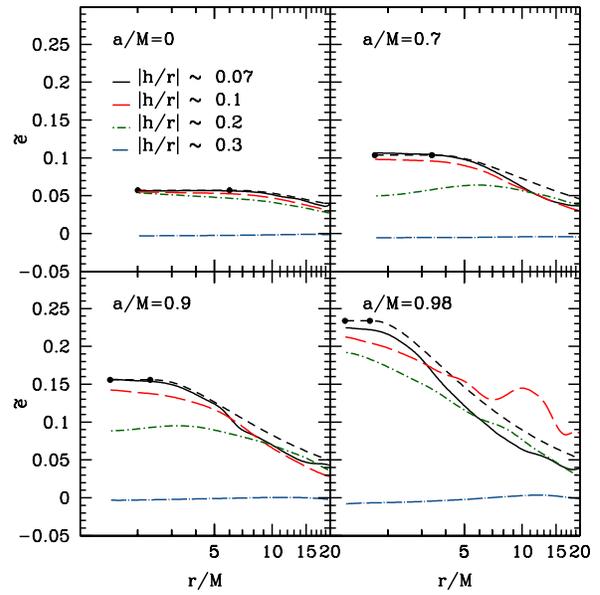}
\caption{Similar to Figure~\ref{jspecvsspin}, but for the nominal efficiency, $\eff$.
For thin ($|h/r|\lesssim 0.1$) disks,
the results are close to NT for all black hole spins.
As expected, the thicker models deviate significantly from NT.
In part this is because the ad hoc cooling function
we use in the simulations
is less accurate for thick disks,
and in part because the models with $|h/r|\approx 0.3$ have no cooling
and start with marginally bound/unbound gas that implies $\eff\sim 0$.
The $a/M=0.98$ models show erratic behavior at large radii
where the flow has not achieved inflow equilibrium.
}
\label{effvsspin}
\end{figure}

Figure~\ref{effvsspin} shows the nominal efficiency, $\eff$, as a function
of radius for the $4\times 4$ grid of models.
Our thickest disk models ($|h/r|\approx 0.3$) do not include
cooling, so the efficiency shown is only due to losses by a wind-jet.
We see that the efficiency is fairly close to the NT
value for all four thin disk simulations with $|h/r|\sim0.07$; even in the worst case,
viz., $a/M=0.98$, the
deviation from NT is only $\sim5\%$.
In the case of thicker disks, the efficiency shows larger deviations
from NT and the profile as a function of radius also looks different.
For models with $|h/r|\approx0.3$, there is no cooling so large deviations are
expected.

\begin{figure}
\centering
\includegraphics[width=3.3in,clip]{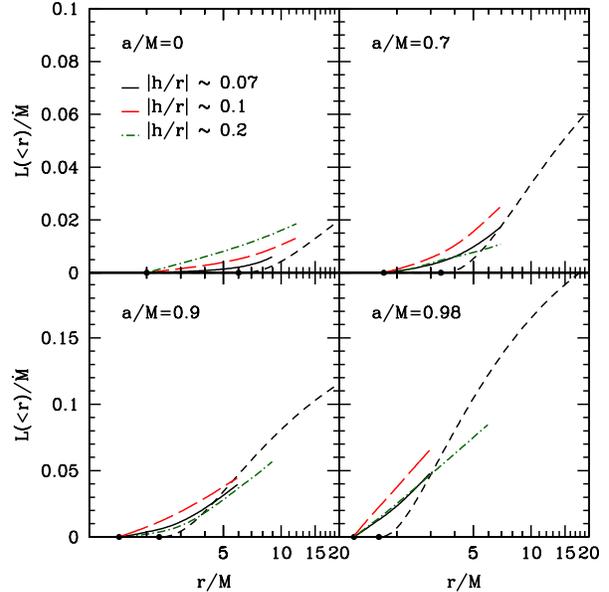}
\caption{Similar to Figure~\ref{jspecvsspin}, but for the normalized
  luminosity, $L(<r)/\dot{M}$.
  For thin disks, the luminosity deviates only slightly from
  NT near and inside the ISCO.  There is no strong evidence for any
  dependence on the black hole spin.
  The region at large radii has not reached
  inflow equilibrium and is not shown.
  }
\label{lumvsspin}
\end{figure}

Figure~\ref{lumvsspin} shows the luminosity, $L(<r)/\dot{M}$,
vs. radius for our $4\times 4$ grid of models, focusing just on the
region that has reached inflow equilibrium.
The luminosity is estimated by integrating over all $\theta$ angles.
Our thickest disk models ($|h/r|\approx 0.3$) do not include
cooling and so are not plotted.
The various panels show that,
as $|h/r|\to 0$, the luminosity becomes progressively closer to the NT
result in the steady state region of the flow near and inside the ISCO.
Thus, once again, we conclude that the NT luminosity profile is valid for
geometrically thin disks even when the accreting gas is magnetized.

A figure (not shown) that is similar to Figure~\ref{jspecvsspin} but
for the specific magnetic flux
indicates that $\Upsilon\le 1$ within $\pm 2|h/r|$
near the ISCO for all black hole spins and disk thicknesses.
For our thinnest models, $\Upsilon\le 0.45$,
for which the model of \citet{gammie99}
predicts that the specific angular momentum will deviate from NT
by less than $1.9\%, ~3.0\%, ~3.8\%, ~4.2\%$ for black hole spins
$a/M=0, ~0.7, ~0.9, ~0.98$, respectively (see Appendix~\ref{sec_gammie}).
The numerical results from the simulations
show deviations from NT that are similar to these values.
Thus, overall, our results indicate that
electromagnetic stresses are weak inside the ISCO for geometrically thin disks.

Finally, for all models
we look at plots (not shown) of $M(<r)$ (mass enclosed within radius),
$\dot{M}(r)$ (total mass accretion rate vs. radius),
and $[h/r](r)$ (disk scale-height vs. radius).
We find that these are consistently flat to the same degree
and to the same radius as the quantity $\jmath(r)$ is constant as shown
in Figure~\ref{jspecvsspin}.
This further indicates that our models are in inflow equilibrium
out to the expected radius.

\subsection{Scaling Laws vs. $a/M$ and $|h/r|$}\label{scalinglaws}

\begin{table*}
\caption{Black Hole Spin and Disk Thickness Study}
\begin{center}
\begin{tabular}[h]{|r|r|r|r|r|r|r|r|r|r|r|r|r|r|r|}
\hline
Model Name & $|h/r|$ & $\dot{M}$ & $\eff$ & $D[\eff]$ & $\jmath$ & $D[\jmath]$ & $\jmath_{\rm in}$ & $D[\jmath_{\rm in}]$ & $s$ &  $\tilde{L}_{\rm in}$  & $\tilde{L}_{\rm eq}$ &  $100\tilde{\Phi}_r$ & $\Upsilon$ \\
\hline
\underline{$\pm 2|h/r|$} & \\
A0HR07  & 0.064 &  0.066 &    0.058 &   -0.829  &   3.363 &   -2.913  &   3.355 &   -3.153 &    3.363 &   0.035   &   0.080   & 1.355   &   0.450  \\
A7HR07  & 0.065 &  0.050 &    0.107 &   -2.899  &   2.471 &   -4.465  &   2.527 &   -2.294 &    1.220 &   0.048   &   0.084   & 0.919   &   0.393  \\
A9HR07  & 0.054 &  0.045 &    0.156 &   -0.157  &   2.042 &   -2.762  &   2.074 &   -1.213 &    0.523 &   0.041   &   0.082   & 0.455   &   0.218  \\
A98HR07 & 0.059 &  0.013 &    0.225 &    3.897  &   1.643 &   -2.335  &   1.679 &   -0.199 &    0.124 &   0.069   &   0.127   & 0.276   &   0.228  \\
A0HR1   & 0.12  &  4.973 &    0.056 &    1.470  &   3.138 &   -9.424  &   3.162 &   -8.724 &    3.138 &   0.084   &   0.134   & 2.976   &   0.871  \\
A7HR1   & 0.09  &  2.443 &    0.099 &    4.808  &   2.446 &   -5.447  &   2.524 &   -2.409 &    1.184 &   0.060   &   0.108   & 0.909   &   0.406  \\
A9HR1   & 0.13  &  2.133 &    0.142 &    9.014  &   1.947 &   -7.261  &   2.124 &    1.157 &    0.402 &   0.068   &   0.107   & 1.064   &   0.466  \\
A98HR1  & 0.099 &  2.372 &    0.213 &    8.810  &   1.626 &  -3.393   &   1.703 &    1.200 &    0.084 &   0.062   &   0.112   & 0.451   &   0.254  \\
A0HR2   & 0.18  & 48.286 &    0.055 &    4.134  &   2.774 &  -19.916  &   2.872 &  -17.098 &    2.774 &   0.167   &   0.235   & 2.518   &   1.235  \\
A7HR2   & 0.16  & 31.665 &    0.049 &   52.330  &   2.412 &   -6.736  &   2.576 &   -0.425 &    1.081 &   0.050   &   0.034   & 0.919   &   0.631  \\
A9HR2   & 0.21  & 40.603 &    0.090 &   41.922  &   1.946 &   -7.315  &   2.155 &    2.624 &    0.309 &   0.011   &  -0.026   & 0.795   &   0.557  \\
A98HR2  & 0.18  & 29.410 &    0.191 &   18.496  &   1.588 &   -5.650  &   1.870 &   11.117 &    0.001 &   0.052   &   0.068   & 0.651   &   0.459  \\
A0HR3   & 0.350 & 44.066 &   -0.003 &  104.582  &   2.309  & -33.331  &   2.408 &  -30.473 &    2.309 &   0.000   &  -0.049   & 3.039   &   1.182  \\
A7HR3   & 0.34  & 41.045 &   -0.007 &  106.384  &   1.967 &  -23.970  &   2.236 &  -13.549 &    0.557 &   0.000   &  -0.060   & 1.956   &   0.746  \\
A9HR3   & 0.341 & 35.852 &   -0.001 &  100.582  &   1.722  & -17.999  &   1.998 &   -4.832 &   -0.080 &   0.000   &  -0.073   & 1.437   &   0.543  \\
A98HR3  & 0.307 & 24.486 &   -0.015 &  106.206  &   1.783  &   5.987  &   1.886 &   12.102 &   -0.205 &   0.000   &  -0.104   & 0.369   &   0.246  \\
\underline{All $\theta$} &    \\
A0HR07  & 0.064 &    0.074  &   0.054 &    4.723  &   3.266 &   -5.717  &   3.281 &   -5.275 &    3.266  & 0.035   &  0.053   &  6.677  &  0.863  \\
A7HR07  & 0.065 &    0.065  &   0.093 &   10.132  &   2.282 &  -11.776  &   2.511 &   -2.933 &    1.012  & 0.040   &  0.040   &  8.496  &  1.156  \\
A9HR07  & 0.054 &    0.060  &   0.135 &   13.294  &   1.860 &  -11.404  &   2.209 &    5.222 &    0.303  & 0.035   &  0.031   &  8.945  &  1.299  \\
A98HR07 & 0.059 &    0.021  &   0.171 &   26.735  &   1.559 &   -7.348  &   1.799 &    6.905 &   -0.065  & 0.048   &  0.039   &  2.460  &  0.626  \\
A0HR1   & 0.12  &    6.036  &   0.053 &    6.579  &   2.908 &  -16.046  &   2.980 &  -13.961 &    2.908  & 0.087   &  0.110   &  8.880  &  1.247  \\
A7HR1   & 0.09  &    2.907  &   0.093 &   10.343  &   2.344 &   -9.376  &   2.457 &   -4.988 &    1.074  & 0.068   &  0.093   &  2.295  &  0.525  \\
A9HR1   & 0.13  &    2.777  &   0.128 &   17.577  &   1.823 &  -13.164  &   2.069 &   -1.449 &    0.254  & 0.066   &  0.075   &  4.256  &  0.735  \\
A98HR1  & 0.099 &    3.235  &   0.197 &   15.950  &   1.425 &  -15.291  &   1.880 &   11.744 &   -0.149  & 0.078   &  0.094   & 6.599   &  1.349  \\
A0HR2   & 0.18  &   59.025  &   0.050 &   12.631  &   2.465 &  -28.830  &   2.596 &  -25.067 &    2.465  & 0.164   &  0.197   &  5.798  &  1.771  \\
A7HR2   & 0.16  &   41.327  &   0.046 &   55.244  &   2.186 &  -15.499  &   2.394 &   -7.453 &    0.851  & 0.045   &  0.015   &  2.679  &  0.923  \\
A9HR2   & 0.21  &   53.746  &   0.085 &   45.564  &   1.739 &  -17.164  &   2.058 &   -2.001 &    0.092  & 0.012   & -0.031   &  3.799  &  1.093  \\
A98HR2  & 0.18  &   43.815  &   0.154 &   34.174  &   1.411 &  -16.120  &   1.876 &   11.497 &   -0.247  & 0.045   &  0.033   &  2.072  &  0.887  \\
A0HR3   & 0.350 &   49.207  &  -0.004 &  106.180  &   2.128 &  -38.572  &   2.220 &  -35.919 &    2.128  & 0.000   & -0.049   &  4.724  &  1.331  \\
A7HR3   & 0.34  &   47.146  &  -0.007 &  107.191  &   1.788 &  -30.878  &   2.065 &  -20.166 &    0.377  & 0.000   & -0.060   &  3.433  &  0.952  \\
A9HR3   & 0.341 &   42.733  &  -0.004 &  102.370  &   1.530 &  -27.117  &   1.869 &  -10.977 &   -0.276  & 0.000   & -0.073   &  2.652  &  0.914  \\
A98HR3  & 0.307 &   30.293  &  -0.014 &  105.873  &   1.597 &   -5.105  &   1.655 &   -1.624 &   -0.390  & 0.000   & -0.104   &  0.668  &  0.273  \\
\hline
\end{tabular}
\end{center}
\label{tbl_thickspin}
\end{table*}

\begin{figure}
  \begin{center}
      \includegraphics[width=3.3in,clip]{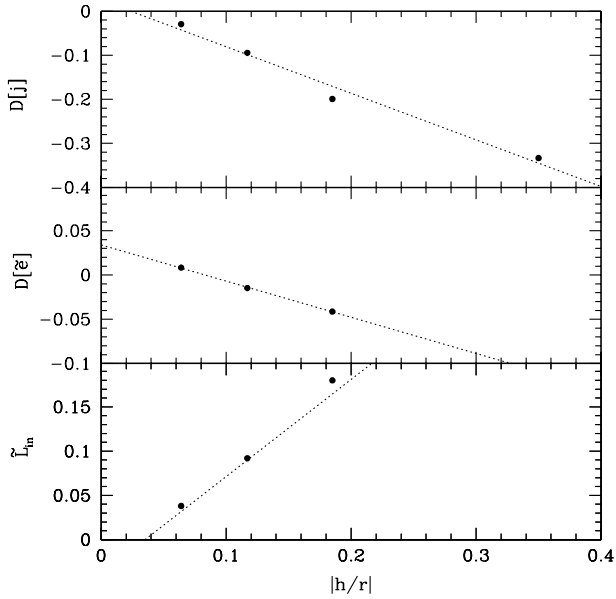}
  \end{center}
  \caption{The relative difference between $\jmath$ in the simulation
  and in the NT model (top panel), the relative difference between the
  nominal efficiency, $\eff$, and the NT value (middle panel),
  and the luminosity inside the ISCO normalized by the
  net radiative efficiency of the NT model,
  $\tilde{L}_{\rm in}$ (bottom panel),
  where $\eff[{\rm NT}]$ has been evaluated at the horizon (equivalently at the ISCO).
  There is a rough linear dependence on $|h/r|$ for all quantities,
  where a linear fit is shown as a dotted line in each panel.
  Note that the thicker disk models are not expected to behave like NT,
  and actually have $\jmath$ roughly similar across all spins.
  For $|h/r|\approx 0.07$, the excess luminosity from within the ISCO
  is less than $4\%$ of the total NT efficiency.
}
  \label{jandeffandlumscale}
\end{figure}

We now consider how the magnitude of $\jmath$, $\eff$, $L(<r_{\rm ISCO})$,
and $\Upsilon$ scale with disk thickness and black hole spin.
Table~\ref{tbl_thickspin} lists numerical results corresponding to
$\theta$ integrations
over $\pm 2|h/r|$ around the midplane and over all angles\footnote{Some
thicker disk models without cooling show small or slightly negative efficiencies, $\eff$,
which signifies the accretion of weakly unbound gas.
This can occur when a magnetic field is inserted
into a weakly bound gas in hydrostatic equilibrium.}.
Figure~\ref{jandeffandlumscale} shows selected results corresponding to
models with a non-rotating black hole for quantities integrated
over $\pm 2|h/r|$.
We see that the deviations of various diagnostics
from the NT values scale roughly as $|h/r|$.  In general, the deviations are quite small for
the thinnest model with $|h/r|\approx 0.07$.

Next, we consider fits of our simulation data as a function
of black hole spin and disk thickness to reveal if, at all,
these two parameters control how much the flow deviates from NT.
In some cases we directly fit the simulation results instead of their deviations from NT,
since for thick disks the actual measurement values can saturate independent of thickness
leading to large non-linear deviations from NT.
Before making the fits, we ask how quantities might scale with $a/M$ and $|h/r|$.
With no disk present, the rotational symmetry forces any scaling
to be an even power of black hole spin \citep{mg04}.
However, the presence of a rotating disk breaks this symmetry, and any accretion flow
properties, such as deviations from NT's model,
could depend linearly upon $a/M$ (at least for small spins).
This motivates performing a linear fit in $a/M$.
Similarly, the thickness relates to a dimensionless speed: $c_{\rm s}/v_{\rm K}\sim |h/r|$,
while there are several different speeds in the accretion problem
that could force quantities to have an arbitrary dependence on $|h/r|$.
Although, in principle, deviations might scale as some power of $|h/r|$, we
assume here a linear scaling $\propto |h/r|$.  This choice is driven partly by
simplicity and
partly by Figure~\ref{jandeffandlumscale} which shows that the simulation results
agree well with this scaling.

These rough arguments motivate obtaining explicit scaling laws for a quantity's
deviations from NT as a function of $a/M$ and $|h/r|$.
For all quantities we use the full $4\times 4$ set of models,
except for the luminosity and efficiency we exclude the two thickest models
in order to focus on the luminosity for thin disks with cooling.
We perform a linear least squares fit in both $a/M$ and $|h/r|$,
and we report the absolute percent difference between the upper $95\%$ confidence limit ($C_+$)
and the best-fit parameter value ($f$) given by $E = 100|C_+ - f|/|f|$.
Note that if $E>100\%$, then the best-fit value
is no different from zero to $95\%$ confidence (such parameter values are not reported).
After the linear fit is provided,
the value of $E$ is given for each parameter in order of appearance in the fit.
Only the statistically significant digits are shown.

First, we consider how electromagnetic stresses depend upon $a/M$ and $|h/r|$.
\citet{gammie99} has shown that the effects of electromagnetic stresses
are tied to the specific magnetic flux, $\Upsilon$,
and that for $\Upsilon\lesssim 1$
there are weak electromagnetic stresses causing only minor deviations
(less than $12\%$ for $\jmath$ across all black hole spins) from NT.
Let us consider how $\Upsilon$ should scale with $|h/r|$,
where $\Upsilon= \sqrt{2} (r/M)^2 B^r/\left(\sqrt{-(r/M)^2\rho_0 u^r}\right)$
in the equatorial plane and is assumed to be constant from the ISCO to the horizon.
For simplicity, let us study the case of a rapidly rotating black hole.
First, consider the boundary conditions near the ISCO provided by the disk,
where $c_{\rm s}/v_{\rm K}\sim |h/r|$
and the Keplerian rotation speed reaches $v_{\rm K}\sim 0.5$.
This implies $c_{\rm s}\sim 0.5|h/r|$.
Second, consider the flow that connects the ISCO and the horizon.
The gas in the disk beyond the ISCO has $\beta\sim (c_{\rm s}/v_{\rm A})^2\sim 10$,
but reaches $\beta\sim 1$ inside the ISCO
in any GRMHD simulations of turbulent magnetized disks,
which gives that $c_{\rm s}\sim v_{\rm A}$.
Thus, $v_{\rm A}\sim 0.5|h/r|$.
Finally, notice that $\Upsilon\sim 1.4 B^r/\sqrt{\rho_0}$
at the horizon where $u^r\sim -1$ and $r=M$.
The Keplerian rotation at the ISCO leads to a magnetic field
with orthonormal radial ($|B_r|\sim |B^r|$) and toroidal ($|B_\phi|$) components
with similar values near the ISCO and horizon,
giving $|B^r|\sim |B_r|\sim |B_\phi|\sim |b|$ and so $\Upsilon\sim 1.4 |b|/\sqrt{\rho_0}$.
Further, the~\alf 3-speed is $v_{\rm A}=|b|/\sqrt{b^2+\rho_0+u_g+p_g}\sim |b|/\sqrt{\rho_0}$
in any massive disk,
so that $\Upsilon\sim 1.4 v_{\rm A}\sim 0.7|h/r|$ for a rapidly rotating black hole.
Extending these rough arguments to all black hole spins at fixed disk thickness
also gives that $\Upsilon\propto -0.8(a/M)$ for $a/M\lesssim 0.7$.
These arguments demonstrate three points:
1) $\Upsilon\gg 1$ gives $b^2/\rho_0\gg 1$, implying a force-free magnetosphere
instead of a massive accretion disk ;
2) $\Upsilon\propto |h/r|$ ;
and 3) $\Upsilon\propto -(a/M)$.
Since the local condition for the magnetic field ejecting mass is $b^2/\rho_0\gg 1$
(see, e.g., \citealt{kombar09}),
this shows that $\Upsilon\sim 1$ defines a boundary that the disk component of the flow
cannot significantly pass beyond without eventually
incurring disruption by the strong magnetic field within the disk.

We now obtain the actual fit, which for an integration over $\pm 2|h/r|$
gives
\begin{equation}
\Upsilon\approx 0.7 + \left|\frac{h}{r}\right| - 0.6\frac{a}{M} ,
\end{equation}
with $E=33\%,~70\%,~40\%$, indicating a reasonable fit.
There is essentially $100\%$ confidence in the sign of the 1st and 3rd parameters
and $98\%$ confidence in the sign of the 2nd parameter.
This fit is consistent with our basic analytical estimate for the scaling.
Since most likely $\Upsilon\le 0.9$ in the limit that $|h/r|\to 0$ across all black hole spins,
the electromagnetic stresses are weak and cause less than $12\%$ deviation from NT in $\jmath$,
This means that NT solution is essentially recovered for magnetized thin disks.
For an integration over all angles, $\Upsilon\approx 1$ with $E=35\%$,
and there is no statistically significant trend
with disk thickness or black hole spin.
The value of $\Upsilon\sim 1$ is consistent with
the presence of the highly-magnetized corona-wind-jet
above the disk component \citep{mg04}.

Next, we consider whether our simulations can determine the
equilibrium value of the black hole spin as a function of $|h/r|$.
The spin evolves as the black hole accretes mass, energy, and angular momentum,
and it can stop evolving when these come into a certain balance leading to $d(a/M)/dt=0$
(see equation~\ref{spinevolve}).
In spin-equilibrium, the spin-up parameter $s = \jmath - 2(a/M)\emath$
has $s=0$ and solving for $a$ gives the equilibrium spin $a_{\rm eq}/M=\jmath/(2\emath)$.
For the NT solution, $s$ is fairly linear for $a/M>0$ and $a_{\rm eq}/M=1$.
In appendix~\ref{sec_gammie}, we note that for $\Upsilon\sim 0.2$--$1$
that the deviations from NT roughly scale as $\Upsilon$.
Since $\Upsilon\propto |h/r|$, one expects $s$ to also roughly
scale with $|h/r|$.  This implies that deviations from NT in the spin equilibrium
should scale as $|h/r|$.  Hence, one should have $1-a_{\rm eq}/M\propto |h/r|$.

Now we obtain the actual fit.
We consider two types of fits.  In one case, we fit $s$
(with fluxes integrated over all angles)
and solve $s=0$ for $a_{\rm eq}/M$.  This gives
\begin{equation}
s \approx 3.2 - 2.5\left|\frac{h}{r}\right| - 2.9\frac{a}{M}  ,
\end{equation}
with $E=8\%,~36\%,~8\%$, indicating quite a good fit.
There is an essentially $100\%$ confidence for the sign of all parameters,
indicating the presence of well-defined trends.
Solving the equation $s=0$ for $a/M$ shows that the spin equilibrium value, $a_{\rm eq}/M$,
is given by
\begin{equation}
1-\frac{a_{\rm eq}}{M} \approx -0.08 + 0.8\left|\frac{h}{r}\right| .
\end{equation}
In the other case, we fit $\jmath/(2\emath)$ and re-solve for $a_{\rm eq}/M$,
which gives directly
\begin{equation}
1-\frac{a_{\rm eq}}{M} \approx -0.10 + 0.9\left|\frac{h}{r}\right| ,
\end{equation}
with $E=9\%,~38\%$ with a $99.99\%$ confidence in the sign of the $|h/r|$ term.
Both of these procedures give a similar fit (the first fit is statistically better)
and agree within statistical errors, which indicates a linear fit is reasonable.
For either fit, one should set $a_{\rm eq}/M=1$ when the above formula gives $a_{\rm eq}/M>1$
to be consistent with our statistical errors and the correct physics.
Note that the overshoot $a_{\rm eq}/M>1$ in the fit
is consistent with a linear extrapolation of the NT dependence
of $s$ for $a/M>0$, which also overshoots in the same way
due to the progressively non-linear behavior of $s$ above $a/M\approx 0.95$.

These spin equilibrium fits imply that, within our statistical errors,
the spin can reach $a_{\rm eq}/M\to 1$ as $|h/r|\to 0$.
Thus, our results are consistent with NT by allowing maximal black hole
spin for thin disks\footnote{Here, we do not include black hole spin changes
by photon capture, which gives a limit of $a_{\rm eq}/M=0.998$ \citep{thorne74}.}.
Our results are also roughly consistent with the thick
disk 1-loop field geometry study by \citet{gammie_bh_spin_evolution_2004}.
Using our definition of disk thickness,
their model had $|h/r|\sim 0.2$--$0.25$
and they found $a_{\rm eq}/M\sim 0.9$, which is roughly consistent with our scaling law.
The fit is also consistent with results for even thicker disks ($|h/r|\sim 0.4$ near the horizon)
with $a_{\rm eq}/M\sim 0.8$ \citep{ajs78,pg98}.

Overall, the precise scaling relations given
for $\Upsilon$ and $a_{\rm eq}$ should be
considered as suggestive and preliminary.
More work is required to test the convergence
and generality of the actual coefficients.
While we explicitly tested convergence for the $a/M=0$
fiducial model, the other $a/M$ were not tested as rigorously.
A potential issue is that we find
the saturated state has fewer cells per
(vertical magnetic field) fastest growing mode for the axisymmetric MRI
in models with $a/M=0.9,0.98$ than in models with $a/M=0,0.7$
due to the relative weakness of the vertical field in the saturated state
for the high spin models.
However, both the rough analytical arguments
and the numerical solutions imply that
electromagnetic stresses scale somewhat linearly with black hole spin.
This consistency suggests that many measurements for the simulations,
such as $\Upsilon$ and $a_{\rm eq}$,
may be independent of smallness of the vertical field.
This fact could be due to these quantities being only directly related
to the radial and toroidal magnetic field strengths rather than
the vertical magnetic field strength.
Further, our thick disk models resolve
the axisymmetric MRI better than the thinnest disk model.
This suggests that the scaling of $\Upsilon$ and $a_{\rm eq}$
with disk thickness may be a robust result.

Lastly, we consider how the specific angular momentum,
nominal efficiency, and luminosity from within the ISCO
deviate from NT as functions of spin and thickness.
Overall, fitting these quantities does not give
very strong constraints on the actual parameter values,
but we can state the confidence level of any trends.
For each of $\eff$, $\jmath$, $\jmath_{\rm in}$,
and $\tilde{L}_{\rm in}$,
the deviation from NT as $|h/r|\to 0$ is
less than $5\%$ with a confidence of $95\%$.
For $\jmath$ integrated over $\pm 2|h/r|$,
$D[\jmath]$ decreases with $|h/r|$ and increases with $a/M$
both with $99\%$ confidence.
When integrating $\jmath$ over all angles,
$D[\jmath]$ only decreases with $|h/r|$ to $99\%$ confidence.
For $\jmath_{\rm in}$ integrated over $\pm 2|h/r|$,
$D[\jmath_{\rm in}]$ only increases with $a/M$ with $99.8\%$ confidence
and only decreases with $|h/r|$ with $97\%$ confidence.
When integrating $\jmath_{\rm in}$ over all angles,
$D[\jmath_{\rm in}]$ only increases with $a/M$ to essentially $100\%$ confidence
and only decreases with $|h/r|$ to $99.8\%$ confidence.
For $\eff$ integrated over $\pm 2|h/r|$,
$D[\eff]$ only increases with $|h/r|$ with $98\%$ confidence
with no significant trend with $a/M$.
When integrating $\eff$ over all angles,
$D[\eff]$ only increases with $a/M$ with $95\%$ confidence
with no significant trend with $|h/r|$.
For $\tilde{L}_{\rm in}$,
there is a $98\%$ confidence for this to increase with $|h/r|$
with no significant trend with $a/M$.
Overall, the most certain statement that can be made
is that our results are strongly consistent
with all deviations from NT becoming less than a few percent
in the limit that $|h/r|\to 0$ across the full range of black hole spins.

\section{Thin Disks with Varying Magnetic Field Geometry}
\label{sec:magneticfield}

We now consider the effects of varying the initial field geometry.
Since the magnetic field can develop large-scale structures
that do not act like a local scalar viscosity, there could in principle be
long-lasting effects on the
accretion flow properties as a result of the initial field geometry.
This is especially a concern for geometrically thin disks,
where the 1-loop field geometry corresponds to a
severely squashed and highly organized field loop bundle with
long-range coherence in the radial direction, whereas our
fiducial 4-loop model corresponds to nearly circular loops which
impose much less radial order on the MRI-driven turbulence.
To investigate this question we have simulated a model similar
to our fiducial run except that we initialized the gas torus
with a 1-loop type field geometry instead of our usual multi-loop geometry.

\begin{figure}
\centering
\includegraphics[width=3.3in,clip]{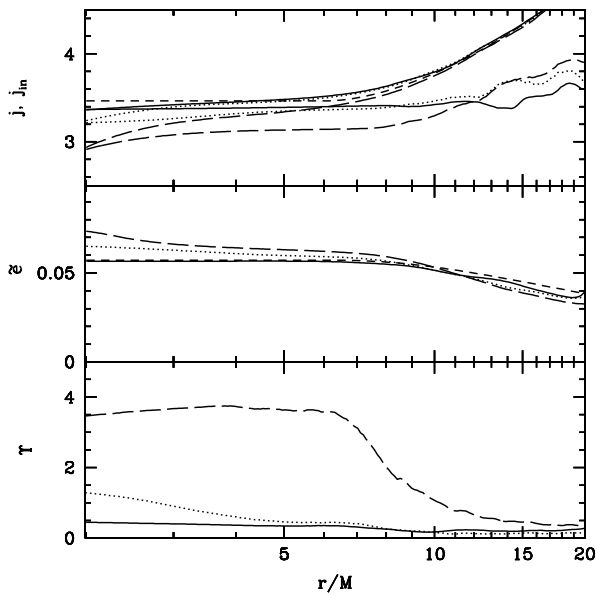}
\caption{Radial profiles of
$\jmath$ and $\jmath_{\rm in}$ (top panel),
$\eff$ (middle panel), and $\Upsilon$ (bottom panel)
are shown for two different initial field geometries.
Results for the fiducial 4-loop field geometry (model A0HR07)
integrated over $\pm 2|h/r|$ around the midplane are shown by solid lines,
for the 1-loop field geometry (model A0HR07LOOP1) integrated over $\pm 2|h/r|$
around the midplane by dotted lines,
and the 1-loop model integrated over all angles by long-dashed lines.
The short-dashed lines in the top two panels show the NT result.
We see that the 1-loop field geometry shows larger deviations from NT
in $\jmath$ and $\Upsilon$ compared to the 4-loop geometry.
The panels also reemphasize the point that including all $\theta$ angles
in the angular integration leads to considerable changes in $\jmath$ and $\Upsilon$
due to the presence of magnetic field in the corona-wind-jet.
}
\label{jfluxvsgeometry}
\end{figure}

Figure~\ref{jfluxvsgeometry} shows the radial dependence
of $\jmath$, $\jmath_{\rm in}$, $\eff$, and $\Upsilon$
for the two field geometries under consideration, and
Table~\ref{tbl_magneticfield} reports numerical estimates
of various quantities at the horizon.
Consider first the solid lines (4-loop fiducial run) and
dotted lines (1-loop run) in Figure~\ref{jfluxvsgeometry},
both of which correspond to integrations in $\theta$ over
$\pm 2|h/r|$ around the midplane.
The simulation with 4-loops is clearly more consistent
with NT than the 1-loop simulation.
The value of $\jmath$ at the horizon in the 4-loop model deviates
from NT by $-2.9\%$.
Between the times of $12900M$ and $17300M$,
the 1-loop model deviates by $-5.6\%$,
while at late time over the saturated period the 1-loop model deviates by
$-7.2\%$.
The long-dashed lines show the effect of integrating over all $\theta$
for the 1-loop model.
This introduces yet another systematic deviation from NT
(as already noted in \S\ref{sec_fluxdiskcorona});
now the net deviation of $\jmath$
becomes $-10.7\%$ for times $12900M$ to $17300M$
and becomes $-15.8\%$ for the saturated state.
Overall, this implies that the assumed initial field geometry
has a considerable impact on the specific angular momentum
profile and the stress inside the ISCO.
This also indicates that the saturated state
is only reached after approximately $17000M$,
and it is possible that the 1-loop model may never properly
converge because magnetic flux of the same sign
(how much flux is initially available is arbitrary
due to the arbitrary position of the initial gas pressure maximum)
may continue to accrete onto the black hole
and lead to a qualitatively different accretion state
(as seen in \citet{igumenshchev03} and \citet{mg04} for their vertical field model).
At early times, the nominal efficiency, $\eff$,
shows no significant dependence on
the field geometry,
and sits near the NT value for both models.
At late time in the 1-loop model, $\eff$ rises somewhat,
which may indicate the start of the formation
of a qualitatively different accretion regime.

\begin{table*}
\caption{Field Geometry Study}
\begin{center}
\begin{tabular}[h]{|r|r|r|r|r|r|r|r|r|r|r|r|r|r|}

\hline
Model Name & $|h/r|$ & $\dot{M}$ & $\eff$ & $D[\eff]$ & $\jmath$ & $D[\jmath]$ & $\jmath_{\rm in}$ & $D[\jmath_{\rm in}]$ & $s$ & $\tilde{L}_{\rm in}$  & $\tilde{L}_{\rm eq}$ & $100\tilde{\Phi}_r$ & $\Upsilon$ \\
\hline
\underline{$\pm 2|h/r|$} & \\
A0HR07      &    0.064 &    0.066 &    0.058 &   -0.829  &   3.363 &   -2.913  &   3.355 &   -3.153 &    3.363 &  0.035    & 0.080   & 1.355   &   0.450    \\
A0HR07LOOP1 &    0.048 &    0.036 &     0.066 &  -14.846 &    3.215 &   -7.193 &    3.234 &   -6.637 &   3.215 & 0.049    &  0.059   & 6.198 &    1.281  \\
A0HR3       &    0.350 &    44.066  & -0.003 &  104.582 &    2.309 &  -33.331 &    2.408 &  -30.473 &    2.309 & 0.000    & -0.049   & 3.039   &   1.182  \\
A0HR3LOOP1  &    0.377 &    32.577  &  0.001 &   98.892 &    1.823 &  -47.389 &    1.984 &  -42.717 &    1.823 & 0.000    &  -0.049   & 7.599   &  2.246 \\
\underline{All $\theta$} & \\
A0HR07          & 0.064 &   0.074  &   0.054 &    4.723  &   3.266 &   -5.717  &   3.281 &   -5.275 &    3.266  & 0.035    & 0.053  &   6.677  &  0.863   \\
A0HR07LOOP1     & 0.048 &   0.040  &   0.075 &  -31.857  &   2.915 &  -15.847  &   2.928 &  -15.478 &    2.915  & 0.046    & 0.048  &   43.935 &  3.464 \\
A0HR3           & 0.350 &   49.207  &  -0.004 &  106.180  &   2.128 &  -38.572  &   2.220 &  -35.919 &    2.128 & 0.000    & -0.049  &   4.724  &  1.331  \\
A0HR3LOOP1      & 0.377 &   39.382 &   -0.001 &  102.499 &   1.575 &  -54.526  &   1.734 &  -49.932 &    1.575  & 0.000    & -0.049  &  11.444 &  2.523 \\
\hline
\end{tabular}
\end{center}
\label{tbl_magneticfield}
\end{table*}

\begin{figure}
\centering
\includegraphics[width=3.3in,clip]{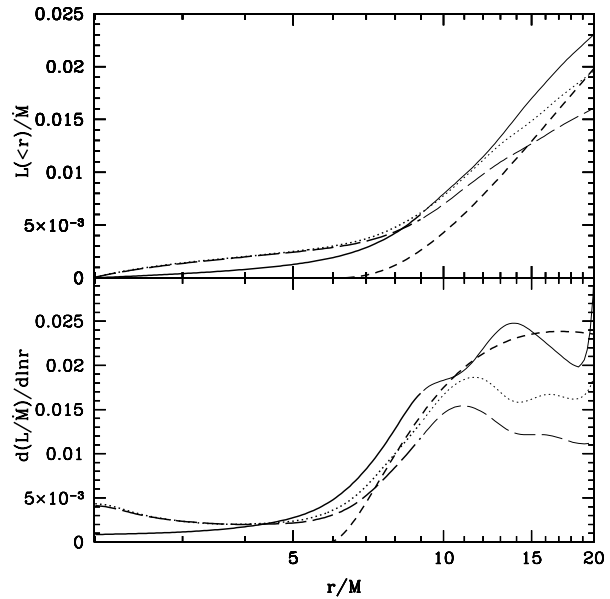}
\caption{Similar to Figure~\ref{jfluxvsgeometry}
for the initial 4-loop and 1-loop field geometries,
but here we show the luminosity (top panel)
and log-derivative of the luminosity (bottom panel).
The luminosity is slightly higher for the 1-loop model
compared to the 4-loop model.
}
\label{lumvsgeometry}
\end{figure}

Figure~\ref{lumvsgeometry} shows the normalized luminosity.
We see that the 1-loop model produces more luminosity
inside the ISCO.
For times $12900M$ to $17300M$,
$\tilde{L}_{\rm in}=5.4\%$ (integrated over all $\theta$)
compared to $3.5\%$ for the 4-loop field geometry.
Thus there is $50\%$ more radiation coming from
inside the ISCO in this model.
At late time during the saturated state,
$\tilde{L}_{\rm in}=4.6\%$ (integrated over all $\theta$).
Thus there is approximately $31\%$ more radiation coming from
inside the ISCO in this model
during the late phase of accretion.

\begin{figure}
\centering
\includegraphics[width=3.3in,clip]{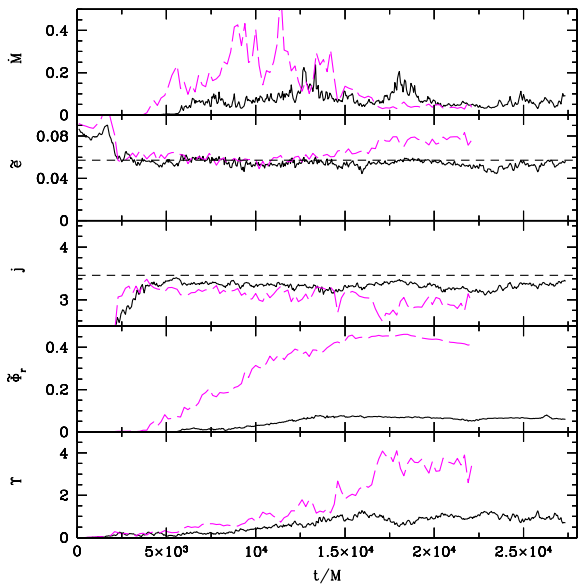}
\caption{Similar to Figure~\ref{5dotpanel},
but here we compare the initial 4-loop fiducial model (black solid lines)
and the 1-loop model (dashed magenta lines).  The horizontal black dashed lines
in the second and third panels show the predictions of the
NT model.
The mass accretion rate, $\dot{M}$, has larger root-mean-squared
fluctuations in the 1-loop model,
which is indicative of more vigorous turbulence.
The nominal efficiency, $\eff$, shows no clear difference.
The specific angular momentum, $\jmath$, is lower in the 1-loop
model compared to the 4-loop model.
This indicates that the 1-loop field leads to larger stress within the ISCO.
The absolute magnetic flux (per unit initial total absolute flux)
on the black hole is larger in the 1-loop model than the 4-loop model.
Since $\tilde{\Phi}_r\sim 1/2$ for the 1-loop model,
essentially half of the initial loop was advected onto the black hole,
while the other half gained angular momentum and has been advected away.
This may indicate that the 1-loop geometry
is a poor choice for the initial field geometry,
since the magnetic flux that ends up on the black hole is determined
by the initial conditions.
For times $12900M$ to $17300M$,
the value of $\Upsilon$ is about twice higher in the 1-loop model,
which implies about twice greater electromagnetic stresses within the ISCO.
}
\label{5dotpanel1loop}
\end{figure}

\begin{figure}
\centering
\includegraphics[width=3.3in,clip]{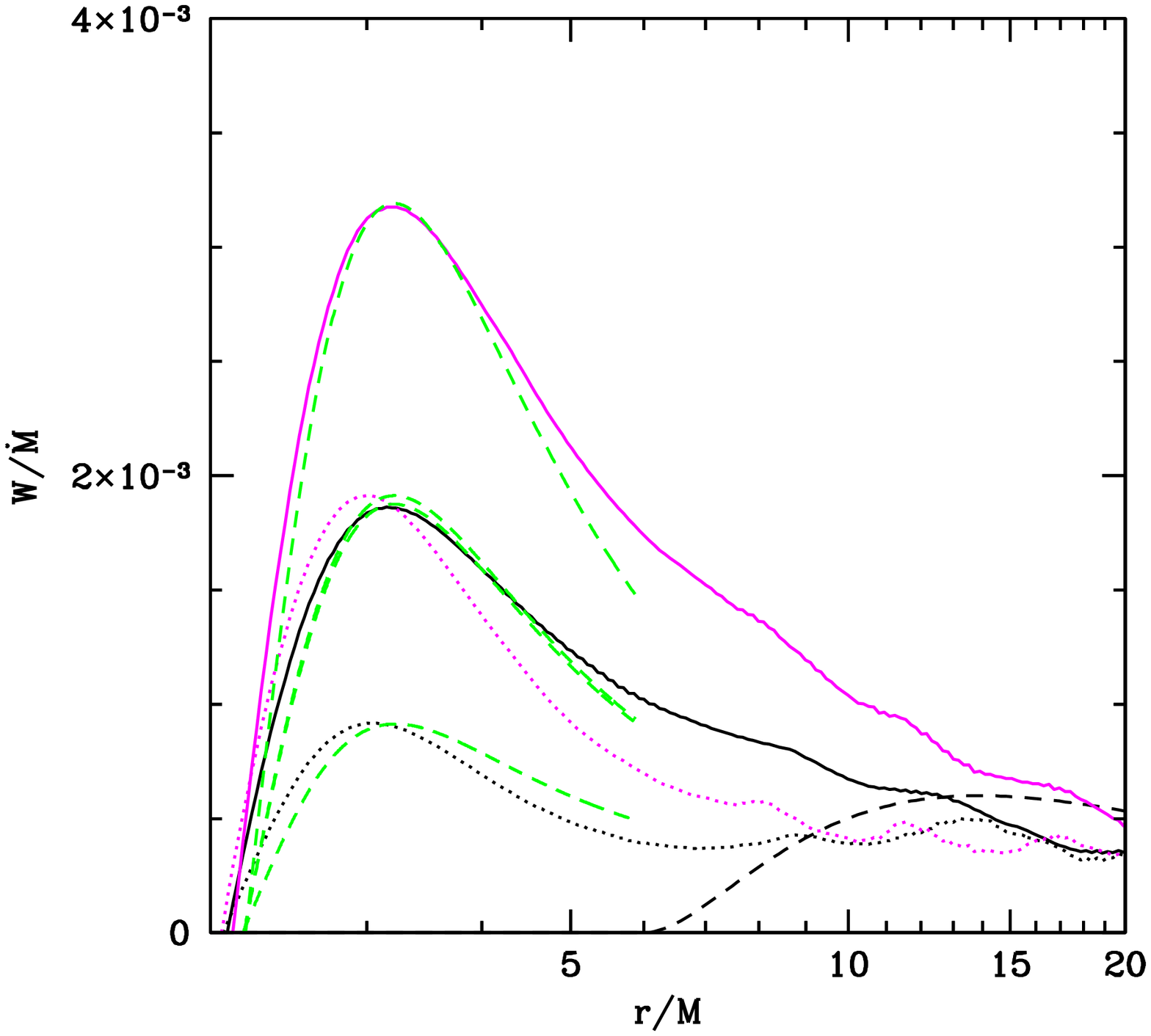}
\caption{Normalized electromagnetic stress, $W/\dot{M}$,
as a function of radius for the fiducial model (black lines)
and the otherwise identical 1-loop model (magenta lines).
The solid lines correspond to a $\theta$ integration over all angles,
while the dotted lines correspond to a $\theta$ integration over $\pm 2|h/r|$.
The dashed line shows the NT result,
while the dashed green lines show the Gammie (1999) result
for $\Upsilon=0.60, 0.89, 0.90, 1.21$ for lines from the bottom to the top.
The Gammie (1999) model gives a reasonable fit to the simulation's
stress profile within the ISCO.
The 1-loop model shows about $50\%$ larger peak normalized stress (integrated over all angles)
compared to the multi-loop model (integrated over all angles),
which is consistent with the 1-loop model leading to larger
deviations from NT (about $50\%$ larger luminosity over the period used
for time averaging).
The large differences between the solid
and dotted lines again highlights the fact that the stress within the disk is much smaller
than the stress over all $\theta$ that includes the corona+wind+jet.
As pointed out in S08, even though such a plot of the electromagnetic stress
appears to indicate large deviations from NT within the ISCO,
this is misleading because one has not specified the quantitative
effect of the non-zero value of $W/\dot{M}$ on physical quantities within the ISCO.
Apparently large values of $W/\dot{M}$ do not
necessarily correspond to large deviations from NT.
For example, quantities such as $\jmath$, $\emath$, and the luminosity
only deviate by a few percent from NT for the multi-loop model.
}
\label{loop1manystress}
\end{figure}

Table~\ref{tbl_magneticfield} also reports the results
for thick ($|h/r|\approx 0.3$) disk models
initialized with the multi-loop and 1-loop field geometries.
This again shows that the deviations from NT are influenced by the
initial magnetic field geometry and scale with $|h/r|$ in a way
expected by our scaling laws.
The 1-loop models show deviations from NT in $\jmath$ are larger
as related to the larger value of $\Upsilon$.
The deviations from NT are less affected
by the initial magnetic field geometry for thicker disks,
because the deviations from NT are also driven
by thermal effects and Reynolds stresses rather
than primarily electromagnetic stresses as for thin disks.

These effects can be partially understood by looking
at the specific electromagnetic stress, $\Upsilon$, shown in Figure~\ref{jfluxvsgeometry}.
We find $\Upsilon\approx 0.45$ for the 4-loop field geometry.
For times $12900M$ to $17300M$,
$\Upsilon\approx 0.71$ in the 1-loop field geometry,
and during the saturated state $\Upsilon\approx 1.28$.
For times $12900M$ to $17300M$,
the $50\%$ larger $\Upsilon$ appears to be the reason for
the $50\%$ extra luminosity inside the ISCO in the 1-loop model.
The magnetized thin disk model of \citet{gammie99}
predicts that, for $a/M=0$, specific magnetic fluxes of $\Upsilon=0.45, ~0.71$
should give deviations from NT of $-D[\jmath]=1.9, ~3.9$, respectively.
These are close to the deviations seen in the simulations,
but they are not a perfect match for reasons we can explain.
First, the details of how one spatially-averages quantities
(e.g., average of ratio vs. ratio of averages)
when computing $\Upsilon$ lead to moderate changes in its value,
and, for integrations outside the midplane,
comparisons to the Gammie model can require
slightly higher $\Upsilon$ than our diagnostic reports.
Second, the finite thermal pressure at the ISCO
leads to (on average over time) a deviation already at the ISCO
that is non-negligible compared to the deviation introduced
by electromagnetic stresses between the ISCO and horizon.
This thermal component is not always important,
e.g., see the comparison in Figure~\ref{gammie4panel}.
Still, as found in \citet{mg04} for thick disks at least,
the deviations from NT contributed by the thermal pressure
are of the same order as the deviations predicted by the Gammie model.
These results motivate extending the \citet{gammie99} model
to include a finite (but still small) thermal pressure
such that the boundary conditions at the ISCO lead to a non-zero radial velocity.

Within the ISCO, we find that the time-averaged
and volume-averaged comoving field strength for the 4-loop geometry
roughly follows $|b|\propto r^{-0.7}$ within $\pm 2|h/r|$ of the disk midplane,
while at higher latitudes we have a slightly steeper scaling.
For times $12900M$ to $17300M$,
the 1-loop geometry has $|b|\propto r^{-1.1}$ within $\pm 2|h/r|$ of the disk midplane,
and again a slightly steeper scaling in the corona.
Other than this scaling, there are no qualitative differences in the distribution
of any comoving field component with height above the disk.
While the \citet{gammie99} solution does not predict a power-law
dependence for $|b|$, for a range between $\Upsilon=0.4$--$0.8$,
the variation near the horizon is approximately $|b|\propto r^{-0.7}- r^{-0.9}$,
which is roughly consistent with the simulation results.
The slightly steeper slope we obtain for the 1-loop geometry is consistent
with a higher specific magnetic flux,
although the variations in $\Upsilon$ for integration over different ranges of angle
imply stratification and a non-radial flow
which the \citet{gammie99} model does not account for.
This fact and the rise in $\Upsilon$ with decreasing radius
seen in Figure~\ref{jfluxvsgeometry}
indicate a non-trivial degree of angular compression
as the flow moves towards the horizon.
Overall, our results suggest that deviations from NT
depend on the assumed field geometry,
and that the \citet{gammie99} model roughly fits the simulations.

Figure~\ref{5dotpanel1loop}
shows the same type of plot as in Figure~\ref{5dotpanel},
but here we compare the fiducial 4-loop model with
the 1-loop model.
As mentioned above, the 1-loop geometry has a
larger deviation in $\jmath$ from the NT value,
corresponding to larger stresses inside the ISCO.
The absolute magnetic flux (per unit initial total absolute magnetic flux)
on the black hole $\tilde{\Phi}_r$ is of order $1/2$,
suggesting that the inner half of the initial field bundle
accreted onto the black hole,
while the other half was advected to larger radii.
This is consistent with what is seen in simulations of thick tori \citep{mg04,bhk08}.
This suggests that using the 1-loop geometry leads to results
that are sensitive to the initial absolute magnetic flux,
while the multiple loop geometry leads to results that are insensitive
to the initial absolute magnetic flux.
Such dependence of the electromagnetic stress on initial magnetic field geometry
has also been reported in 3D pseudo-Newtonian simulations by \citet{ra01}
and in 3D GRMHD simulations by \citet{bhk08}.

Figure~\ref{loop1manystress}
shows the electromagnetic stress as computed by equation~(\ref{stress})
for the multiple loop fiducial model (A0HR07)
and the otherwise identical 1-loop model (A0HR07LOOP1).
We only show the electromagnetic part of the stress,
and within the ISCO this is to within a few percent the same
as the total stress obtained by including all terms in the stress-energy tensor.
Outside the ISCO, the total stress agrees more with the NT model.
The figure shows the full-angle integrated electromagnetic stress,
the electromagnetic stress integrated over only $\pm 2|h/r|$,
the NT stress,
and the \citet{gammie99} electromagnetic stress for $\Upsilon=0.60, 0.89, 0.90, 1.21$
(we choose $\Upsilon$, the only free parameter of the model,
such that the peak magnitude of the stress agrees with the simulation).
The chosen $\Upsilon$ values are close to our diagnostic's value
of $\Upsilon$ for these models, which demonstrates that the \citet{gammie99} model
is consistently predicting the simulation's results with a single free parameter.
The stress is normalized by the radially dependent $\dot{M}(r)$
that is computed over the same $\theta$ integration range.
We do not restrict the integration to bound material as done in S08
(in S08, the stress is integrated over $\pm 2|h/r|$ and only for bound material,
while in N10 the stress\footnote{
N10's figures 12 and 13 show stress vs. radius,
but some of the integrals they computed were not re-normalized
to the full $2\pi$ when using their simulation $\phi$-box size of $\pi/2$,
so their stress curves are all a constant factor of $4$ times larger
than the actual stress (Noble, private communication).}
is only over bound material).
The stress for the fiducial model was time-averaged over
the saturated state, while the 1-loop model was time-averaged
from time $12900M$ to $17300M$ in order to best compare
with the early phase of accretion for the 1-loop model studied in N10.

Figure~\ref{loop1manystress} shows that
the simulation and NT stress do not agree well,
and it suggests there is an {\it apparently} large stress within the ISCO.
However, as first pointed out by S08 and discussed in section~\ref{magneticfluxdiag},
this stress does not actually correspond to a large deviation from NT in
physically relevant quantities such as the
specific angular momentum, specific energy, and luminosity.
This point is clarified by making a comparison to the \citet{gammie99} model's stress,
which agrees reasonably well with the simulation stress inside the ISCO.
Even though the stress may appear large inside the ISCO,
the stress corresponding to the Gammie model with (e.g.) $\Upsilon=0.60$
only translates into a few percent deviations from NT.
This figure also demonstrates that the initial magnetic field
geometry affects the amplitude of the stress in the same direction as it
affects other quantities and is reasonably well predicted by the \citet{gammie99} model.
The initial magnetic field sets the saturated value of $\Upsilon$,
which is directly related to the electromagnetic stresses within the ISCO.
The 1-loop model leads to a peak stress (integrated over all angles)
within the ISCO that is about $50\%$ larger than the multi-loop model (integrated over all angles),
which is likely related to the extra $50\%$ luminosity in the 1-loop
model compared to the multi-loop model.
The fact that the stress normalization changes with initial field geometry
is consistent with other 3D GRMHD simulations of thick disks by \citet{bhk08}.
This figure again shows how the stress within the disk ($\pm 2|h/r|$)
is much smaller than the total disk+corona+wind+jet (all $\theta$).

Finally, we discuss previous results obtained for
other field geometries using an energy-conserving
two-dimensional GRMHD code \citep{mg04}.
While such two-dimensional simulations are unable to sustain turbulence,
the period over which the simulations do show turbulence agrees quite well
with the corresponding period in three-dimensional simulations.
This implies that the turbulent period in the
two-dimensional simulations may be qualitatively correct.
The fiducial model of \citet{mg04} was of a thick ($|h/r|\sim 0.2$--$0.25$)
disk with a 1-loop initial field geometry around an $a/M=0.9375$
black hole.
This model had $\Upsilon\sim 1$ near the midplane within the ISCO
and $\Upsilon\sim 2$ when integrated over all $\theta$ angles.
Their measured value of $\jmath\approx 1.46$ integrated over all angles,
$|b|\propto r^{-1.3}$ within the ISCO within the disk midplane \citep{mck07b},
along with $\Upsilon\sim 1$--$2$
are roughly consistent with the \citet{gammie99} model prediction
of $\jmath\approx 1.5$.
Similarly, the strong vertical field geometry model they studied had $\Upsilon\sim 2$
near the midplane within the ISCO and $\Upsilon\sim 6$ integrated over all $\theta$ angles.
Their measurement of $\jmath\approx -1$ integrated over all angles
is again roughly consistent with
the model prediction of $\jmath\approx -1.2$ for $\Upsilon\sim 6$.
Note that in this model, $\Upsilon$ rises (as usual to reach saturation)
with time, but soon after $\Upsilon\gtrsim 2$ in the midplane,
the disk is pushed away by the black hole and then $\Upsilon$ is forced to be even larger.
Evidently, the accumulated magnetic flux near the black hole pushes
the system into a force-free magnetosphere state -- not an accretion state.
This shows the potential danger of using strong-field initial conditions
(like the 1-loop field geometry),
since the results are sensitive to the assumed initial flux
that is placed on (or rapidly drops onto) the black hole.
Even while the disk is present, this particular model
exhibits net angular momentum extraction from the black hole.
This interesting result needs to be confirmed using
three-dimensional simulations of both thick and thin disks.

\section{Comparisons with Other Results}
\label{sec:comparison}

The results we have obtained in the present work are consistent with
those of \citet{arc01} and \citet{rf08}, who carried out pseudo-Newtonian studies,
and with the results of S08, who did a full GRMHD simulation.
Both of these studies found only minor deviations from NT for thin accretion disks
with a multi-loop initial field geometry.  However, more recently,
N09 and N10 report {\it apparently} inconsistent results,
including factors of up to five larger deviations from NT
in the specific angular momentum ($2\%$ in S08 versus $10\%$ in N10)
for the same disk thickness of $|h/r|\sim 0.07$.
They also find a $50\%$ larger deviation
from NT in the luminosity ($4\%$ in S08 versus $6\%$ in N09).
Furthermore, in N10 they conclude that the electromagnetic stresses
have no dependence on disk thickness or initial magnetic field geometry,
whereas we find that the electromagnetic stresses have a statistically significant dependence
on both disk thickness and magnetic field geometry.

We have considered several possible explanations for these differences,
as we now describe.
We attempt to be exhaustive in our comparison
with the setup and results by N09 and N10,
because our works and their works seek accuracies much better than order two
in measuring deviations from NT.
Thus, any deviations between our results by factors of two or more
must be investigated further in order to ensure a properly understood and accurate result.

First, we briefly mention some explanations that N10 propose
as possible reasons for the
discrepant results, viz., differences in
1) numerical algorithm or resolution;
2) box size in $\phi$-direction: $\Delta\phi$;
3) amplitude of initial perturbations;
4) accuracy of inflow equilibrium;
and 5) duration of the simulations.
Our algorithms are similar except that their PPM
interpolation scheme assumes primitive quantities are cell averages (standard PPM),
while ours assumes they are point values (as required to be applied in a higher-order scheme).
They used LAXF dissipative fluxes,
while we used HLL fluxes that are about twice more accurate for shocks
and may be more accurate in general.
On the other hand, they used parabolic interpolation for the Toth electric field,
while we use the standard Toth scheme.
Given these facts, we expect that the accuracy of our algorithms is similar.
Overall, our convergence testing and other diagnostics (see \S\ref{sec:convergence})
confirm that none of their proposed issues can be the cause of differences between S08 and N10.

We have shown that inflow equilibrium must
include saturation of the specific magnetic flux, $\Upsilon$,
which generally saturates later in time than other quantities.
By running our fiducial model A0HR07 to a time of nearly $30000M$,
we ensure that we have a long period of steady state conditions
to compute our diagnostic quantities. The fact that we need to run our fiducial thin disk simulation for such a long time
to reach inflow equilibrium up to a radius $r\sim 9M$ is completely consistent with our
analytical estimate of the time scale calculated using Eq. \ref{eq:inflow} of Appendix~\ref{sec_inflow} (see
the earlier discussion in \S\ref{sec_infloweq}
and Fig. \ref{velvsr}).  In the comparison between the numerical and analytical
results shown in Figure~\ref{velvsr}, we found agreement
by setting $\alpha |h/r|^2\approx 0.00033$ which,
for our disk with $|h/r|\approx 0.064$,
corresponds to $\alpha\approx 0.08$.
With this value of $\alpha|h/r|^2$, we would have to run
the simulation until $t\sim83000M,~160000M$, to reach inflow
equilibrium out to $15M,~20M$, respectively,
corresponding to a couple viscous timescales at that radius.
N10 state that they reach inflow equilibrium within $r\sim 15M$--$20M$
in a time of only $t\sim 10000M$.
Since their disk thickness is $|h/r|\approx 0.06$,
even a single viscous timescale would require
their simulations to have $\alpha\sim0.38$ to reach inflow equilibrium up to $r\sim 15M$,
and an even larger value of $\alpha$ for $r\sim20M$.  This seems unlikely.
We can partially account for their result by considering our 1-loop model,
which up to $t\sim 17000M$
has $\alpha |h/r|^2$ twice as large and $\alpha$
about $70\%$ larger than in the fiducial 4-loop run.
However, this still falls far short by a factor of roughly $3$
of what N10 would require for inflow equilibrium up to $15M$--$20M$.
Further, our A0HR07LOOP1 model, which is similar to their model,
only reaches a saturated state
by $17000M$, and only the $\Upsilon$ quantity indicates
that saturation has been reached.
If we were to measure quantities from $10000M$ to $15000M$
as in N10, we would have underestimated the importance of magnetic field geometry
on the electromagnetic stresses.

Since all these simulations are attempting to obtain accuracies
better than factors of two in the results,
this inflow equilibrium issue should be explored further.
A few possible resolutions include:
1) N10's higher resolution leads to a much larger $\alpha$;
2) their disk has a larger ``effective'' thickness, e.g. $|h/r|\sim 0.13$,
according to Eq. 5.9.8 in NT (see Eq. \ref{eq:inflow} of Appendix~\ref{sec_inflow});
3) some aspects of their solution have not yet reached inflow equilibrium
within a radius much less than $r\sim 15M$,
such as the value of $\Upsilon$ vs. time that saturates much later than other quantities;
or 4) they achieve constant fluxes vs. radius due to transient non-viscous effects
-- although one should be concerned that the actual value of such fluxes
continues to secularly evolve in time and one still requires evolution
on the longer viscous (turbulent diffusion) timescale to reach true inflow equilibrium.

Second, we considered various physical setup issues, including differences in:
1) range of black hole spins considered;
2) range of disk thicknesses studied;
3) ad hoc cooling function;
and 4) equation of state.
We span the range of black hole spins and disk thicknesses studied by N10,
so this is unlikely to explain any of the differences.
Some differences could be due to the disk thickness vs. radius
profiles established by the ad hoc cooling functions in the two studies.
N10's cooling function is temperature-based and
allows cooling even in the absence of any dissipation,
while ours is based upon the specific entropy and
cools the gas only when there is dissipation.
Both models avoid cooling unbound gas.
In S08 and in the present paper, we use an ideal gas equation of
state with $\Gamma=4/3$,
while N09 and N10 used $\Gamma=5/3$.
The properties of the turbulence do appear to depend on the equation
of state \citep{mm07}, so it is important to investigate further
the role of $\Gamma$ in thin disks.

Third, the assumed initial field geometry may introduce
critical differences in the results.
Issues with the initial field geometry include
how many field reversals are present,
how isotropic the field loops are in the initial disk,
how the electromagnetic field energy is distributed in the initial disk,
and how the magnetic field is normalized.
In S08 and here, we have used a multi-loop geometry in the initial torus
consisting of alternating polarity poloidal field bundles stacked radially.
We ensure that the field loops are roughly isotropic within the initial torus.
We set the ratio of maximum gas pressure to maximum magnetic pressure
to $\beta_{\rm maxes}=100$, which gives us a volume-averaged mean $\beta$ within
the dense part of the torus ($\rho_0/\rho_{0,\rm max}\ge 0.2$) of $\bar{\beta}\sim 800$.
Our procedure ensures that all initial local values of $\beta$ within the disk
are much larger than the values in the evolved disk, i.e., there is
plenty of room for the magnetic field to be amplified by the MRI.

We have also studied a 1-loop geometry that is
similar to the 1-loop geometry used in N09 and N10.
Their initial $\phi$-component of the vector potential is
$A_\phi\propto {\rm MAX}(\rho_0/\rho_{0,\rm max} - 0.25,0)$
(Noble, private communication).
They initialize the magnetic field geometry
by ensuring that the volume-averaged gas pressure divided by
the volume-averaged magnetic pressure is $\beta_{\rm averages}=100$
(Noble, private communication).
(They stated that the mean initial plasma $\beta$ is $\bar{\beta}=100$.)
For their thin disk torus model parameters,
this normalization procedure leads to a portion of the
inner radial region of the torus to have a local value of $\beta\sim 3-8$,
which may be a source of differences in our results.
Such a small $\beta$ is lower than present in the saturated disk.
N10 make use of an older set of simulations from a different non-energy-conserving
code \citep{hk06,bhk08} to investigate the effect of other field geometries.
The results from this other code have strong outliers, e.g., the KD0c model,
and so we are unsure if these other simulations can be used for such a study.

N10 state that they find no clear differences in the electromagnetic stresses
for different initial field geometries.
As shown in their figures 12 and 13, the \citet{ak00} model
captures the smoothing of the stress outside the ISCO,
but it is not a model for the behavior of the stress inside the ISCO.
We find that electromagnetic stresses have a clear dependence
on both disk thickness and the initial
magnetic field geometry, with a trend that agrees
with the \citet{gammie99} model of a magnetized thin disk.
Our Figure~\ref{loop1manystress} shows that the stress within the ISCO
is reasonably well modelled by the \citet{gammie99} model.
Our 1-loop thin disk model gives a peak normalized stress (integrated over all angles)
of about $3.2\times 10^{-3}$ for times $12900M$ to $17300M$,
which is comparable to the 1-loop thin disk model
in N10 with peak normalized stress (integrated over all angles) of about $2.5\times 10^{-3}$
(after correcting for their $\phi$-box size).
Hence, we are able to account for the results of their 1-loop model.

In addition, we used the specific magnetic flux, $\Upsilon$,
an ideal MHD invariant that is conserved within the ISCO,
to identify how electromagnetic stresses scale with disk thickness and magnetic field geometry.
In the saturated state, the value of $\Upsilon$,
which controls the electromagnetic stresses,
is different for different initial magnetic field geometries.
We find that $\jmath$ within the disk ($\pm 2|h/r|$ from the midplane)
deviates from NT by
$-3\%$ in our 4-loop model and $-6\%$ in our 1-loop model
for times $12900M$ to $17300M$.
Integrating over all angles, $\jmath$ deviates by $-6\%$ for the 4-loop
model and $-11\%$ for the 1-loop model for times $12900M$ to $17300M$.
Thus, we find a clear factor of two change, depending on
the assumed initial field geometry and the range of integration.
The excess luminosity is $3.5\%$ for the 4-loop model
and $5.4\%$ for the 1-loop model for times $12900M$ to $17300M$.
Recalling that N10 find a deviation from NT of about $-10\%$ in $\jmath$
(integrated over all angles) and a luminosity excess beyond NT of about $6\%$,
this shows we can completely account for the {\it apparent}
inconsistencies mentioned by N10 by invoking
dependence of the results on the initial field geometry
and the presence of extra stress beyond the disk component of the accretion flow.

Fourth, let us consider measurement and interpretation differences.
Our ultimate goal is to test how well the NT model
describes a magnetized thin accretion disk.
The primary quantity that is used to measure this effect in S08 and N10
is the specific angular momentum $\jmath$.  However, the
measurements are done differently in the two studies.
In S08 as well as in this paper, we focus on the disk gas
by limiting the range of $\theta$ over which we compute the
averaging integrals ($\pm2|h/r|$ from the midplane).
In contrast, N10 compute their integrals
over a much wider range of $\theta$ which includes
both the disk and the corona-wind-jet
(Noble, 2010, private communications).
We have shown in \S~\ref{sec_fluxdiskcorona}
that the disk and corona-wind-jet contribute
roughly equally to deviations of $\jmath$ from the NT value.
In principle, the luminosity from the corona-wind-jet could be important,
but we have shown that the excess luminosity of bound gas
within the ISCO is dominated by the disk.
This means that the measure used by N10,
consisting of integrating over all gas to obtain $\jmath$,
cannot be used to infer the excess luminosity of bound gas within the ISCO.
Further, the corona would largely emit non-thermal radiation,
so for applications in which one is primarily interested in the thermal component
of the emitted radiation, one should evaluate the accuracy of the NT model by
restricting the angular integration range to the disk component within $\pm 2|h/r|$.

Fifth, let us consider how the results from N10 scale with disk
thickness for the specific case of a non-spinning ($a/M=0$) black
hole.  We have performed a linear least squares fit of
their simulation results, omitting
model KD0c which is a strong outlier.  For $\jmath$ integrated
over all $\theta$, their relative difference follows
$D[\jmath]\approx -7 - 45|h/r|$
with confidence of $95\%$ that these coefficients, respectively,
only deviate by $\pm 67\%$ and $\pm 89\%$.
These fits imply that, as $|h/r|\to 0$, the
relative deviation of $\jmath$ from the NT value is about $-7\%$,
but they could easily be as low as $-2\%$.
Their results do not indicate a statistically significant
large deviation from NT as $|h/r|\to 0$.
Since the total deviation in $\jmath$ from NT includes the effects of
electromagnetic (and all other) stresses, this implies that
their models are consistent with weak electromagnetic stresses as $|h/r|\to 0$.

Further, we have already established that the 1-loop geometry gives
(at least) twice the deviation from NT compared to the 4-loop geometry,
plus there is another factor of two arising from including
the corona-wind-jet versus not including it.
This net factor of 4 applied to N10's results implies
that $\jmath$ would deviate by about $-2\%$ or even as low as $-0.5\%$
from NT in the limit $|h/r|\to 0$ if they were to consider a 4-loop field geometry
and focus only on the disk gas.
Thus, their models show no statistically significant
large deviations from NT.
In addition, our results in section~\ref{scalinglaws}
show that, whether we consider
an integral over all angles or only over the disk, there is
no statistically significant large deviation from NT as $|h/r|\to 0$.

In summary, we conclude that the apparent differences between
the results obtained in S08 and the present paper on the one hand,
and those reported
in N09 and N10 on the other, are due to
1) dependence on initial magnetic field geometry (multi-loop vs 1-loop);
2) dependence upon the initial magnetic field distribution and normalization;
and 3) measurement and interpretation differences
(disk vs. corona-wind-jet).
Note in particular that the 1-loop initial field geometry is
severely squashed in the vertical direction and elongated
in the radial direction for thin disks,
and it is not clear that such a geometry would ever arise naturally.
There are also indications from our simulation that the 1-loop geometry
may actually never reach a converged state due to the arbitrary
amount of magnetic flux accreted onto the black hole
due to the single polarity
of the initial magnetic field.
Finally, if one is trying to test how well
simulated thin accretion disks compare with NT,
then it is important to restrict the comparison to disk
gas near the midplane.
One should not expect the gas in the corona-wind-jet
to agree with the NT model.

\section{Discussion}
\label{sec:discussion}

We now discuss some important consequences of our results
and also consider issues to be addressed by future calculations.
First, we discuss the relevance to black hole spin measurements.

In recent years, black hole spin parameters have been measured in
several black hole x-ray binaries by fitting their x-ray continuum
spectra in the thermal (or high-soft) spectral
state \citep{zcc97,shafee06,mcc06,ddb06,liu08,gmlnsrodes09,gmsncbo10}.
This method is based on several
assumptions that require testing \citep{nms2008proc},
the most critical being the assumption
that an accretion disk in the radiatively-efficient thermal state is
well-described by the Novikov-Thorne model of a thin disk.  More
specifically, in analyzing and fitting the spectral data, it is
assumed that the radial profile of the radiative flux, or equivalently the
effective temperature, in the accretion disk
closely follows the prediction of the NT model.

Practitioners of the continuum-fitting method generally restrict their
attention to relatively low-luminosity systems below $30\%$ of the
Eddington luminosity.  At these luminosities, the maximum height of
the disk photosphere above the midplane is less than $10\%$ of the
radius, i.e., $(h/r)_{\rm photosphere} \leq 0.1$ \citep{mcc06}.
For a typical disk, the photospheric disk thickness is approximately
twice the mean absolute thickness $|h/r|$ that we consider in
this paper.  Therefore, the disks that observers consider for
spin measurement have $|h/r| \lesssim 0.05$, i.e., they are thinner
than the thinnest disk ($|h/r|_{\rm min} \sim 0.06$) that we (S08,
this paper) and others (N09, N10) have simulated.

The critical question then is the following: Do the flux profiles of
very thin disks match the NT prediction?  At large radii the two will
obviously match very well since the flux profile is determined simply
by energy conservation\footnote{This is why the formula for the flux as a function
of radius in the standard thin disk model does not depend on details like the
viscosity parameter $\alpha$ \citep{fkr92}.}.  However,
in the region near and inside the ISCO, analytic models have to apply a boundary
condition, and the calculated flux profile in the inner region of the disk
depends on this choice.  The conventional choice is
a ``zero-torque'' boundary condition at the ISCO.  Unfortunately, there is disagreement on
the validity of this assumption.  Some authors have argued that
the magnetic field strongly modifies the zero-torque condition and that,
therefore, real disks might behave very differently from the NT model near the ISCO
\citep{krolik99,gammie99}.  Other authors, based either on heuristic arguments or on
hydrodynamic calculations, find that the NT model is accurate even near
the ISCO so long as the disk is geometrically thin \citep{pac00,ap03,shafee08,abr10}.
Investigating this question was the primary
motivation behind the present study.

We described in this paper GRMHD simulations of geometrically thin
($|h/r|\sim0.07$) accretion disks around black holes with a range of
spins: $a/M=0, ~0.7, ~0.9, ~0.98$.  In all cases, we find that the
specific angular momentum $\jmath$ of the accreted gas as measured at
the horizon (this quantity provides information on the dynamical
importance of torques at the ISCO) shows only minor deviations at the
level of $\sim 2\%$--$4\%$ from the NT model.  Similarly, the
luminosity emitted inside the ISCO is only $\sim 3\%$--$7\%$ of the
total disk luminosity.  When we allow for the fact that a large fraction of this
radiation will probably be lost into the black hole because of
relativistic beaming as the gas plunges inward (an effect ignored in
our luminosity estimates), we conclude that the region inside
the ISCO is likely to be quite unimportant.  Furthermore, our
investigations indicate that
deviations from the NT model decrease with decreasing $|h/r|$.
Therefore, since the disks of interest to observers are
generally thinner than the thinnest disks we have simulated, the NT model
appears to be an excellent approximation for modeling the spectra of
black hole disks in the thermal state.

One caveat needs to be mentioned.  Whether or not the total luminosity
of the disk agrees with the NT model is not important since, in
spectral modeling of data, one invariably fits a normalization (e.g.,
the accretion rate $\dot{M}$ in the model KERRBB; \citealt{lznm05})
which absorbs any deviations in this quantity.  What is important is the
{\it shape} of the flux profile versus radius.  In particular, one is
interested in the radius at which the flux or effective temperature is
maximum \citep{nms2008proc,mnglps09}.  Qualitatively, one imagines that the
fractional  shift in this radius
will be of order the
fractional torque at the ISCO, which is likely to be of order the
fractional error in $\jmath$.  We thus guess that, in the systems
of interest, the shift is nearly always below $10\%$.  We plan to
explore this question quantitatively in a future study.

Another issue is the role of the initial magnetic field topology.  We
find that, for $a/M=0$, starting with a 1-loop field geometry gives an
absolute relative deviation in $\jmath$ of $7.1\%$, and an excess luminosity
inside the ISCO of $4.9\%$, compared to $2.9\%$ and $3.5\%$ for our
standard 4-loop geometry.  Thus, having a magnetic field distribution
with long-range correlation in the radial direction seems to increase
deviations from the NT model, though even the larger effects we find in this case are
probably not a serious concern for black hole spin measurement.  Two
comments are in order on this issue.  First, the 4-loop geometry is
more consistent with nearly isotropic turbulence in the poloidal plane
and, therefore, in our view a more natural initial condition.
Second, the 1-loop model develops a stronger field inside the ISCO and
around the black hole and might therefore be expected to produce a
relativistic jet with measurable radio emission.  However, it is
well-known that black hole x-ray binaries in the thermal state have no
detectable radio emission.  This suggests that the magnetic field is
probably weak, i.e., more consistent with our 4-loop
geometry.

Next, we discuss the role of electromagnetic stresses on the
dynamics of the gas in the plunging region inside the ISCO.
In order to better understand this issue, we have
extracted for each of our simulations the radial profile of the
specific magnetic flux, $\Upsilon$.  This quantity appears
as a dimensionless free parameter (called $F_{\theta\phi}$) in the
simple MHD model of the plunging region developed by \citet{gammie99}.
The virtues of the specific magnetic flux are its well-defined
normalization and its constancy with radius for stationary flows
\citep{tntt90}. In
contrast, quantities like the stress $W$ or the viscosity
parameter $\alpha$ have no well-defined normalization; $W$ can be
normalized by any quantity that has an energy scale, such as $\rho_0$, $\dot{M}$,
or $b^2$, while $\alpha$ could be defined with respect to the total pressure, the gas
pressure, or the magnetic pressure.  The numerical values of $W$ or
$\alpha$ inside the ISCO can thus vary widely, depending on which
definition one chooses.  For instance, although S08 found $\alpha\sim
1$ inside the ISCO, the specific angular momentum flux, $\jmath$,
deviated from NT by no more than a few percent.
Further, Figure~\ref{loop1manystress} shows that (even for the multi-loop model)
the stress appears quite large within the ISCO,
but this is misleading because the effects of the stress
are manifested in the specific angular momentum, specific energy,
and luminosity -- all of which agree with NT to within less than $10\%$
for the multi-loop model.
Since $W$ and $\alpha$ do not have a single value within the ISCO
or a unique normalization,
we conclude that they are not useful
for readily quantifying the effects of the electromagnetic stresses within the ISCO.

Gammie's (1999) model shows how the value of $\Upsilon$
is directly related to the electromagnetic stresses within the ISCO.
Unfortunately, the actual value of $\Upsilon$ is a free parameter
which cannot be easily determined from first principles.
It is possible that accretion disks might have $\Upsilon\gg 1$,
in which case, the model predicts large deviations from NT.
For example, if $\Upsilon=6$, then for an $a/M=0$ black hole
$\jmath$ is lowered by $56\%$ relative to the NT model.
We have used our 3D GRMHD simulations which include self-consistent MRI-driven turbulence
to determine the value of $\Upsilon$ for
various black hole spins, disk thicknesses, and field geometries.
For the multiple-loop field geometry, we find that
the specific magnetic flux varies with disk thickness and spin as
\begin{equation}
\Upsilon\approx 0.7 + \left|\frac{h}{r}\right| - 0.6\frac{a}{M} ,
\end{equation}
within the disk component,
which indicates that electromagnetic stresses are weak
and cause less than $8\%$ deviations in $\jmath$
in the limit $|h/r|\to 0$ for all black hole spins.
Our rough analytical arguments for how $\Upsilon$ should scale
with $|h/r|$ and $a/M$ are consistent with the above formula.
Even for the 1-loop field geometry, $\Upsilon\lesssim 1$ for thin disks,
so electromagnetic stresses cause only minor deviations from NT
for all black hole spins (for $\Upsilon\lesssim 1$, less than $12\%$ in $\jmath$).
Not all aspects of the \citet{gammie99} model agree with our simulations.
As found in \citet{mg04},
the nominal efficiency, $\eff$, does not match well
and for thin disks is quite close to NT.
Since the true radiative efficiency is limited to no more than $\eff$,
this predicts only weak deviations from NT in the total luminosity
even if $\jmath$ has non-negligible deviations from NT.
Also, this highlights that the deviations from NT in $\jmath$ are due to non-dissipated
electromagnetic stresses and cannot be used to directly predict the excess luminosity within the ISCO.
The assumption of a radial flow in a split-monopole field
is approximately valid, but the simulations do show some non-radial
flow and vertical stratification, a non-zero radial velocity at the
ISCO, and thermal energy densities comparable to magnetic energy
densities.
Inclusion of these effects is required
for better consistency with simulation results inside the ISCO.

Next, we consider how our results lend some insight into the spin evolution of black holes.
Standard thin disk theory with photon capture predicts that
an accreting black hole spins up until it reaches
spin equilibrium at $a_{\rm eq}/M\approx 0.998$ \citep{thorne74}.
On the other hand, thick non-radiative accretion flows
deviate significantly from NT and reach equilibrium at
$a_{\rm eq}/M\sim 0.8$ for a model with $\alpha\sim 0.3$
and $|h/r|\sim 0.4$ near the horizon \citep{pg98}.
GRMHD simulations of moderately thick
($|h/r|\sim 0.2$--$0.25$) magnetized accretion flows
give $a_{\rm eq}/M\approx 0.9$ \citep{gammie_bh_spin_evolution_2004}.
In this paper, we find from our multi-loop field geometry models
that spin equilibrium scales as
\begin{equation}
\frac{a_{\rm eq}}{M} \approx 1.1 - 0.8\left|\frac{h}{r}\right| ,
\end{equation}
where one should set $a_{\rm eq}/M=1$ if the above formula gives $a_{\rm eq}/M>1$.
This gives a result consistent with the above-mentioned studies of thick disks,
and it is also consistent with our rough analytical prediction
based upon our scaling of $\Upsilon$ and using the Gammie model
prediction for the spin equilibrium.
This result also agrees with the NT result in the limit $|h/r|\to 0$
within our statistical errors,
and shows that magnetized
thin disks can approach the theoretical limit of $a_{\rm eq}/M\approx 1$,
at least in the multi-loop case.
In the single-loop field geometry, because of the presence of
a more radially-elongated initial poloidal field,
we find a slightly stronger torque on the black hole.
However, before a time of order $17000M$,
the deviations in the equilibrium spin parameter, $a_{\rm eq}/M$,
between the 4-loop and 1-loop field geometries appear to be
minor, so during that time the scaling given above roughly holds.
Of course, it is possible (even likely) that
radically different field geometries or anomalously
large initial field strengths will lead
to a different scaling.

Lastly, we mention a number of issues which we have neglected but are
potentially important.  A tilt between the angular momentum vector of
the disk and the black hole rotation axis might significantly affect
the accretion flow \citep{fragile07}.  We have not accounted for any
radiative transfer physics, nor have we attempted to trace photon
trajectories (see, e.g. N09 and \citealt{noblekrolik09}).
In principle one may require the simulation to be evolved for
hundreds of orbital times at a given radius in order to completely
erase the initial conditions \citep{sorathia10},
whereas we only run the model for a couple of viscous time scales.
New pseudo-Newtonian simulations show that convergence may require
resolving several disk scale heights with high resolution
\citep{sorathia10}, and
a similar result has been found also
for shearing box calculations with no net flux and no stratification
(Stone 2010, private communication).  In contrast,
we resolve only a couple of scale heights.
Also, we have only studied two different types of initial field geometries.
Future studies should consider whether alternative field geometries
change our results.

\section{Conclusions}
\label{sec:conclusions}

We set out in this study to test the standard model of thin accretion
disks around rotating black holes as developed by \citet{nt73}.  We
studied a range of disk thicknesses and black hole spins and found
that magnetized disks are consistent with NT to within a few percent
when the disk thickness $|h/r|\lesssim 0.07$.  In addition, we noted
that deviations from the NT model decrease as $|h/r|$ goes down.
These results suggest that black
hole spin measurements via the x-ray continuum-fitting method
\citep{zcc97,shafee06,mcc06,ddb06,liu08,gmlnsrodes09,gmsncbo10}, which
are based on the NT model, are robust to model uncertainties so long
as $|h/r|\lesssim 0.07$.  At luminosities below $30\%$ of Eddington,
we estimate disk thicknesses to be $|h/r|\lesssim0.05$, so the NT
model is perfectly adequate.

These results were obtained by performing global 3D GRMHD simulations
of accreting black holes with a variety of disk thicknesses, black
hole spins, and initial magnetic field geometries in order to test how these
affect the accretion disk structure, angular momentum transport,
luminosity, and the saturated magnetic field.  We explicitly tested
the convergence of our numerical models by considering a range of
resolutions, box sizes, and amplitude of initial perturbations.
As with all numerical studies, future calculations should continue to clarify what
aspects of such simulations are converged by performing more parameter
space studies and running the simulations at much higher resolutions.
For example, it is possible that models with different black hole
spins require more or less resolution than the $a=0$ models,
while we fixed the resolution for all models and only tested convergence
for the $a=0$ models.

We confirmed previous results by S08 for a non-spinning ($a/M=0$) black
hole, which showed that thin ($|h/r|\lesssim 0.07$) disks initialized
with multiple poloidal field loops agree well with the NT
solution once they reach steady state.  For the fiducial model
described in the present paper, which has similar parameters as the
S08 model, we find $2.9\%$ relative deviation in the specific angular
momentum accreted through the disk, and $3.5\%$ excess luminosity from
inside the ISCO.  Across all black hole spins that we have considered,
viz., $a/M=0, ~0.7, ~0.9, ~0.98$, the relative deviation from NT in
the specific angular momentum is less than $4.5\%$, and the luminosity
from inside the ISCO is less than $7\%$ (typically smaller, and
much of it is likely lost to the hole).  In addition, all
deviations from NT appear to be roughly proportional to $|h/r|$.

We found that the assumed initial field geometry modifies the
accretion flow.  We investigated this effect by considering two
different field geometries and quantified it by measuring the specific
magnetic flux, $\Upsilon$, which is an ideal MHD invariant (like the specific
angular momentum or specific energy).  The specific magnetic flux can
be written as a dimensionless free parameter that enters the
magnetized thin disk model of \citet{gammie99}.  This
parameter determines how much the flow deviates from NT as a result of
electromagnetic stresses.
We found that $\Upsilon$ allows a quantitative understanding
of the flow within the ISCO, while the electromagnetic stress ($W$)
has no well-defined normalization and varies widely within the ISCO.
While a plot of the stress may appear to show large stresses
within the ISCO, the actual deviations from NT can be small.
This demonstrates that simply plotting $W$ is not a useful diagnostic
for measuring deviations from NT.
We found that the specific magnetic flux of the
gas inside the ISCO was substantially larger when we used a single
poloidal magnetic loop to initialize the simulation compared to our
fiducial 4-loop run.  For $a/M=0$ and $|h/r|\lesssim 0.07$, the
early saturated phase (times $12900M$ to $17300M$)
of the evolution for the 1-loop
geometry gave $5.6\%$ relative deviation in the specific angular
momentum and $5.8\%$ excess luminosity inside the ISCO.  These
deviations are approximately twice as large as the ones we found for
the 4-loop simulation.
At late times, the 1-loop model generates significant
deviations from NT, which is a result similar to that found
in a vertical field model in \citet{mg04}.
However, we argued that the multiple loop geometry we used
is more natural than the single loop geometry, since
for a geometrically thin disk the magnetic field in the 1-loop model
is severely squashed vertically and highly elongated radially.
The 1-loop model is also likely to produce a strong radio jet.

More significant deviations from NT probably
occur for disks with strong ordered magnetic field, as found in 2D
GRMHD simulations by \citet{mg04}.  Of course, in the limit that the
magnetic field energy density near the black hole exceeds the local
rest-mass density, a force-free magnetosphere will develop and
deviations from the NT model will become extreme.
We argued that this corresponds to when the specific magnetic flux
$\Upsilon\gtrsim 1$ near the disk midplane.
Our 1-loop model appears to be entering such a phase
at late time after accumulation of a significant amount of magnetic flux.
Such situations likely produce powerful jets that are
not observed in black hole x-ray binaries in the thermal state.
However, transition between the thermal state
and other states with a strong power-law component \citep{fend04a,remm06}
may be partially controlled by the accumulation of magnetic flux
causing the disk midplane (or perhaps just the corona)
to breach the $\Upsilon\sim 1$ barrier.
Such a behavior has been studied in the non-relativistic regime \citep{nia03,ina03,igu09},
but more work using GRMHD simulations is required to validate the behavior.

We also found that the apparently different results obtained
by N10 were mostly due to measurement and interpretation differences.
We found that both the disk and the corona-wind-jet contribute nearly
equally to deviations in the total specific angular momentum relative
to the NT model.  However, the corona-wind-jet
contributes much less to the luminosity than the disk component.
Therefore, if one is interested in comparing the luminous portion
of the disk in the simulations against the NT model,
the only fair procedure is to consider only the disk gas,
i.e., gas within a couple of scale heights of the midplane.  This is
the approach we took in this study (also in S08).  N10 on the other
hand included the corona-wind-jet gas in their calculation of the specific angular
momentum.  The dynamics of
the coronal gas differs considerably from the NT model.  Therefore,
while it does not contribute to the luminosity of bound gas,
it doubles the deviation of the specific angular momentum from the NT model.
In addition, N10 used a 1-loop initial field geometry for their work which,
as discussed above, further enhanced deviations.

\section*{Acknowledgments}

We thank Phil Armitage, the referee, for
comments that greatly improved the paper's presentation,
and we thank Scott Noble for detailed discussions
about his work.
We thank Charles Gammie, Chris Reynolds, Jim Stone,
Kris Beckwith, John Hawley, Julian Krolik,
Chris Done, Chris Fragile, Martin Pessah, and Niayesh Afshordi
for useful discussions.
This work was supported in part by NASA grant
NNX08AH32G (AT \& RN), NSF grant AST-0805832 (AT \& RN),
NASA Chandra Fellowship PF7-80048 (JCM),
an NSF Graduate Research Fellowship (RFP),
and by the NSF through TeraGrid resources provided by
NCSA (Abe), LONI (QueenBee), NICS (Kraken)
under grant numbers TG-AST080025N and TG-AST080026N.

\appendix

\section{Example Solutions and Scalings for the Gammie (1999) Model}
\label{sec_gammie}

Table~\ref{tbl_gammie} gives representative solutions for the \citet{gammie99} model
of a magnetized thin accretion flow.
The columns correspond to the black hole spin, $a$;
the specific magnetic flux, $\Upsilon$;
the nominal efficiency, $\eff$;
percent deviation of $\eff$ from the NT value;
the specific angular momentum, $\jmath$;
percent deviation of $\jmath$ from NT;
and the normalized rate of change of the dimensionless black hole spin, $s$ (see Eq.~\ref{spinevolve}).
For $\Upsilon\lesssim 0.5$ and across all black hole spins,
the relative change in the specific angular momentum is less than $5\%$
and the relative change in the efficiency is less than $9\%$.
For small values of  $\Upsilon\lesssim 1$, the
deviations of $\jmath$ and $\eff$
from NT behave systematically and one can derive simple fitting functions.
For $\jmath$ we find
\begin{eqnarray}
&{}& \log_{10}\left[-D[\jmath]\right] \nonumber \\
&\approx& 0.79 + 0.37 (a/M) + 1.60 \log_{10}\Upsilon \\
&\sim& (4/5) + (1/3)(a/M) + (8/5)\log_{10}\Upsilon  ,
\end{eqnarray}
with an L2 error norm of $0.7\%,0.7\%$, respectively,
for the first and second relations, while for
$\eff$ we find
\begin{eqnarray}
&{}&\log_{10}\left[D[\eff]\right]\nonumber \\
&\approx& 1.44 + 0.12 (a/M) + 1.60 \log_{10}\Upsilon \\
&\sim& (3/2) + (1/10)(a/M) + (8/5)\log_{10}\Upsilon ,
\end{eqnarray}
with an L2 error norm of $0.9\%,1\%$, respectively,
for the first and second relations.
These results indicate that the deviations from NT scale as $\Upsilon^{8/5}$
for $\Upsilon\lesssim 1$.  For $\Upsilon\gtrsim 1$,
the index on $\Upsilon$ depends on the spin parameter.
In the span from $\Upsilon\sim 0.2$ to $\Upsilon\sim 1$,
a linear fit across all black hole spins
gives $-D[\jmath]\sim -1+11\Upsilon$ and $D[\eff]\sim -4+33\Upsilon$,
which are rough, though reasonable looking, fits.

\begin{table}
\caption{Thin Magnetized Inflow Solutions}
{\small
\begin{center}
\begin{tabular}{lllllll}
\hline
$\frac{a}{M}$ & $\Upsilon$  & $\eff$ & $D[\eff]$ & $\jmath$ & $D[\jmath]$ & $s$  \\
\hline
      0 &    0.1 & 0.0576 &  0.709 &   3.46 &    -0.172 &      3.46 \\
      0 &    0.2 & 0.0584 &   2.14 &   3.45 &     -0.52 &      3.45 \\
      0 &    0.3 & 0.0595 &   4.08 &   3.43 &    -0.991 &      3.43 \\
      0 &    0.5 & 0.0624 &   9.17 &   3.39 &     -2.23 &      3.39 \\
      0 &      1 & 0.0727 &   27.1 &   3.24 &     -6.57 &      3.24 \\
      0 &    1.5 & 0.0859 &   50.2 &   3.04 &     -12.2 &      3.04 \\
      0 &      6 &   0.19 &    232 &   1.51 &     -56.4 &      1.51 \\
    0.7 &    0.1 &  0.105 &   1.03 &   2.58 &    -0.286 &      1.33 \\
    0.7 &    0.2 &  0.107 &   3.07 &   2.56 &    -0.853 &      1.31 \\
    0.7 &    0.3 &   0.11 &    5.8 &   2.54 &     -1.61 &       1.3 \\
    0.7 &    0.5 &  0.117 &   12.8 &   2.49 &     -3.57 &      1.26 \\
    0.7 &      1 &  0.142 &   36.7 &   2.32 &     -10.2 &      1.12 \\
    0.7 &    1.5 &  0.172 &   66.3 &   2.11 &     -18.5 &      0.95 \\
    0.7 &      6 &  0.477 &    360 & -0.00714 &      -100 &     -0.74 \\
    0.9 &    0.1 &  0.157 &   1.17 &   2.09 &    -0.386 &     0.576 \\
    0.9 &    0.2 &  0.161 &   3.37 &   2.08 &     -1.11 &     0.567 \\
    0.9 &    0.3 &  0.165 &   6.29 &   2.06 &     -2.07 &     0.555 \\
    0.9 &    0.5 &  0.177 &   13.7 &   2.01 &      -4.5 &     0.524 \\
    0.9 &      1 &  0.215 &   38.3 &   1.84 &     -12.6 &     0.423 \\
    0.9 &    1.5 &  0.262 &   68.3 &   1.63 &     -22.5 &       0.3 \\
    0.9 &      6 &  0.845 &    443 & -0.958 &      -146 &     -1.24 \\
   0.98 &    0.1 &  0.236 &  0.949 &   1.68 &    -0.397 &     0.179 \\
   0.98 &    0.2 &  0.241 &   2.86 &   1.66 &      -1.2 &     0.174 \\
   0.98 &    0.3 &  0.246 &   5.36 &   1.64 &     -2.25 &     0.168 \\
   0.98 &    0.5 &  0.261 &   11.6 &    1.6 &      -4.9 &     0.152 \\
   0.98 &      1 &  0.309 &   32.2 &   1.45 &     -13.6 &       0.1 \\
   0.98 &    1.5 &  0.368 &   57.1 &   1.28 &     -24.1 &    0.0379 \\
   0.98 &      6 &   1.21 &    416 &  -1.27 &      -175 &    -0.862 \\
  0.998 &    0.1 &  0.319 &  -0.63 &    1.4 &     0.344 &    0.0374 \\
  0.998 &    0.2 &  0.327 &   2.02 &   1.38 &     -1.11 &    0.0342 \\
  0.998 &    0.3 &  0.332 &   3.66 &   1.36 &        -2 &    0.0322 \\
  0.998 &    0.5 &  0.345 &   7.73 &   1.33 &     -4.22 &    0.0273 \\
  0.998 &      1 &  0.388 &   20.9 &   1.23 &     -11.4 &    0.0113 \\
  0.998 &    1.5 &  0.439 &     37 &   1.11 &     -20.2 &  -0.00819 \\
  0.998 &      6 &   1.19 &    272 & -0.675 &      -148 &    -0.292 \\
\hline
\end{tabular}
\end{center}
}
\label{tbl_gammie}
\end{table}

\section{Inflow Equilibrium Timescale in the Novikov-Thorne Model}
\label{sec_inflow}

The radius out to which inflow equilibrium is achieved in a given
time may be estimated by calculating the mean radial velocity $v_r$
and then deriving from it a viscous timescale $-r/v_r$.

When the flow has achieved steady state, the accretion rate,
\begin{equation}\label{eq:mdot}
\dot{M}=-2\pi r\Sigma v_r \mathcal{D}^{1/2},
\end{equation}
is a constant independent of time and position.  Here we derive an
expression for $v_r$ corresponding to the general relativistic NT thin
disk model.  In what follows, capital script letters denote standard
functions of $r$ and $a$ (c.f. eqns. (14) and (35) in \citet{pt74})
which appear as relativistic corrections in otherwise Newtonian
expressions.  They reduce to unity in the limit
$r/M\rightarrow\infty$.

The vertically-integrated surface density may be defined as $\Sigma=2
h\rho$, where $h$ is the disk scale-height and $\rho$ is the rest-mass
density at the midplane.  In equilibrium, density is related to
pressure by
\begin{align}
\frac{dp}{dz}&=\rho \times (\mbox{``acceleration of gravity''})\\
             &=\rho\frac{Mz}{r^3}\frac{\mathcal{B}^2\mathcal{D}\mathcal{E}}{\mathcal{A}^2\mathcal{C}},
\end{align}
the vertically-integrated solution of which is
\begin{equation}
h=(p/\rho)^{1/2}/|\Omega| \mathcal{A}\mathcal{B}^{-1}\mathcal{C}^{1/2}\mathcal{D}^{-1/2}\mathcal{E}^{-1/2}.
\end{equation}
The pressure may be parameterized in terms of the viscous stress,
$|t_{\hat{r}\hat{\phi}}|=\alpha p$, which is a known function of $r$ and
$a$:
\begin{equation}
W=2ht_{\hat{r}\hat{\phi}}=\frac{\dot{M}}{2\pi}\Omega\frac{\mathcal{C}^{1/2}\mathcal{Q}}{\mathcal{B}\mathcal{D}}.
\end{equation}
The surface density is then
\begin{equation}
\Sigma=\frac{1}{2\pi}\frac{\dot{M}}{\alpha h^2|\Omega|}\mathcal{A}^{2}\mathcal{B}^{-3}\mathcal{C}^{3/2}\mathcal{D}^{-2}\mathcal{E}^{-1}\mathcal{Q}.
\end{equation}
Substituting this in Eq. (\ref{eq:mdot}), the radial velocity is
\begin{equation}\label{eq:inflow}
v_r=-\alpha|h/r|^2|\Omega| r \mathcal{A}^{-2}\mathcal{B}^{3}\mathcal{C}^{-3/2}\mathcal{D}^{3/2}\mathcal{E}\mathcal{Q}^{-1}.
\end{equation}
This result is independent of the exact form of the pressure and
opacity and so is valid in all regions of the disk.  The inflow
equilibrium time may be estimated as $t_{\rm ie}\sim
-2r/v_r$.

\bibliographystyle{mnras}
\bibliography{mybib}

\label{lastpage}
\end{document}